\documentclass[aps,prb,twocolumn,amsmath,amssymb,nofootinbib,superscriptaddress,floatfix]{revtex4-1}
\usepackage{amssymb}
\usepackage[hypertex]{hyperref}

\usepackage{mathrsfs}

\begin{document}

\title{
  Symmetry-protected Topological Phases, Generalized Laughlin Argument and Orientifolds
   }

\author{Chang-Tse Hsieh} 
 \affiliation{
 Department of Physics, University of Illinois
 at Urbana-Champaign, 
 1110 West Green St, Urbana IL 61801
             }

\author{Olabode Mayodele Sule}
 \affiliation{
 Department of Physics, University of Illinois
 at Urbana-Champaign, 
 1110 West Green St, Urbana IL 61801
             }

\author{Gil Young Cho}
 \affiliation{
 Department of Physics, University of Illinois
 at Urbana-Champaign, 
 1110 West Green St, Urbana IL 61801
             }

 \author{Shinsei Ryu}
 \affiliation{
 Department of Physics, University of Illinois
 at Urbana-Champaign, 
 1110 West Green St, Urbana IL 61801
             }

 \author{Robert G. Leigh}
 \affiliation{
 Department of Physics, University of Illinois
 at Urbana-Champaign, 
 1110 West Green St, Urbana IL 61801
             }

\date{\today}

\begin{abstract}
We generalize Laughlin's flux insertion argument, 
originally discussed in the context of the quantum Hall effect,  
to topological phases protected by non-on-site unitary symmetries,
in particular by parity symmetry
or parity symmetry combined with an on-site unitary symmetry. 
As a model, we discuss fermionic or bosonic systems in two spatial 
dimensions with CP symmetry, 
which are, by the CPT theorem, related to time-reversal symmetric
topological insulators (e.g., the quantum spin Hall effect). 
In particular, we develop the stability/instability 
(or ``gappability''/``ingappablity'') criteria for   
non-chiral conformal field theories with parity symmetry
that may emerge as an edge state of a symmetry-protected topological phase. 
A necessary ingredient, as it turns out, is to consider 
the edge conformal field theories on unoriented surfaces, such as 
the Klein bottle, which arises naturally from
enforcing parity symmetry by a projection operation. 
\end{abstract}


\maketitle

\tableofcontents

\section{Introduction}
\label{Introduction}

One of the most fundamental and defining properties
of the quantum Hall effect (QHE)
is the charge pumping discussed in
Laughlin's thought experiment.
\cite{Laughlin1981, review_QHE}
This non-perturbative argument explains the extreme robustness of the QHE
against disorder and interactions.
In the language of quantum field theories,
the charge pumping in Laughlin's gauge argument is a manifestation of a {\it quantum anomaly}, {\it i.e.}, the breakdown of a classical symmetry caused by quantum effects. This  is an extreme case where quantum mechanical effects
completely betray our expectations from classical physics.
To be more precise, 
the quantum Hall state supports, in the presence of 
a boundary (an edge), a chiral edge state.
If we focus on an edge, 
the total charge is not conserved within the edge, 
{\it i.e.}, 
the U(1) symmetry associated with the particle number conservation is 
violated, 
as the charge leaks
into the bulk precisely because of the QHE. 
This well-known {\it bulk-boundary correspondence}
of the QHE relates the bulk topological properties and the gauge anomaly
(non-conservation of charge) at the edge.

The connection to a quantum anomaly
gives  the conceptual backbone of the QHE. 
In fact, 
it is desirable to connect topological phases of {\it any kind}, 
not just the QHE, 
to a quantum anomaly
for the following reasons. 
First, quantum anomalies often provide a way to detect a
non-trivial topological nature of the state 
({\it e.g.}, charge pumping in the QHE described above),
and hence gives an operational definition of a topological phase. 
Second, once a topological phase is characterized 
in terms of a quantum anomaly,  
it is most likely to be stable against interactions and disorder. 

There are, however, a class of topological phases that have emerged more recently and that are
less well understood from this perspective. 
These are the so-called symmetry-protected topological (SPT) phases. 
(For recent works on SPT phases and in particular 
their connection to anomalies, 
see, for example,  
Refs.\ \onlinecite{
HasanKane2010,
QiZhang2011,
HasanMoore2011,
Schnyder2008,
SRFLnewJphys, 
Kitaev2009, 
Pollmann2010, 
Chen2011, 
Schuch2011, 
Fidkowski2010, 
Fidkowski2011, 
Turner2011, 
Tang2012,
Chen2013,
Chen2012,
Gu2012,
Qi2013,
Ryu2012,
Yao2012,
GuLevin2013,
Lu2012,
RyuMooreLudwig2012, 
Ringel2012,
Koch-Janusz2013, 
Cappelli13,
Wang2013}.)
In typical situations, they do not have any topological properties in the absence of symmetry conditions such as time-reversal.
In particular, they are adiabatically connected to a topologically trivial state if there is no symmetry. In the
presence of a certain set of symmetries, however, 
SPT phases are sharply distinct from a trivial state and host a
number of interesting topological properties. 
Canonical examples of SPT 
phases include the Haldane phase realized in one-dimensional quantum spin chains, the
quantum spin Hall effect (QSHE) in two dimensions, 
and the three-dimensional time-reversal symmetric
topological insulator.
\cite{HasanKane2010,
QiZhang2011,
HasanMoore2011}

One of our main goals in this paper is to provide a general scheme 
that allows us to judge if a given phase is a SPT phase or not, 
or, to be more precise, to diagnose 
under which symmetry condition a given phase can possibly be a SPT phase. 
\footnote{
As a clarifying remark,
SPT phases are not topologically-ordered phases of matter in that they  
do not have defining properties of topological
order such as non-trivial ground state degeneracy, fractional statistics, etc. 
Nevertheless, such phases are topologically non-trivial
in the sense that they are not adiabatically
connected to a trivial phase. On the other hand,
there is a class of topologically-ordered phases
that have some interesting properties in the
presence of symmetries -- they are called
symmetry-enriched topological phases. 
The methodology proposed in this paper
(a generalization of Laughlin's argument) 
works both for SPT phases
and symmetry-enriched topological phases.
}
While a fairly general topological 
classification of non-interacting fermion systems 
is possible,
judging if a given state can adiabatically be deformable to
a trivial state of matter or not 
is in general a non-trivial question
in the presence of strong interactions.  
In fact,   
there are known examples where a would-be SPT phase, for which one can define a topological invariant of some 
sort at the level of single-particle wave functions, 
can actually be adiabatically connected to a topologically 
trivial phase once one includes interactions.
\cite{Fidkowski2010,Fidkowski2011,Turner2011,Qi2013, Ryu2012, Yao2012}

More specifically, 
we will focus on SPT/non-SPT phases in (2+1) dimensions,
and discuss ``gappability''/``ingappability'' of their 
(1+1)-dimensional [(1+1)D] edge states
in the presence of symmetry conditions. 
We consider a given (1+1)D gapless (conformal) field theory,
which may emerge as an edge state of a bulk theory,
and ask if its gapless nature can be protected by some symmetry
conditions. 
Once the ingappability of the edge state is established, 
the corresponding bulk theory cannot adiabatically 
be connected to a topologically trivial state that do not support a gapless edge theory --
the state in question is in a SPT phase protected by the symmetries. 
On the other hand, if the edge theory turns out to be gappable
in the presence of the symmetries, the bulk theory may be deformable 
to a topologically trivial state.

To diagnose gappability/ingappability of an edge theory,
a generalization of Laughlin's gauge argument was proposed 
in Ref.\ \onlinecite{Ryu2012}, in a way that can be applied to edge states of SPT phases.
The purpose of this paper is to extend the scheme proposed in Ref.\ \onlinecite{Ryu2012}
to 
study topological phases
protected by unitary {\it non-on-site} symmetries, {\it e.g.}, parity symmetries. 
(We will mainly be interested in unitary symmetries,
but {\it anti-unitary} symmetries such as time-reversal symmetry 
are also relevant to our discussion.)

A key ingredient of the strategy suggested 
in Ref.\ \onlinecite{Ryu2012}
is the strict enforcement of symmetry conditions 
by a projection operation, 
or, to be more precise, an ``orbifolding'' procedure 
in the edge conformal field theory (CFT).
An orbifold of a theory,
which is invariant under 
a global unitary on-site symmetry,  
is given by averaging the partition function 
over 
boundary conditions twisted by a group element in 
the symmetry group. 
Roughly speaking, 
this procedure removes states that are not invariant
under the symmetry group. 
One can then study an adiabatic evolution 
of the projected (``orbifolded'') partition function. 
For example, 
if a U(1) symmetry is conserved, 
as in Laughlin's original argument,
one can ask if the orbifolded partition function 
is invariant or not under 
a large U(1) gauge transformation. 
While an original non-projected (non-orbifolded) theory
may be anomaly free, 
once orbifolded, the edge theory may fail to perform an anomaly-free
adiabatic process. 
Non-invariance of the orbifolded edge theory
under a large U(1) gauge transformation
signals non-trivial topological properties of the corresponding
bulk state.
One can also ask, perhaps more fundamentally, 
the invariance/non-invariance of 
the orbifolded system under large coordinate transformations,
such as modular transformations 
on a space-time manifold with non-zero genus, {\it e.g.}, a torus.

This scheme is demonstrated to work for various examples.
\cite{Sule13, Levin2013}
A similar projection or ``gauging'' procedure is
also employed in Ref.\ \onlinecite{LevinGu12},
where a criterion 
for the ingappability/gappability of an edge theory 
is derived using fractional statistics of the defects obtained from gauging some unitary on-site global symmetry in a gapped bulk theory. 


In this paper, we extend the scheme proposed 
in Ref.\ \onlinecite{Ryu2012} to 
SPT phases that are protected by unitary 
spatial symmetries, in particular to parity symmetry
(denoted by $\mathscr{P}$ in the following). 
An interplay between spatial (and, more generally, crystal)
symmetries and topological properties of electronic states 
have been studied extensively recently. 
\cite{Turner11,Turner12,Hughes11,Fu11, 
Slager13,Fang12a,Fang12b,Hsieh12,Teo08, 
Tanaka12,Dziawa12,Chiu13,Morimoto13,Zhang13}
Following the general strategy described above,
we consider the orbifolding or gauging
procedure by a parity symmetry. 
Unlike orbifolding a unitary on-site symmetry,
orbifolding parity symmetry
naturally leads to a change of the topology of the spacetime manifold 
of the edge theory. 
Once orbifolding parity symmetry, 
an edge theory is defined 
on an {\it unoriented} (1+1)D spacetime surface,
such as the Klein bottle 
instead of a spacetime torus.  
\cite{Callan1987, 
PolchinskiCai1988, 
Angelantonj02}
We refer to these conformal field theories
as {\it orientifold}\cite{Sagnotti1988,Dai1989} conformal field theories.  
\footnote{
 For a connection between orientifolds and time-reversal symmetric
 topological insulators and superconductors from 
 the spacetime point of view (as opposed to the worldsheet point of view
 presented in this paper), see Refs.\ \onlinecite{RyuTakayanagi2010a, RyuTakayanagi2010b}. 
}

In the presence of yet another symmetry 
(represented by a symmetry group $\mathscr{G}$ --
the total symmetry group is $\mathscr{P} \rtimes \mathscr{G}$)
in addition to parity, 
there is a simple consequence of the topology change 
from the torus to the Klein bottle, 
which 
can be inferred by comparing their fundamental 
groups, {\it i.e.}, the space of non-contractible loops
on these surfaces. 
On the torus, there are two independent cycles and
one can assign a group element to each cycle,
{\it e.g.}, a U(1) phase factor; these two group elements
($g_1$ and $g_2$, say) 
represent boundary conditions along each cycle. 
On the Klein bottle, on the other hand,
because of its unoriented nature, 
there is a certain restriction on the group elements 
that one can assign for cycles; 
while one of the group elements, $g_1$, say, can be
any element in $\mathscr{G}$, the other group element $g_2$ 
must satisfy $g^{\ }_2= g^{-1}_2$ (see discussion near eq.\ (\ref{Klein part fn general}) below for more details).

In this work,
we focus on the cases where $\mathscr{G}$ is a U(1) symmetry
(either charge U(1) or ``spin'' U(1) symmetry).  
One of our main observations
is a crucial role played by the $\mathbb{Z}_2$ flux 
satisfying $(g_2)^2=1$, {\it i.e.}, 
along this cycle, the only allowed
boundary conditions are periodic or antiperiodic.  
While $g_1$ can be used as an adiabatic parameter 
that we can use as a ``knob'' to implement a Laughlin
argument ({\it i.e.}, an adiabatic evolution of the partition function
as we change $g_1 \in \mathscr{G}$), 
$g_2$ turns out to be fixed by the type of parity symmetry $\mathscr{P}$. 
We will show in this paper that 
these distinctions by $g_2$-flux are closely related to the
topological classification of parity symmetric systems.

While our methodology is applicable to 
a wider class of systems with parity symmetry, 
in this paper, 
we choose to work with 
systems with parity combined with charge conjugation 
symmetry (CP symmetry).
\cite{Hsieh13}
Specifically, we consider three examples: 
(i) fermionic systems with 
conserved charge U(1) symmetry
and CP symmetry 
(CP symmetric topological insulators);
\footnote{
The fermionic models with CP and charge U(1) symmetries 
can also be interpreted/realized as a BdG system with
parity and spin U(1) symmeries 
({\it e.g.}, $z$-component of spin, $S_z$, is conserved). }
(ii) bosonic systems with 
conserved charge U(1) and CP symmetries;
and 
(iii) K-matrix theory with CP and U(1) symmetries.   
We discuss the topological classification of these systems
by using the method of the generalized Laughlin argument. 

The systems with CP symmetry are our canonical examples
in the sense that they are closely related to SPT phases protected by time-reversal symmetry through the CPT theorem.
For systems with Lorentz invariance, the
CPT theorem tells us that 
any perturbation (mass terms and interactions)
prohibited by T (CP) symmetry
is also excluded by CP (T) symmetry. 
Thus, for Lorentz invariant systems,
SPTs protected by time-reversal 
are automatically protected by CP symmetry as well. 

For condensed matter systems, Lorentz invariance is not a prerequisite.
However, for non-interacting fermions,
it is known that the general topological classification can be
obtained solely from the topological classification of 
Lorentz invariant Dirac Hamiltonians. 
When available, topological field theory descriptions 
of topological phases are also Lorentz invariant. 
In addition, 
it is known that,
if one considers {\it the entanglement spectrum}
as a tool to study SPT phases,
CP symmetry of a physical Hamiltonian
is translated to an effective time-reversal symmetry 
of the corresponding entanglement Hamiltonian
if one bipartites the system into two subsystems 
that are related by CP symmetry. 
\cite{Turner11,Chang13,Hsieh13}
For these reasons, our method also provides a new insight  
into time-reversal symmetric topological systems, 
including the QSHE, 
by relating them to orientifold conformal field theories.
Thus we provide a method for ``twisting'' or ``gauging'' 
time-reversal symmetry. 
(See recent discussion in Refs.\ \onlinecite{Chen2014, Kapustin2014}.)

\subsection{The outline and main results of the paper}

The main results and outline of the paper can be summarized as follows: 
for the remainder of Sec.\ \ref{Introduction}, 
we will review 
gauge and chiral anomalies in (1+1) dimensions
and their connection to topological phases in (2+1) dimensions. 
In particular, we rephrase the original Laughlin argument in 
a quantum field theory language, which we will use  
for our later discussion.     

In Sec.\
\ref{2D fermionic topological phases protected by symmetries},
we begin our discussion by introducing 
a free fermion model
with CP and electromagnetic U(1) symmetries.
We consider two kinds of CP symmetries,
one which protects gapless edge states
and hence leads to a non-trivial bulk 
symmetry-protected topological phase, and the other
which does not lead to a topological insulator. 
An anomaly (``CP anomaly'') is identified in  
the non-chiral edge states of  
the CP-symmetric topological insulator. 
We then present, by using the CP-symmetric topological insulator
as an example, a generalization of Laughlin's argument 
to systems with parity symmetry (CP symmetry, in this case).  
By considering the partition function of the non-chiral edge
theory with CP projection, it is shown that the distinction between
the two cases shows up as the presence/absence of 
a $\mathbb{Z}_2$ flux on the Klein bottle
(``$g_2$'' in the above notation).  
Under an adiabatic insertion of electromagnetic U(1) flux 
(``$g_1$'' in the above notation),   
the projected partition function is anomalous/anomaly-free
when the edge theory is ingappable/gappable. 

In Sec.\ 
\ref{Bosonic CP symmetry protected topological phase}, 
the generalized Laughlin argument is applied to
bosonic SPT phases with a single-component non-chiral boson 
edge theory with CP symmetry.  
The results are consistent with 
microscopic stability analysis of CP symmetric edge theories
given in Ref.\ \onlinecite{Hsieh13}.  

In Sec.\
\ref{Symmetry protected topological phase: K-matrix theory}, 
we consider a broader range of edge theories
described by the K-matrix theory with CP symmetry.
With the generalized Laughlin argument,
we derive the stability criterion 
for the edge theories,
which agrees 
with the stability criterion of 
the K-matrix theory with time-reversal symmetry   
\cite{LevinStern2009, LevinStern2012, Neupert2011}
as expected from the CPT theorem.

We conclude in Sec.\ \ref{Discussion}. 
In Appendix \ref{Discussion on the CP eigenvalue of the ground state},
we discuss the eigenvalue of the CP transformation of the ground state
of edge CFTs, and in particular its evolution under 
an adiabatic evolution of the background flux.  
Once we choose to preserve the U(1) symmetry, 
the CP eigenvalue must be independent of the background flux, 
which we assume for the bulk of the paper. 
On the other hand, an alternative point of view is possible where
we strictly enforce the U(1) symmetry. 
Once this is done, CP symmetry may be anomalous, and hence 
the ground state CP eigenvalue may be dependent on the background flux. 
This issue is discussed  
in Appendix \ref{Discussion on the CP eigenvalue of the ground state}
by making use of the state-operator correspondence of CFTs. 
Appendix 
\ref{Statistical phase factor of the chiral boson field under symmetry transformation}
explains a technical detail
that arises when diagnosing the stability of  
K-matrix edge theories
in Sec.\ \ref{Symmetry protected topological phase: K-matrix theory}.

\subsection{The integer QHE and gauge anomaly}

For chiral topological phases in two spatial dimensions,
their edge states (which are chiral) are anomalous.
When there is the electromagnetic U(1) symmetry,  
the chiral edge states are anomalous under infinitesimal as well as large 
U(1) gauge transformations.
Even in the absence of the electromagnetic U(1) invariance,
the edge states are still anomalous under 
infinitesimal as well as large 
diffeomorphisms.

For later use, let us review the anomaly
under large U(1) gauge transformations at
the edge of the integer QHE.
(See, for example, 
Refs.\ 
\onlinecite{Cappelli01, Cappelli96, Cappelli09, Cappelli09b}, 
for discussion on the edge theory of various 
quantum Hall states).
(We will follow notations in Ref.\ \onlinecite{Mirror}.)
The chiral edge mode of the integer QHE is described by
the action
\begin{align}
 \label{chiral edge}
  S
  =\frac{1}{2\pi}
  \int dt dx\, 
  {i} \psi^{\dag}_R (\partial_t + \partial_x) \psi^{\ }_R, 
\end{align}
where $(t,x)$ is the spacetime coordinate of the edge theory,
and chirality is chosen, say, to arrive at the right-moving fermions. 

Following Laughlin's thought experiment,
we now insert magnetic flux into the system of a cylindrical shape.
In terms of the fermion field in the edge theory, 
this amounts to imposing  
the following twisted boundary conditions
both for space and time directions:
\begin{align}
 \label{twisting bc}
  \psi^{\ }_R (t, x+2\pi)
&=  
e^{2\pi {i} a}
\psi^{\ }_R (t, x),
\nonumber \\
\psi^{\ }_R (t+ 2\pi \tau_2, x+2\pi\tau_1)
&=  
e^{2\pi {i} b}
  \psi^{\ }_R (t, x).
\end{align}
Here the edge theory is defined on a spatial circle of radius $2\pi$,
and $\tau=\tau_1 + i\tau_2$ is the modular parameter 
of the spacetime torus. 
Under these boundary conditions,
the right-moving partition function is computed to be
\begin{align}
 \label{chiral part fn Dirac}
  Z_{[a,b]}(\tau)
  &=
 q^{-\frac{1}{24} +\frac{1}{2} (a-1/2)^2 }
 e^{-2\pi i (b-1/2) (a-1/2)}
 \nonumber \\
 &\quad
 \times 
 \prod_{n=0}^{\infty}
 \left[
 1+ e^{-2\pi i(b-1/2)} q^{n+a}
 \right]
 \nonumber \\
 &\quad \times 
 \prod_{n=-1}^{-\infty}
 \left[
 1+ e^{+2\pi i (b-1/2)} q^{-n-a}
 \right]
\nonumber \\
&=
  \frac{1}{\eta(\tau)}
  \vartheta
  \left[
    \begin{array}{c}
      a-1/2 \\
      -b+1/2
      \end{array}
    \right]
  (0,\tau), 
\end{align}
where $q=\exp (2\pi i \tau)$
and the theta function with
characteristics is defined by  
\begin{align}
 \label{theta fn}
\vartheta\left[
    \begin{matrix}
      \alpha\\
     \beta
    \end{matrix}
  \right] (\nu, \tau)\equiv 
  \sum_{n \in \mathbb{Z}}
  e^{\pi i\tau (n+\alpha)^2}e^{2\pi i(\nu+\beta)(n+\alpha)}.
\end{align}
While the classical theory, 
defined in terms of the action  
(\ref{chiral edge})
together with the boundary condition 
(\ref{twisting bc}),
is invariant under
large gauge transformations
$a\to a+1$
and
$b\to b+1$, 
the partition function violates this invariance:   
\begin{align}
  Z_{[a,b]}
  =
  Z_{[a+1,b]}
  =
  -e^{ 2\pi {i} a}
  Z_{[a,b+1]}
  =
  Z_{[-a,1-b]},  
\end{align}
and thus the edge theory is anomalous under this transformation. 
\footnote{
In the above calculations,
the invariance is violated only for 
the temporal boundary condition.
One in fact has a choice: 
by redefining $Z_{[a,b]}\to Z_{[a,b]}e^{-2\pi {i}ab}$, 
the partition function is now anomalous for 
spatial boundary condition.
Such multiplication of the phase 
is related to (re-) assignment the U(1) charge 
to the ground state. 
See later discussion for more details. 
}

\subsection{The $S_z$ conserving QSHE and chiral anomaly}

As yet another exercise,
let us consider a bulk topological insulator 
characterized by non-zero {\it spin} Chern-number. 
We require both charge U(1) and spin U(1) symmetries. 
The edge state, if it exists, also respects these symmetries, 
at least classically. 
However, either one of these U(1) symmetries must be 
spoiled by quantum mechanical effects.
Let us for now insist on  the conservation of electromagnetic U(1) charge. 
One can then introduce an electromagnetic vector potential $A$.
We then consider 
a non-chiral fermion coupled with the electromagnetic U(1) gauge field,
\begin{align}
S =
\int d^2 z\,
\big(
i \psi^{\dag}_R D_z \psi^{\ }_{R}
+
i \psi^{\dag}_{L} D_{\bar{z}} \psi^{\ }_{L}
\big), 
\end{align}
where $D_{z}$ is a covariant derivative with the electromagnetic U(1) 
gauge field $A$. 
As is well known,
the theory is not invariant under chiral gauge transformations,
which in the present context are gauge transformations
associated to the spin U(1) symmetry. 
This has to do with the presence of 
a non-trivial bulk topological phase 
protected by charge U(1) and spin U(1) symmetries.

The chiral anomaly comes about since 
the path integral measure is not invariant under 
chiral transformations. 
Let us consider the case where 
the Chern number
associated to the vector potential $A$ is non-zero:
\begin{align}
 {Ch}&=
 \frac{i}{2\pi} \int \mathrm{tr}\, F
>0, 
\end{align}
where $F$ is the field strength of the external U(1)
gauge potential $A$. 
Then, by the index theorem,
the number of
$\psi_L$ zero modes 
(= the number of $\psi^{\dag}_R$ zero modes)
is larger by ${Ch}$ than the number of
$\psi^{\dag}_L$ zero modes
(= the number of $\psi^{\ }_R$ zero modes). 
The path integral measure is given by
\begin{align}
 \mathcal{D}\left[\psi^{\dag},\psi\right]
 =
 \prod_{\alpha=1}^{{Ch}}
 da_{\alpha}
 da^*_{\alpha}
 \prod_{n=1}^{\infty}
 db^{\ }_n
 dc^{\ }_n
 db^*_n
 dc^*_n,
\end{align}
where 
 $
 \prod_{n=1}^{\infty}
 db^{\ }_n
 dc^{\ }_n
 db^*_n
 dc^*_n
 $
represents the ``oscillator'' part of the measure,
whereas 
$\prod_{\alpha=1}^{{Ch}}
 da_{\alpha}
 da^*_{\alpha}
 $
 represents the measure associated to the zero modes. 
While the measure
$
 db^{\ }_n
 dc^{\ }_n
 db^*_n
 dc^*_n$
is
invariant under both 
electromagnetic and spin $\mathrm{U}(1)$ global transformations,
the zero mode part
$da^{\ }_{\alpha} da^*_{\alpha}$
has electromagnetic (vector) charge zero,
but axial charge 2. 
Thus, the path integral measure is not invariant
under the axial (spin) U(1) rotation.
In fact, in the presence of 
nonzero flux with the Chern number
${Ch}$,
the axial U(1) is broken 
down to its $\mathbb{Z}_{2 {Ch}}$ subgroup. 

To summarize, once we demand the electromagnetic U(1) symmetry
to be preserved 
then, 
the chiral anomaly tells us that the spin U(1) [axial U(1)]
must be broken at the edge -- this is nothing but 
the QSHE, {\it i.e.}, the spin quantum number is pumped by an adiabatic threading
of the electromagnetic flux.

In fact, one has a choice -- 
if one decides to preserve spin U(1) symmetry, 
instead of charge U(1), 
one could thread ``spin flux'' and consider
the corresponding spin vector potential. 
Going through the above argument, 
one then concludes the charge is not conserved. 
This has to do with charge pumping by 
insertion of spin flux.

\section{2D fermionic topological phases protected by CP symmetry}
\label{2D fermionic topological phases protected by symmetries}

In this section, 
we describe our methodology (a generalization of Laughlin's argument)
in terms of a simple two-dimensional fermionic system
(although the method applies to a wider class of systems). 
The system of interest
conserves the electromagnetic U(1) charge 
and respects a discrete symmetry, CP,
that is a combination of  
parity, P: $(x,y) \rightarrow (-x, y)$ in two spatial dimensions, 
and charge conjugation, C, 
which is 
a unitary $Z_2$ on-site symmetry.

By the CPT theorem, 
the CP symmetric system (the CP symmetric topological insulator)
is related to the time-reversal symmetric topological insulator. 
(In fact, they are equivalent when there is Lorentz invariance.)
As two-dimensional insulators 
with time-reversal symmetry that squares to $-1$
are classified in terms of the Kane-Mele $\mathbb{Z}_2$ topological invariant, 
so are CP symmetric insulators. 
The CP symmetric fermionic system can also be interpreted as
a topological superconductor that conserves 
parity and 
the $z$-component of spin (this is an example of ``T-duality''). 
See Ref.\ \onlinecite{Hsieh13} for more details of 
superconducting systems equivalent (dual) to CP symmetric insulators.

\subsection{CP symmetric insulators}

\paragraph{The bulk tight-binding model}

A lattice model of the topological insulator 
with CP symmetry can be constructed on the 
two-dimensional square lattice by 
taking two copies of the above two-band Chern insulator 
with opposite chiralities.
Consider the Hamiltonian in momentum space, 
\begin{align}
H 
&=
\sum_{k\in \mathrm{BZ}}
\Psi^{\dag}(k)
\mathcal{H}(k)
\Psi(k), 
\label{chiral+R}
\end{align}
where 
$\Psi(k)$ is a four-component fermion field
with momentum $k$,
$\mathrm{BZ}$ represents the first Brillouin zone of the 
two-dimensional square lattice, 
and 
the single particle Hamiltonian in momentum 
space is given in terms of the $4\times 4$ matrix as, 
\begin{align}
\mathcal{H}
(k)
=
n_x(k) \tau_z \sigma_x
+
n_y(k) \tau_0 \sigma_y
+
n_z(k) \tau_0 \sigma_z,  
\end{align}
where $\sigma_{\mu}$ and $\tau_{\mu}$
$(\mu=0,1,2,3)$
are two sets of Pauli matrices
with $\sigma_{0}$ and $\tau_0$ being a $2\times 2$ unit matrix.  
The $k$-dependent three-component vector is given by 
\begin{align}
\vec{n}(k)
&=
\left[
\begin{array}{c}
-  \sin k_x \\
-  \sin k_y \\
 (\cos k_x + \cos k_y) + \mu
\end{array}
\right]. 
\end{align}
We will focus on the region
$-2<\mu<0$ or 
$0<\mu<+2$. 

The Hamiltonian is invariant under 
the following two CP transformations
\begin{align}
 (\mathcal{CP})
\Psi(r)
(\mathcal{CP})^{-1}=
U_{\mathrm{CP}} \Psi^{\dag}(\tilde{r}), 
\end{align}
where $r=(x,y)$ labels sites on the square lattice,
$\tilde{r}:=(-x, y)$ and
the $2\times 2$ unitary matrix $U_{\mathrm{CP}}$ is given by either of
\begin{align}
 U_{\mathrm{CP}}&=
\tau_x \sigma_x
\quad
U_{\mathrm{CP}}^T
=
+U_{\mathrm{CP}},  
\quad 
(\eta=+1),
\nonumber \\
U_{\mathrm{CP}}&=
\tau_y \sigma_x, 
\quad 
U_{\mathrm{CP}}^T
= 
-U_{\mathrm{CP}},  
\quad
(\eta=-1). 
\end{align}
To distinguish these two cases,
we have introduced an index $\eta$;
$\eta=\pm 1$ refers to the first/second case. 
We will also use the notation
$\eta = e^{2\pi i \epsilon}$
where 
$\epsilon=0, 1/2$
for 
$\eta = 1, -1$, 
respectively. 
The distinction between these two CP symmetries can be
summarized as
\begin{align}
 (\mathcal{CP})^2 = e^{2\pi i \epsilon N_f} 
\end{align}
where CP acts on states with 
$N_f$ fermions. 

It turns out that imposing $U_{\mathrm{CP}}=\tau_x\sigma_x$ ($\eta=1$) leads to
CP symmetric topological insulators. This can be seen by looking at the stability of the edge mode that can appear when we terminate the system
in the $y$-direction ({\it i.e.}, the edge is along the $x$-direction.)
One can check, numerically, and also in terms of the continuum edge theory
(see below), $U_{\mathrm{CP}}=\tau_x\sigma_x$
protects the edge state while $U_{\mathrm{CP}}=\tau_y\sigma_x$ 
does not in the presence of the electromagnetic U(1) symmetry.

\paragraph{The edge theory}

We now develop a continuum theory for the edge modes along the $x$-direction. 
The edge theory is described by, at low-energies,  
the free fermion Hamiltonian with relativistic dispersion:
\begin{align}
  H &=
  \frac{v}{2\pi}
  \int dx
  \big(
    \psi^{\dag}_L {i}\partial_x \psi^{\ }_L
     -\psi^{\dag}_R {i}\partial_x \psi^{\ }_R 
    \big),
\end{align}
where the single-component complex fermion field operators 
$\psi_L$ and $\psi_R$ represent  
left-moving and right-moving electrons,
and 
$v$ is the Fermi velocity.
The Hamiltonian conserves the U(1) charge 
\begin{align}
 F_V = \int dx\, \left[ 
  \psi^{\dag}_R \psi^{\ }_R
+
\psi^{\dag}_L \psi^{\ }_L  
 \right]. 
\end{align}
The subscript ``$V$'' here represents the fact that
this is a vector U(1) charge, as opposed to an axial U(1) charge,
\begin{align}
F_A = \int dx\, 
\left[
 \psi^{\dag}_R \psi^{\ }_R - \psi^{\dag}_L \psi^{\ }_L
\right]. 
\end{align}

Since the edge runs along $x$-direction,
the Hamiltonian with the edge preserves (is consistent with) CP symmetry, 
{\it i.e.}, CP transformation is closed within the edge. 
Corresponding to the bulk CP transformations,
we consider the following two types of CP symmetry operations
that act within the edge theory:\footnote{
In the presence of both U(1) and CP symmetries, by combining these two symmetries, one can generate a series of CP transformations, 
\cite{Weinberg}
$$
\begin{array}{lll}
\mathcal{U}_{\alpha} \psi^{\ }_L(x) 
\mathcal{U}^{-1}_{\alpha} &=& e^{i\alpha} \psi^{\dag}_{R}(-x), 
\\
\mathcal{U}_{\alpha} \psi^{\ }_R(x) 
\mathcal{U}^{-1}_{\alpha} &=& \eta e^{i\alpha} \psi^{\dag}_{L}(-x),  
\end{array}
$$
where
$\mathcal{U}_{\alpha} = e^{i \alpha F_V} \mathcal{CP}$. 
While these transformations are each qualified to be called
CP, the inclusion of such a U(1) phase factor does not play
any essential role in our discussion.
}
\begin{align}
 \label{CP action on fermions}
 (\mathcal{CP})
 \psi^{\ }_L(x)
 (\mathcal{CP})^{-1}
&=
\psi^{\dag}_{R}(-x), 
\nonumber \\
 (\mathcal{CP})
 \psi^{\ }_R(x)
(\mathcal{CP})^{-1}
&=
\eta \psi^{\dag}_{L}(-x). 
\end{align}
The sign $\eta$ is 
$+/-$ respectively for topological/non-topological cases;
there are two uniform fermion mass bilinears 
that are consistent with the charge U(1) symmetry, 
$M_1= \psi^{\dag}_{L} \psi^{\ }_R + \psi^{\dag}_{R}\psi^{\ }_L$,  
and 
$M_2 = -i\big(
\psi^{\dag}_{L} \psi^{\ }_R - \psi^{\dag}_{R}\psi^{\ }_L 
\big)$. 
These masses are odd under CP and prohibited when $\eta=+1$,
whereas they are even under CP with $\eta=-1$. 
Thus, 
\begin{align}
\eta &
= e^{2\pi i \epsilon}
=
\left\{
 \begin{array}{ll}
  +1& \mbox{"topological"} \\
  -1& \mbox{"trivial"} \\
 \end{array}
 \right.
\end{align}
We conclude that the gapless edge theory is, 
at least at the {\it quadratic level},
stable (ingappable); 
the stability/instability of the edge theory
in the presence of interactions is one of our main focuses in the
following sections.

Let us consider, in addition, 
quadratic but spatially inhomogeneous perturbations. 
The two uniform mass terms $M_1$ and $M_2$ are not allowed 
in the presence of CP symmetry with $\epsilon=0$. 
However, one can still consider
$\int dx\, M(x)$ where, 
$
 M (x)= a_1(x) M_1 + a_2(x) M_2,
$
as a perturbation to the edge theory. 
The perturbation is not allowed by CP symmetry if 
$a_1(x)$ is constant but allowed 
if 
 $\vec{a}(-x) = - \vec{a}(x)$. 
This perturbation gaps out most of the edge theory,
but not completely. At the point $x=0$ which is invariant under CP symmetry, it leaves a zero energy mode. This is a type of zero energy mode
akin to the zero energy bound state in a soliton in polyacetyline,
and carries 1/2 charge. This is also similar to the mass domain wall in the helical edge mode of the QSHE discussed previously. 
\cite{QiHughesZhang2007}
The difference, however, is that for the time-reversal symmetric 
quantum spin Hall effect, the mass domain wall breaks TRS;
the only exception being the location of the kink. 
In the CP symmetric case, the mass domain wall, as a whole,
preserves CP symmetry.

\subsection{CP anomaly}
\label{Generalized Laughlin's argument} 

Although the edge theory cannot be gapped at the quadratic level,
whether or not it is gappable under arbitrary symmetry-preserving
perturbations is not clear;
a state that appears to be non-deformable to a trivial state
may actually be deformable to a trivial state once one includes
perturbations beyond the quadratic level.
For the CP symmetric topological phase
defined above, 
we now try to develop
a generalized Laughlin argument. 
To put it differently, we will ask if there is a quantum anomaly
or not that may guarantee the gapless nature of the edge theory. 

Quite often, non-chiral theories are anomaly-free,
and hence are not qualified as a topological phase
without symmetry condition.
However, there may be a tension between
symmetry conditions and an attempt to make the theory
self-consistent (anomaly-free). 
To be more precise, if one insists on the symmetry conditions,
one might not be able to achieve anomaly-freedom. 
In the scheme proposed in Ref.\ \onlinecite{Ryu2012}, 
the symmetry conditions are strictly enforced by
the projection operations. Subsequently, we ask if the projected theory is anomaly-free or not.

We now show that there is indeed a tension between
CP and charge conservation symmetries
when $\epsilon=0$. That is, if one enforces one of them, 
the other will be violated by quantum effects. 
In the following, as a warm-up, 
we first enforce  charge conservation
and then in this case see that CP will be violated.
After that, we show that enforcing CP symmetry
will violate the charge conservation. 
The latter can be thought of as a generalization of 
Laughlin's argument to symmetry-protected topological phases
with charge U(1) symmetry. 

The QSHE described in the introduction is an SPT phase 
protected by on-site [$\mathrm{U}(1)\times \mathrm{U}(1)$] symmetry
and characterized by an integer topological invariant. 
We now move on to a symmetry-protected topological insulator
protected by non-on-site symmetry, 
and characterized by a $\mathbb{Z}_2$ topological invariant. 
Similar to the case of the $S_z$ conserving QSHE, 
we can identify an anomaly in the edge theory.  
It is the parity anomaly discussed in Ref.\ \onlinecite{Brunner2003}.

The argument goes as follows: as in the case of the chiral anomaly,
we consider a background field with non-zero Chern number
${Ch}$. When non-zero and positive, there are zero modes
in $\psi^{\ }_L$ and $\psi^{\dag}_R$, and the path-integral measure has a factor
\begin{align}
 \prod^{{Ch}}_{\alpha=1}
 d a^{\ }_{\alpha}
 d a^{*}_{\alpha}. 
\end{align}
Observe now that the Chern number ${Ch}$ flips its sign under parity, P. 
It also flips its sign under the charge conjugation, C.  
Thus, the Chern number remains invariant under the combination of CP. 
Let us first consider the case of $\eta=+1$,
where CP transformation is given by
$(\mathcal{CP})
 \psi^{\ }_L(x)
(\mathcal{CP})^{-1} 
=
 \psi^{\dag}_{R}(-x)$,
 $(\mathcal{CP})
 \psi^{\ }_R(x)
(\mathcal{CP})^{-1}
=
 \psi^{\dag}_{L}(-x). 
$
Thus, 
by CP, 
the $(\psi_L, \psi^{\dag}_R)$
zero modes
in the background $A(x)$
are sent to
the $(\psi^{\dag}_R, \psi^{\ }_L)$
zero modes
in the background 
$-A(\tilde{x})$. 
Because of the Fermi statistics,
the measure is transformed as
\begin{align}
 \prod^{{Ch}}_{\alpha=1}
 d a^{\ }_{\alpha}
 d a^{*}_{\alpha}
 \to
 (-1)^{{Ch}}
 \prod^{{Ch}}_{\alpha=1}
 d a^{\ }_{\alpha}
 d a^{*}_{\alpha}.
\end{align}
Since the field configurations $A(x)$ and $-A(\tilde{x})$ are smoothly connected, 
there is no way to define the measure so that it is invariant under CP. As we have seen, this case corresponds to topologically non-trivial
bulk. On the other hand, when $\eta=-1$,
 $(\mathcal{CP})
 \psi^{\ }_L(x)
(\mathcal{CP})^{-1}
=
 \psi^{\dag}_{R}(-x)$, 
$(\mathcal{CP})
 \psi^{\ }_R(x)
(\mathcal{CP})^{-1}
=
- \psi^{\dag}_{L}(-x) 
$, 
with the extra minus sign, the measure is invariant.
In the next section, we make contact between
these two cases (the case with and without parity anomaly)
and topologically non-trivial and trivial insulators. 

\subsection{Generalized Laughlin's argument}

In the above considerations,
we have (implicitly) assumed that
the electromagnetic charge U(1) is strictly conserved. 
However, it would be possible to instead demand 
CP symmetry to be strictly conserved. 
Given a conflict between CP symmetry and charge conservation
(when $\epsilon=0$)
suggested by the above argument, 
it would not be
possible to preserve electromagnetic U(1) charge symmetry
once we demand CP symmetry. 
This suggests the following: 
let us  twist the boundary condition
by the conserved electromagnetic U(1) charge 
(denoted by $a$ and $b$ as before in our discussion in
the QH edge).
The partition function depends on these twisting angles. 
One can then enforce CP symmetry by 
performing a projection on to a
space with definite CP eigenvalue. 
In the path integral picture, 
the enforcement of CP symmetry
leads 
to a conformal field theory
defined on an unoriented spacetime, 
{\it i.e.}, 
the spacetime of the edge theory has the topology
of the Klein bottle. 
Once we insist on CP symmetry,
one may not be able to achieve gauge invariance under (large) U(1)
gauge transformations.
Equivalently, 
the partition function would not be invariant
under $a\to a+1$ or $b\to b+1$.
We view this conflict between
the charge U(1) and CP symmetries 
as a signal for the existence of a
bulk topological phase.

\paragraph{Twisted boundary conditions}
Let us now canonically quantize the fermion theory
in the presence of the following spatial boundary condition:
\begin{align}
 \psi^{\ }_L(x+\ell_1)&=
 e^{ 2\pi i \nu_L} 
 \psi^{\ }_L(x),
 \nonumber \\
 \psi^{\ }_R(x+\ell_1)
 &=
 e^{ 2\pi i \nu_R}
 \psi^{\ }_R(x). 
 \label{space twisted bc}
\end{align}
where the edge theory is put on a circle of circumference $\ell_1$. 
A discrete symmetry (CP, in our example) may be compatible/incompatible
with the boundary condition. 
By acting with CP on the boundary condition
(\ref{space twisted bc}), 
\begin{align}
 &\quad
 (\mathcal{CP})
 \psi^{\ }_L(x+\ell_1)
 (\mathcal{CP})^{-1}
 =
 e^{2\pi i \nu_L}
 (\mathcal{CP})
 \psi^{\ }_L(x)
 (\mathcal{CP})^{-1}
\nonumber \\
&\Rightarrow 
\psi^{\dag}_R(-x-\ell_1)
 =
 e^{2\pi i \nu_L}
 \psi^{\dag}_R(-x)
\nonumber \\
&\Rightarrow e^{2\pi i \nu_L} \psi^{\ }_R(x-\ell_1) =
 \psi^{\ }_R(x),
 \label{CP and twist}
\end{align}
we conclude that CP symmetry is consistent with the twisted boundary condition
when $\nu_L=\nu_R$, 
{\it i.e.},
only charge twist is allowed,
 $\psi^{\ }_L(x+\ell_1)
 =
 e^{  2\pi i \nu} 
 \psi^{\ }_L(x)$,
 $\psi^{\ }_R(x+\ell_1)
 =
 e^{ 2\pi i \nu}
 \psi^{\ }_R(x)$.
By similar considerations, P symmetry
is consistent with the 
twisted boundary condition only
when 
$\nu_L=-\nu_R$
({\it i.e.}, only the spin twist is allowed),
and C symmetry is consistent with the twisted boundary 
condition only when $\nu_L = 0,1/2$ and $\nu_R=0,1/2$. 

\paragraph{The torus partition function}

For the CP symmetric case, 
we thus consider
the spatial boundary condition with 
$\nu_L=\nu_R=a$. 
The corresponding partition function 
on the torus is 
\begin{align}
 Z(\tau,\bar{\tau})
 &=
 \mathrm{Tr}_{a \otimes a}\,
 \left[
 e^{-2\pi i (b-1/2) F_V}
 q^{L_R}
 \bar{q}^{L_L}
\right]
\nonumber \\
 &=
 Z_{[a,b]}(\tau) \overline{Z_{[a,b]} ({\tau})}, 
\end{align}
where
$\overline{\cdots}$ denotes
complex conjugation,
and the Hamiltonian $H = H_R + H_L$ is given in terms of
the left- and right-moving parts as
\begin{align}
 L_R &= L_0 - \frac{c_R}{24},
 \quad
 L_L = \bar{L}_0 - \frac{c_L}{24},
 \nonumber \\
 H_R&= \frac{2\pi v}{\ell_1} L_R,
 \quad
 H_L= \frac{2\pi v}{\ell_1} L_L, 
\end{align}
with $c_L=c_R=1$. 
We have introduced the modular parameter through 
\begin{align}
 q &= e^{2\pi i \tau},
 \quad
 \tau= \tau_1 + i \tau_2,
 \quad
 \tau_2
 = 
 \frac{v \ell_2}{\ell_1}.  
\end{align}
Here, $\ell_2$ represents the inverse temerature 
and we have included,
in addition to the (imaginary) time translation generated 
by the Hamiltonian,
the space translation generated by the momentum
with the corresponding periodicity $\tau_1$. 
(As we will see, $\tau_1$ will not play any role 
once we impose CP symmetry.)

Written as a product
$Z_{[a,b]}(\tau) \overline{Z_{[a,b]} ({\tau})}$,
where 
$Z_{[a,b]}(\tau)$
is given by 
Eq.\ (\ref{chiral part fn Dirac}), 
the partition function is large gauge invariant
under $b\to b+1$ and $a\to a+1$. 
One can also check that the partition 
function is modular invariant. 

\paragraph{The Klein bottle partition function with CP symmetry}

Let us now consider the partition function with CP projection: 
\begin{align}
 Z^{\mathrm{Proj}}_{[a]}(\tau)
 =
 \mathrm{Tr}_{a \otimes a}\,
 \left[
 \frac{1+\mathcal{CP}}{2}
 e^{-2\pi i (b-1/2) F_V}
 q^{L_R}
 \bar{q}^{L_L}
\right] 
\end{align}
where we have inserted a projection operator, $(1+\mathcal{CP})/2$.
The first term in the projection gives nothing but the
torus partition function. The second term
can be interpreted as a path integral over the fermion
fields on the Klein bottle
(with twisted boundary condition in the time direction.)
To develop this picture, we first perform the Wick rotation $t=-i x_2$.
The insertion of CP operator into the trace has the effect that 
by translating a fermion field $\psi_R$, say, 
once around the time direction, 
it comes back as $(\mathcal{CP}) \psi_R (\mathcal{CP})^{-1}$.  
Thus,
the time direction boundary condition is 
($\ell_2=2\pi\tau_2$)
\begin{align}
 \psi^{\ }_{R}(x_1, x_2) &= -(\mathcal{CP}) \psi^{\ }_{R}(x_1,x_2+\ell_2) (\mathcal{CP})^{-1}, 
 \nonumber \\
 \psi^{\ }_{L}(x_1,x_2) 
 &=
 -(\mathcal{CP})
 \psi^{\ }_L(x_1,x_2+\ell_2)
 (\mathcal{CP})^{-1}. 
\end{align}
where the factor $-1$
comes from 
the antiperiodic boundary condition
of the fermion fields (we have set $b=1/2$ for simplicity).
{\it I.e.,}
\begin{align}
 \psi^{\ }_{R}(x_1, x_2) &= -\eta 
 \psi^{\dag}_{L}(-x_1, x_2+\ell_2), 
 \nonumber \\
 \psi^{\ }_{L}(x_1, x_2)
 &=
 -
 \psi^{\dag}_{R}(-x_1, x_2+\ell_2). 
\end{align}
(Observe that $\tau_1$ is ``projected out'' by CP -- see below.)
The fermion fields are defined 
on the Klein bottle
$(x_1,x_2)\equiv (x_1+\ell_1, x_2) \equiv (-x_1, x_2+\ell_2)$
with periodic boundary condition along $x_1$
(possibly twisted by $a$),
but with the CP-twisted boundary condition along $x_2$ direction. 

There is a simple consequence of the topology change 
from the torus to the Klein bottle, 
induced by CP projection. 
The partition function on a Riemann surface of genus $g$
(denoted by $\Sigma_g$)
is given as the sum over all possible monodromies
on $\Sigma_g$
\cite{Horava1996}: 
\begin{align}
 Z^{\Sigma_g}(\tau)
 =
 \frac{1}{|\mathscr{G}|^g}
 \sum_{
  \alpha: \pi_1(\Sigma_g)\to \mathscr{G}}
 Z^{\Sigma_g}(\alpha; \tau),
 \label{Klein part fn general}
\end{align}
where $\tau$ are the moduli of $\Sigma_g$,
$\pi_1(\Sigma_g)$ is the fundamental group
of $\Sigma_g$,
and
$Z^{\Sigma_g}(\alpha; m)$
denotes the partition function 
calculated with the particular set of
monodromies $\alpha$. For the torus, the fundamental group is 
$\pi_1(T^2)=\langle \alpha,\beta| \alpha \beta \alpha^{-1}\beta^{-1}=1\rangle$,
{\it i.e.}, $\alpha\beta = \beta\alpha$.
This means that, 
for any Abelian group $\mathscr{G}$
by considering 
a correspondence $\alpha\to g_1$,
$\beta \to g_2$ where $g_{1,2}\in \mathscr{G}$,
the summation is  
$\mathrm{Hom}\, [\pi_1(T^2), \mathscr{G}]=(\mathscr{G})^2$. 
On the other hand, for the Klein bottle, 
the fundamental group is given
by 
$\pi_1(K)=\langle \alpha,\beta| \alpha \beta =\beta^{-1}\alpha\rangle$. 
For an Abelian group, this means $g_2 = (g_1)^{-1}$.   
As we will show below,
the $\mathbb{Z}_2$ flux $g_2$ distinguishes 
topological and non-topological CP symmetric insulators.

Let us work out the effects of the projection explicitly. 
We first mode-expand the fermion fields as 
\begin{align}
 \psi^{\ }_R (x) &= 
 \sqrt{\frac{2\pi}{\ell_1}} \sum_{r\in \mathbb{Z}+a} \psi^{\ }_{R,r} e^{i\frac{2\pi}{\ell_1}rx},
 \nonumber \\
 \psi^{\ }_L (x)
 &=
 \sqrt{\frac{2\pi}{\ell_1}}
 \sum_{r\in \mathbb{Z}+a} 
 \psi^{\ }_{L,r} e^{i\frac{2\pi}{\ell_1}rx}.
\end{align}
The CP transformation acts on 
the fermion modes as 
\begin{align}
 (\mathcal{CP}) \psi^{\ }_{Lr}
 (\mathcal{CP})^{-1}
 &=
 \psi^{\dag}_{Rr}, 
 \nonumber \\
(\mathcal{CP}) \psi^{\ }_{Rr}
(\mathcal{CP})^{-1}
 &=
 \eta  \psi^{\dag}_{Lr}. 
\end{align}
For a given $r>0$,
there are four states, 
 $|\mathrm{GS}_{a}\rangle$, 
 $\psi^{\dag}_{Rr}|\mathrm{GS}_{a}\rangle$, 
 $\psi^{\ }_{Lr} |\mathrm{GS}_{a}\rangle$,
 $\psi^{\ }_{Lr} \psi^{\dag}_{Rr}|\mathrm{GS}_{a}\rangle$,
 where $|\mathrm{GS}_{a}\rangle \propto \psi^{\dag}_{Lr}|0\rangle$
is the ground state for the boundary condition specified by $a$. 
On these states, CP acts as,
{\it e.g.}, 
\begin{align}
 (\mathcal{CP})|\mathrm{GS}_{a}\rangle
& = 
P_{[a]} |\mathrm{GS}_{a}\rangle,
 \nonumber \\
 (\mathcal{CP}) \psi^{\ }_{Lr}\psi^{\dag}_{Rr} 
 |\mathrm{GS}_{a}\rangle
 &= 
 -\eta P_{[a]} \psi^{\ }_{Lr}\psi^{\dag}_{Rr}|\mathrm{GS}_{a}\rangle. 
\end{align}
Here, $P_{[a]}$, the CP eigenvalue of the ground state,
is, {\it a priori}, undetermined.
We have demanded that the system is CP invariant,
and hence the first equation follows.
Since CP is unitary, 
the eigenvalue $P_{[a]}$ should be a complex number of
unit modulus. 

For our discussion, it is crucial to know 
the $a$-dependence of the 
CP eigenvalue of the ground state. 
In particular, we need to compare 
the relative phase difference between $P_{[a]}$ and $P_{[a+1]}$. 
Under the assumption of the strict enforcement of 
CP symmetry, 
$P_{[a]}$ should be independent of $a$, 
and in particular, $P_{[a]}=P_{[a+1]}$.
(If $P_{[a]}$ is dependent on $a$, the projection operation
in fact is ill-defined.)

It is also insightful to use an alternative but equivalent picture 
for the effects of the fluxes $a$ and $b$, where  
they are introduced as,
instead of 
twisting angles for twisted boundary conditions,  
constant background gauge fields. 
In this picture, 
the Hamiltonian depends explicitly on $a$
and is given by 
\begin{align}
{H}(a) 
= 
\frac{v}{2\pi}
\int dx 
\big[
{\psi}_L^{\dag}
i \partial_{x} 
{\psi}_L 
-
{\psi}^{\dag}_R
i \partial_{x} 
{\psi}_R 
\big] 
+
\frac{a}{\ell_1}
J_V, 
\end{align}
where $J_V=v F_A$ is the current operator. 
The fermion fields obey boundary conditions that are independent of $a$, 
\begin{align}
 {\psi}_L(x+\ell_1)  
 = 
 {\psi}_L(x),
 \quad
 {\psi}_R(x+\ell_1)  
 = 
 {\psi}_R(x). 
\end{align}
\footnote{
 In addition to the Hamiltonian, 
 there is a chemical potential which appears as an 
 operator insertion 
 $e^{-2\pi i (b-1/2)F_V}$ in the partition function.    
Viewing this operator as a part of the partition 
function, 
the system with the chemical potential is in general
not invariant under CP
since
 $(\mathcal{CP}) {F}_V {(\mathcal{CP})}^{-1}
 =
 -{F}_V$.
The only exceptions are the cases when $b=0,1/2$.   
When $b\neq 0, 1/2$, the system is not invariant 
under CP and so we cannot make a projection by CP symmetry. 
We therefore limit ourselves to $b=0,1/2$.
}
Under an infinitesimal change in the flux $a\to a+\delta a$, 
since the perturbation 
commutes with CP and hence does not mix
states with different eigenvalues of CP,
the CP eigenvalue $P_{[a]}$ should be constant. 
For a similar discussion,
see Refs.\ \onlinecite{Oshikawa2000a, Oshikawa2000b, Oshikawa2003}. 
\footnote{
 The equivalence of the two pictures,
 one in terms of twisting boundary conditions,
 and the other in terms of background gauge fields,
 can be established by a gauge transformation
 that ``unwinds'' the boundary conditions, and vice versa. 
 When the electromagnetic U(1) symmetry happens to be anomalous, 
 care may be required in invoking such equivalence.  
 (See, for example, Ref.\ \onlinecite{Landsteiner2013}.)
 In our approach, when an ambiguity such as the CP eigenvalue 
 of the ground state arises, we follow what we expect in the absence
 of anomalies. 
 We test the consistency of such an assumption
 arising from enforcement of CP symmetry with the electromagnetic 
 U(1) symmetry by inspecting the behavior of the partition function
 under the adiabatic process of flux insertion. 
}
(See also discussion in 
Appendix \ref{Discussion on the CP eigenvalue of the ground state}.)

Finally, the projected partition function can be calculated in 
a straightforward fashion, leading to 
\begin{align}
& Z^{\mathrm{Klein}}_{[a]} =
 \mathrm{Tr}_{a \otimes a}\,
 \left[
 (\mathcal{CP})
 e^{-2\pi i (b-1/2) F_V}
 q^{L_R}
 \bar{q}^{L_L}
\right] 
\nonumber \\
&\quad =
 \frac{
  P_{[a]} e^{ 2\pi i (a-1/2)(\epsilon-1/2)}
}{\eta(q\bar{q})}
\vartheta
\left[
 \begin{array}{c}
  a - 1/2\\
 -(\epsilon-1/2)
 \end{array}
\right]
(0,q\bar{q}),  
\label{fermionic CP projected part fn}
\end{align}
where we note
 $q \bar{q}=
 e^{2\pi i \tau}
 e^{-2 \pi i \bar{\tau}}
 =
 e^{-4\pi\tau_2}
 =
 e^{- 4\pi\frac{v\ell_2}{\ell_1} } 
$
and 
$
\eta(q\bar{q})
=\eta(2i \mathrm{Im}\, \tau)
$.
Observe that the partition function is independent of $\tau_1$,
{\it i.e.,}
it is projected out by CP. 
Similarly, the chemical potential $b$ also does not show up
in the projected partition function. 

With 
$P_{[a]}= P_{[a+1]}$,
which we enforce by CP symmetry, 
the partition function is invariant
under $a\to a+1$
for the topologically trivial case
($\epsilon=1/2$)
whereas it is not 
for the topologically non-trivial case
($\epsilon=0$),
\begin{align}
 Z^{\mathrm{Klein}}_{[a+1]}
 = e^{2\pi i(\epsilon-1/2)} Z^{\mathrm{Klein}}_{[a]}. 
\end{align}
By comparison 
with the chiral partition function (\ref{chiral part fn Dirac}), 
we observe 
the distinction between
topologically trivial ($\epsilon=1/2$)
and nontrivial ($\epsilon=0$)
cases shows up as a fictitious chemical
potential 
($\pi$ flux in time direction). 
For the topological case
the fermion effectively feels periodic 
boundary condition in time direction,
whereas for the trivial case,
the fermion effectively feels antiperiodic boundary condition.
This anomaly vanishes if we consider two copies (more generally, an even number of copies)
of the fermion theory, which suggests a $\mathbb{Z}_2$ classification of CP symmetric topological insulators. 
\footnote{
It is instructive to compare the CP projected partition function
with the partition function with P projection.
Parity transformation acts on fermion fields as
$$
\begin{array}{ccc}
\mathcal{P}\psi_L(x) \mathcal{P}^{-1}&=& \eta e^{i\alpha} \psi_R(-x), \\
\mathcal{P}\psi_R(x) \mathcal{P}^{-1}&=& e^{i\alpha} \psi_L(-x).
\end{array}
$$
In our fermionic edge theory, by analyzing mass terms, 
one can check that there is no topological phase protected by 
parity symmetry (of any kind) and the electromagnetic U(1) symmetry. 
The absence of topological phases  
can be seen from the fact that the Klein bottle partition 
function with parity projection is anomaly-free.   
First recall that P symmetry is consistent with the 
twisting boundary condition only when 
$\nu_L=-\nu_R$.
As we require only
the U(1) charge conservation,
this means only
$\nu_{L}=\nu_{R}=0$
(periodic boundary condition)
or
$\nu_{L}=-\nu_R=1/2$
(antiperiodic boundary condition)
are allowed. 
With this in mind, 
the projection works,
for a given $r>0$,
as
$$
\begin{array}{l}
 \displaystyle 
 \mathrm{Tr}\,
 \left[
 \mathcal{P}\,
 e^{-2\pi i (b-1/2) F_V} 
  q^{H_R} \bar{q}^{H_L}
 \right]
 \\
\propto  
 \displaystyle 
\prod_r 
  \left[
  1 +\eta e^{4\pi i (b-1/2)}
  e^{- 2 i \alpha}
  (q \bar{q})^r 
 \right]. 
\end{array}
$$
Thus, the phase $\alpha$ as well as
$\eta$ simply shifts
the chemical potential $b$. 
In the case of P,
we can freely change the time boundary condition $b$,
but not the spatial boundary condition. 
Observe that this situation is opposite to what we had for CP symmetry.
In the case of CP symmetry, we can freely change the space boundary condition,
but not the time boundary condition. 
As before, we change $b\to b+1$ and 
ask if the theory is invariant under
this large gauge transformation or not. 
Depending on the spatial boundary conditions, 
(periodic/antiperiodic), 
the partition function may pick up 
an anomalous phase. 
However, observe that the 
chemical potential enters in the partition
function as $e^{ 4\pi i (b-1/2)}$
not $e^{2\pi i (b-1/2)}$. 
Due to this doubling, there are no
anomalous phases.
}

\section{2D bosonic topological phases protected by CP symmetry}
\label{Bosonic CP symmetry protected topological phase}

\subsection{The edge theory}

Armed with insights from
the fermionic symmetry protected topological phases,
we now discuss the bosonic topological phases. 
Below, we study the partition function of the
edge of the bosonic CP symmetric topological insulator. 
We start from the single-component free boson theory
on a ring of circumference $\ell$
defined by 
$
Z=\int\mathcal{D}[\phi] \exp {(iS)}
$
with the action
\begin{align}
S=\frac{1}{4\pi\alpha'}\int dt \int_{0}^{\ell} dx
\left[
\frac{1}{v}( \partial_{t}\phi )^{2}
-v( \partial_{x}\phi )^{2}
\right],
\end{align}
where the $\phi$-field is compactified 
with the compactification radius $R$ as
$\phi \equiv \phi + 2\pi R$;
$\alpha'$ is the coupling constant of the boson theory.
The canonical commutation relation is
\begin{align}
&
\left[
 \phi(t,x),\partial_t {\phi}(t,x^{\prime})
\right]
=2\pi i \alpha' v \sum_{m\in\mathbb{Z}}\delta(x-x^{\prime}-m\ell).
\end{align}
The theory can be quantized 
and decomposed into the left- and
right-moving sectors. 
We introduce the chiral decomposition of the boson field
$\phi$ as 
\begin{align}
\phi(t,x)=\varphi_L(x^+)+\varphi_R(x^-),
\quad
x^{\pm}:=vt\pm x. 
\end{align}
and also the dual boson field as 
\begin{align}
\theta(t,x)=\varphi_L(x^+)-\varphi_R(x^-). 
\end{align}

As in the fermionic CP symmetric topological insulator,
we consider two kinds of CP symmetries 
specified by $\epsilon=0,1/2$ as follows:
\begin{align}
 \label{def: CP acting of boson}
 (\mathcal{CP})\phi(t,x)(\mathcal{CP})^{-1}&= -\phi(t,-x),   
 \nonumber \\
 (\mathcal{CP})\theta(t,x)(\mathcal{CP})^{-1}&= +\theta(t,-x) 
 + 2\pi \epsilon \alpha'/R.  
\end{align}
The single-component boson model 
with these CP symmetries is studied in 
Ref.\ \onlinecite{Hsieh13},
and it was demonstrated, based on 
microscopic analysis of gapping potentials,
that the case with $\epsilon=0$ is gappable 
while the case with $\epsilon=1/2$ is not. 
We will reproduce this result from the generalized Laughlin
argument with CP symmetry.

\paragraph{Quantization}

The mode expansions for the left- and right-moving boson fields 
is given by 
\begin{align}
\varphi_{L}(x^{+})
&=
x_L
+
\pi \alpha' p_L
\frac{ x^{+}}{\ell}
+
i\sqrt{ \frac{\alpha'}{2} }
\sum_{n\neq 0}
 \frac{\alpha_{n}}{{n}}e^{-\frac{2\pi i nx^{+}}{\ell}}
, 
\nonumber \\
\varphi_{R}(x^{-})
&=
x_R
+
\pi \alpha'  p_R
\frac{x^{-}}{\ell}
+
i \sqrt{ \frac{\alpha'}{2} }
\sum_{n\neq 0}
 \frac{\tilde{\alpha}_{n}}{{n}}e^{-\frac{2\pi i nx^{-}}{\ell}}
, 
\nonumber \\
\mbox{where} 
&\quad  
 \left[\alpha_m, \alpha_{-n}\right] =
 \left[\tilde{\alpha}_m, \tilde{\alpha}_{-n}\right] =
 m \delta_{m,n}, 
 \quad 
(n,m>0) 
\nonumber \\
&\quad 
\left[x_L, p_L\right]= \left[x_R, p_R\right]= {i}, 
\end{align}
and  all the other commutators vanish. 
With the periodic boundary condition
\begin{align}
\phi(t,x+\ell)= \phi(t,x) + 2\pi R w,
\quad
w\in 
\mathbb{Z}, 
\end{align}
the allowed momentum values are
\begin{align}
&p_L
=
+\frac{R}{\alpha'}w 
+\frac{k}{R},
\quad
p_R
=
-\frac{R}{\alpha'}w 
+\frac{k}{R},
\nonumber \\
&p=\frac{1}{2}\left(p_L + p_R\right)
=
\frac{k}{R},
\quad
\tilde{p}
=
\frac{1}{2}
\left(
p_L - p_R
\right)
=
\frac{R}{\alpha'}w, 
\end{align}
where $k$ and $w$ are integers. 
Correspondingly, the chiral boson fields
and
the dual boson field obey 
\begin{align}
\varphi_L(x+\ell) - \varphi_L(x)&=+\pi \alpha' p_L,
\nonumber \\
\varphi_R(x+\ell) - \varphi_R(x)&= -\pi \alpha' p_R,
\nonumber \\
 \phi (x+\ell) - \phi(x)&= \pi \alpha'(p_L-p_R)=2\pi R w,
 \nonumber \\
 \theta(x+\ell) - \theta(x)&=
 \pi \alpha'(p_L+p_R) = 2\pi \frac{\alpha'}{R}k. 
\end{align}
The set of bosonic exponents consistent with
the boundary conditions are
$\exp
i
\left[
 \frac{k}{R}\phi+ \frac{wR}{\alpha'}\theta 
\right]
=
\exp
i\left[
 p_L \varphi_L + p_R\varphi_R
\right].  
$
The Hamiltonian is given by
\begin{align}
&H=H_L + H_R
=
\frac{2\pi v}{\ell}
\left(
L_L
+
L_R 
\right)
,
\nonumber \\
&
L_L
=
\frac{\alpha'p^2_L}{4}
+
\sum_{n=1}^{\infty}
\alpha_{-n}\alpha_{n}
-\frac{1}{24}
,
\nonumber \\
&
\bar{L}_R 
=
\frac{\alpha' p^2_R}{4}
+
\sum_{n=1}^{\infty}
\tilde{\alpha}_{-n}\tilde{\alpha_{n}}
-\frac{1}{24}. 
\end{align}
Observe that the spectrum depends only on
 $R/\sqrt{\alpha'}$ and is invariant under $R\to\alpha'/R$ as expected.

\subsection{Twisted boundary conditions and twisted partition function}

Two conserved U(1) charges, one for each left- and right-moving
sector, can be introduced as follows:
\begin{align}
 & N_{L,R}= \int^{\ell}_0  dx\, \partial_x \varphi_{L,R} =
 \alpha' \pi p_{L,R},
\end{align}
which satisfy 
$
\left[
\varphi_L, N_L
\right]
=
\left[
\varphi_R, N_R
\right]
=
\alpha' \pi {i}.  
$
The operator  
\begin{align}
 \mathcal{G}(a_c, a_s) &=
\exp
 i \left[ 2\pi a_cR p + 
 2\pi a_s (\alpha'/R) \tilde{p}
\right]. 
\end{align}
generates translations in $\phi$ and $\theta$ as 
\begin{align}
 \phi &\to \phi + 2\pi a_c R,   
\quad 
\theta \to 
\theta +
2\pi a_s (\alpha'/R).
\end{align}

By using the $\mathrm{U}(1)\times \mathrm{U}(1)$ 
symmetry generators, 
it is possible to twist the spatial boundary condition as 
\begin{align}
\phi(x+\ell) 
&=
\phi(x)
+
2\pi R a_c  
+ 
2\pi R w,
\nonumber \\
\theta(x+\ell) 
&=
\theta(x)
+
2\pi
\frac{\alpha'}{R}
a_s 
+
2\pi \frac{\alpha'}{R}k.
\end{align}
With the twisted boundary condition,
the momenta are given by
\begin{align}
  \alpha' \tilde{p}
&=
 R(a_c   + w), 
\quad
 \alpha' p
=
\frac{\alpha'}{R}
\left(
a_s + k 
\right). 
\end{align}
As compared to the original quantization 
conditions, in the presence of 
the twist, the quantization conditions
on $p$ and $\tilde{p}$ are ``shifted''
by $a_c$ and $a_s$. 
Below, we will focus on the twist by 
the diagonal U(1) symmetry, $a_s=0$.

Let us now consider the partition function 
twisted both in time and space directions.
It can be written as 
\begin{align}
 \label{boson part fn with twist}
 Z_{[a_c, b_c]}
&=
\mathrm{Tr}_{a_c}
\left[
 \mathcal{G}(b_c)  
e^{ 2\pi {i}\tau (L_0 - c/24)
 -2\pi {i}\bar{\tau} (\bar{L}_0 - c/24)
}
\right]
\end{align}
where the trace is taken
with the quantization conditions, 
\begin{align}
\tilde{p}
&=
\Delta \tilde{p}
 +  \frac{R}{\alpha'}w,
\quad
\Delta \tilde{p} = 
\frac{R a_c}{\alpha'}, 
\nonumber \\
 p
&=
\Delta p
+
 \frac{k}{R},
 \quad
 \Delta p
 =
 \frac{(\alpha'/R) a_s }{\alpha'}.
\end{align}
On the other hand, the twist in
time direction is implemented as an insertion
of the operator 
$\mathcal{G}(b_c)=\mathcal{G}(b_c, b_s=0)$.

\subsection{The CP projected partition function}

Following the discussion for fermionic CP symmetric
topological insulators, we now project with the CP operator.
Inserting the projection operator in
the partition function (\ref{boson part fn with twist}),
we consider 
\begin{align}
& Z^{\mathrm{Proj}}_{[a_c]} =
 \mathrm{Tr}_{a_c}\, 
 \left[
  \frac{1+\mathcal{CP}}{2} 
  \mathcal{G}(b_c)
  q^{(L_0 -c/24)}
  \bar{q}^{(\bar{L}_0-c/24)}
 \right]. 
\end{align}
We will consider an adiabatic process 
where we change $a_c$ to $a_c+1$ and ask
if the CP projected partition function is invariant 
or not. 

When evaluating the CP projected partition function, 
it is necessary to know the action of CP operators on the states
in the Hilbert space. 
The symmetry transformation on $\phi$ and $\theta$
implies the action of CP on each mode
in the mode expansion of $\phi$ and $\theta$: 
\begin{align}
 (\mathcal{CP}) \alpha_n (\mathcal{CP})^{-1} &=  \tilde{\alpha}_n,
 \nonumber \\
 (\mathcal{CP}) \phi_0(\mathcal{CP})^{-1}  
&= -\phi_0,
\quad
(\mathcal{CP}) p(\mathcal{CP})^{-1}  = -p, 
 \nonumber \\
 (\mathcal{CP}) \theta_0 (\mathcal{CP})^{-1}  
 &= \theta_0 + 2\pi \epsilon ({\alpha'}/{R}),
 \quad
 (\mathcal{CP}) \tilde{p}(\mathcal{CP})^{-1} = 
 \tilde{p}, 
\end{align}
where $\phi_0=x_L+x_R$ and $\theta_0=x_L-x_R$. 
[Since CP flips the sign of $p$, for generic value of $a_s$,
there is no state that is invariant under $a_s$.
For the purpose of the CP projection, 
we thus should set $\Delta {p}=0 \Rightarrow a_s =0$.) 
See discussion near Eq.\ (\ref{CP and twist}).]

For later use, we need to know
the action of CP on 
the states in the zero mode sector. 
We will use the momentum basis
$\{| p,  \tilde{p}\rangle\}$, 
where $| p,  \tilde{p}\rangle$
is the momentum eigen ket. 
Recall that due to the compactification condition, 
$\phi_0 \equiv \phi_0 + 2\pi R$
and
$\theta_0 \equiv \theta_0+2\pi \alpha'/R$, 
the corresponding momenta lie in the BZ.
Since CP transformation sends the momentum operators as  
$p\to -p$ and $\tilde{p}\to \tilde{p}$, 
the momentum eigen ket $\mathcal{CP}|p,\tilde{p}\rangle$
must be equal to $|-p,\tilde{p}\rangle$
up to a phase,
$\mathcal{CP} |p, \tilde{p}\rangle = e^{i A(p,\tilde{p})} |-p, \tilde{p}\rangle$. 
Similarly,
the ket $\mathcal{CP}|\phi_0, \theta_0\rangle$ must be
equivalent to $|-\phi_0, \theta_0 + 2\pi \epsilon (\alpha'/R)\rangle$,
$\mathcal{CP}|\phi_0,\theta_0\rangle
=e^{iB(\phi_0,\theta_0)} |-\phi_0, \theta_0 + 2\pi \epsilon (\alpha'/R)\rangle$. 
We can read off these phases from  
the Fourier representation of the basis ket: 
\begin{align}
 |p, \tilde{p}\rangle
 &=
 \int d\phi_0 d\theta_0\, 
 e^{ i p \phi_0 + i \tilde{p}\theta_0}
 | \phi_0, \theta_0\rangle,
 \nonumber \\
 \mbox{as}
 \quad 
 \mathcal{CP} |p, \tilde{p}\rangle
 &=
 e^{-i 2 \pi \epsilon (w+ a_c)} 
 \nonumber \\
 &\quad 
 \times 
 \int d\phi_0 d\theta_0\, 
 e^{ -i p \phi_0 + i \tilde{p}\theta_0}
e^{iB} 
| \phi_0, \theta_0  \rangle.
\end{align}
In order to have
$\mathcal{CP}|p,\tilde{p}\rangle\propto |-p, \tilde{p}\rangle$, 
we need to take $B(\phi_0,\theta_0)=\mbox{const}.=B$,
and we conclude
\begin{align}
 \mathcal{CP}|p,\tilde{p}\rangle
 =
 e^{-i 2 \pi \epsilon (w+a_c)} 
 e^{iB} 
 | -p, \tilde{p}\rangle,  
\end{align}
{\it i.e.}, 
$
A(p,\bar{p}) = B -  2\pi \epsilon \tilde{p} \alpha' /R
$. 
One can also check, 
from the inverse Fourier representation
\begin{align}
 |\phi_0, \theta_0\rangle
 &= 
 \sum_{ p, \tilde{p}} 
 e^{-i p \phi_0 -i \tilde{p}\theta_0}
 |p,\tilde{p}\rangle, 
 \end{align}
 that $\mathcal{CP}|\phi_0, \theta_0\rangle = 
  e^{iB} |-\phi_0,\theta_0 + 2 \pi \epsilon \alpha'/R \rangle$. 

Summarizing, 
acting with CP operator on the basis ket
$|p,\tilde{p}\rangle$, 
\begin{align}
\mathcal{CP}|p,\tilde{p}\rangle =
P_{[a_c]} e^{-i 2 \pi \epsilon w }
|-p, \tilde{p}\rangle, 
\end{align}
where $P_{[a_c]}$ is independent of 
$p$ and $\tilde{p}$, but may be dependent on 
the adiabatic parameters, $a_c$ and $b_c$. 
While the presence of the phase factor $e^{-i 2\pi \epsilon w }$ 
can directly be seen from the CP transformation laws 
on the zero mode operators,
the phase factor $P_{[a_c]}$ cannot be determined;
this originates from our 
ignorance on 
the CP eigenvalue of the ground state,  
and, in particular, 
on its dependence on the adiabatic parameters.  
Based on our previous discussion, however, 
we enforce CP invariance 
for arbitrary value of the adiabatic parameters. 
This means that 
we demand that our CP operation does not depend on
the inserted flux. 
While the Hilbert space changes adiabatically,
we do not allow the phase factor to be dependent on $a_c$. 
This is the same assumption we had before for the case of fermions. 
We could then choose $B= \pi a_c$,
and hence $P_{[a_c]}=1$.

Having established the CP action on the zero-mode wave functions,
we now calculate the projected partition function explicitly. 
Note also
 $(\mathcal{CP}) 
 \mathcal{G}(b_c) (\mathcal{CP})^{-1}
 =
 \mathcal{G}(-b_c)$. 
This limits a reasonable value of $2\pi R b_c$
to be $0$ and $\pi$. 
Then, the Klein bottle partition function is  
\begin{align}
 \label{bosonic cp projected part fn}
& Z^{\mathrm{Klein}}_{[a_c]} =
 \mathrm{Tr}_{a_c}\, 
 \left[
  (\mathcal{CP}) 
  \mathcal{G}(b_c)
  q^{(L_0 -c/24)}
  \bar{q}^{(\bar{L}_0-c/24)}
 \right]
 \nonumber \\
&\quad 
=
(q\bar{q})^{-\frac{1}{24}}
\prod_{n=1}
\left[
 1-(q\bar{q})^{n}
\right]^{-1}
\nonumber \\
&\quad 
\times 
\sum_{p, \tilde{p}}
\langle p \tilde{p}|
(\mathcal{CP})
e^{ i R b_c \frac{1}{2}(p_L + p_R)}
q^{ \frac{\alpha'}{4} p^2_L}
\bar{q}^{ \frac{\alpha'}{4} p^2_R}
| p \tilde{p}\rangle.  
\end{align}
In order for 
$
\langle p \tilde{p}|
\mathcal{CP} 
| p \tilde{p}\rangle 
$
to be non-zero, 
$p$ must be zero ($k=0$).
Then,
$p_L=-p_R= \tilde{p}$
and hence,
\begin{align}
 Z^{\mathrm{Klein}}_{[a_c]} &=
(q\bar{q})^{-\frac{1}{24}}
\prod_{n=1}
\left[
 1 - (q\bar{q})^n
\right]^{-1}
\nonumber \\
&\quad
\times 
\sum_{w\in \mathbb{Z}}
(q \bar{q})^{ \frac{1}{4} 
\left(\frac{R}{\sqrt{\alpha'}}
\right)^2
\left( w + a_c \right)^2}
e^{-i 2 \pi \epsilon w}. 
\end{align}
When $\epsilon=1/2$,
the partition function
is not invariant under
$a_c \to a_c +1$,
as it picks up an overall minus sign,
$Z^{\mathrm{Klein}}_{[a_c+1]} =
-Z^{\mathrm{Klein}}_{[a_c]}$. 
As in the fermionic CP symmetric topological insulator,
the anomaly is of $\mathbb{Z}_2$ kind
since it vanishes when we consider two copies (or any even number of copies)
of the theory.

\section{K-matrix theories protected by symmetries}
\label{Symmetry protected topological phase: K-matrix theory}

In this section, based upon the previous sections,
we consider edge theories consisting of multiple free bosons
that can describe, in addition to non-interacting topological 
insulators, 
interacting Abelian topologically ordered phases.  
We will develop a criterion for the stability of the edge theories
in the presence of CP and U(1) symmetries. 

\subsection{K-matrix theories}
\label{subsec: K-matrix theory}

Let us consider the $K$-matrix theory
with $N$ component compactified boson fields described by 
the Lagrangian 
\begin{align}
 \mathcal{L}&= \frac{1}{4\pi} \left(K_{IJ}\partial_t\phi^I\partial_x\phi^{J}-V_{IJ}\partial_x\phi^I\partial_x\phi^{J} \right) 
 \nonumber \\
&\quad 
+\frac{e}{2\pi}\epsilon^{\mu\nu} Q_I \partial_{\mu}\phi^I A^c_{\nu}
+\frac{s}{2\pi}\epsilon^{\mu\nu} S_I \partial_{\mu}\phi^I A^s_{\nu}, 
\label{K-matrix}
\end{align}
where 
$K$ is an $N\times N$ symmetric and invertible matrix with 
integer-valued matrix elements, 
$V$ is an $N\times N$ symmetric and positive definite matrix 
that accounts for the (screened) translation-invariant two-body 
interactions between electrons.
The $N$ component vector (``charge vector") $Q_I$, 
together with the unit of electric charge $e$,
describe how the system couples
to an external electromagnetic U(1) gauge potential, 
$A^c_{\mu}$.
Similarly,
the $N$ component vector (``spin vector") $S_I$, 
together with the unit of ``spin" charge $s$,
describe how the system couples
to an external ``spin'' U(1) gauge potential $A^s_{\mu}$  
that couples to the spin-1/2 degrees of freedom 
along some quantization axis, $z$-axis, say.  

The boson fields are compact variables, meaning 
field configurations $\phi^I$ differ by 
an integer multiple of $2\pi$
are identified: 
\begin{align}
\label{compactification}
\phi^I(t,x) \equiv \phi^I(t,x)+2\pi n^I, 
\end{align}
with $n^I \in \mathbb{Z}$ for all $I=1,\ldots,N$.
The equal-time canonical commutation relations of the boson fields are given by 
\footnote{
Here and in the following, the Dirac delta function
 $\delta(x-x')$ and $\mathrm{sgn}\, (x-x')$ in the commutator should be interpreted
as its periodic counter part,
such as $\sum_{m \in \mathbb{Z}} \delta(x-x'-2m\pi )$, 
when the system is put on a circle of circumference $2\pi$.}
\begin{align}
[\phi^I(t, x), \partial_{x}\phi^J(t, x')]=-2\pi i(K^{-1})^{IJ}\delta(x-x'),
\end{align}
or equivalently
\begin{align}
 \label{CCR K matrix 1}
[\phi^I(t, x), \phi^J(t, x')]=-i\pi\left[
 (K^{-1})^{IJ}\text{sgn}(x-x')+\Theta^{IJ} \right],
\end{align}
where the Klein factor
\begin{align}
\Theta^{IJ}:= (K^{-1})^{IK} \left[\text{sgn}(K-L)(K_{KL}+Q_KQ_L) \right](K^{-1})^{LJ}
\end{align}
is included to ensure 
that local excitations satisfy the proper commutation relations.

The goal of this section is to develop,
in the presence of either charge or spin U(1) symmetry, 
together with a discrete symmetry (such as CP or parity symmetry), 
a stability (``ingappability'') criterion of the edge theory (\ref{K-matrix})
against interactions. 

\paragraph{The rotated basis}

We start our discussion by quantizing the $K$-matrix theory
with the (untwisted) compactificataion
condition (\ref{compactification}). 
We introduce, starting from the original 
boson fields $\{\phi^I\}_{I=1,\ldots,N}$, 
a new basis $\{\varphi^i\}_{i=1,\ldots,N}$
that is obtained by a rotation matrix 
$e^{i}_I$ and its inverse $e^{\star J}_j$
as 
\begin{align}
 \varphi^i \equiv e^{i}_I \phi^I, 
\quad \phi^J \equiv e^{\star J}_j \varphi^j,
\nonumber \\
e^{i}_I e^{\star J}_i=\delta_I^J, 
\quad 
e^{i}_I e^{\star I}_j=\delta^i_j.
\end{align}
The ``vielbein'' $e^{i}_I$ and $e^{\star J}_j$
diagonalize the $K$-matrix as
\begin{align}
 K_{IJ}=e^{i}_I\eta_{ij}e^{j}_J, 
\quad 
\eta_{ij}=e^{\star I}_i K_{IJ} e^{\star J}_j=\eta_i\delta_{ij}
\end{align}
where $\eta_{ij}=\eta_i \delta_{ij}$ is a diagonal matrix.
We also note 
\begin{align}
 (K^{-1})^{IJ}=e^{\star I}_i (\eta^{-1})^{ij} e^{\star J}_j. 
\end{align}
In the following, by choosing 
$e^{i}_I$ and
$e^{\star I}_i$ 
properly,
we assume that
$\eta_i$'s are $\pm 1$. 
In the rotated basis $\varphi$, the Lagrangian can be written as 
\begin{align}
\mathcal{L}&= \frac{1}{4\pi} \left(\eta_{ij}\partial_t\varphi^i\partial_x\varphi^{j}-v\delta_{ij}\partial_x\varphi^i\partial_x\varphi^{j} \right)
\nonumber \\
&\quad +\frac{e}{2\pi}\epsilon^{\mu\nu} \tilde{Q}_i \partial_{\mu}\varphi^i A_{\nu}
+\frac{s}{2\pi}\epsilon^{\mu\nu} \tilde{S}_i \partial_{\mu}\varphi^i B_{\nu},
\end{align}
where we have introduced the charge and spin vectors in the rotated 
basis as
\begin{align}
\tilde{Q}_i \equiv e^{\star I}_i Q_I, 
\quad 
\tilde{S}_i \equiv e^{\star I}_i S_I, 
\end{align}
and assumed,  
for simplicity,
$e^{\star I}_i V_{IJ}{e^{\star J}_j}=v_{ij}=v_i\delta_{ij}$. 
The compactification condition in the original basis 
(\ref{compactification})  
is translated into, 
in the rotated basis, 
\begin{align}
\label{compactification_2}
&
\varphi^i(t, x)
\equiv \varphi^i(t,x)
+2\pi e^{i}_I {n}^I.  
\end{align}
%
%

\paragraph{Quantization without the background fields}

As a warm up,
we first quantize canonically the theory without the background fields
on the spatial circle of radius $2\pi$:
\begin{align}
\mathcal{L}_0=& 
\frac{1}{4\pi} \left(\eta_{ij}\partial_t\varphi^i\partial_x\varphi^{j}
-v_{ij}\partial_x\varphi^i\partial_x\varphi^{j} \right).
\end{align}
The equal-time commutation relations are 
\begin{align}
\label{commutator_2}
[\varphi^i(t, x), \partial_{x}\varphi^j(t, x')]= 
-2\pi i(\eta^{-1})^{ij}
\delta(x-x'). 
\end{align}
The mode expansion of $\varphi$ is given by
\begin{align}
\label{expansion}
&\varphi^i(t,x)
=
\varphi^i_0
-
p_j 
\left[ (\eta^{-1})^{jk} v_{kl} (\eta^{-1})^{li} t+ (\eta^{-1})^{ji} x\right]
\nonumber \\
&\qquad 
+
i\sum_{n \neq 0} b_{nj} 
e^{-in\left[ (\eta^{-1})^{jk} v_{kl} (\eta^{-1})^{li} t+ (\eta^{-1})^{ji} x\right]  },
\end{align}
%
%
%
together with the commutation relation
\begin{align}
&[\varphi^i_0, p_j]=i\delta^{i}_j, 
\quad 
[b_{ni}, b_{mj}]=\frac{1}{m}\delta_{ij}\delta_{n+m}. 
\end{align}
(All other commutators vanish.) 

As 
$p_i$ is conjugate to 
$\varphi^i(t,x)$, 
which obeys the compactification condition
$
\varphi^i(t,x)\simeq 
\varphi^i(t,x)
+
2\pi e^{i}_I n^I 
$, 
the quantization condition of $p_i$ is given 
in terms of the reciprocal lattice vectors $e^{\star I}_i$ as
\begin{align}
 p_i=e^{\star I}_i m_I,
\quad
m_I
\in 
\mathbb{Z}^N.   
\end{align}
{\it I.e.,} 
while the coordinates 
$\varphi^i$ are compactified 
on a lattice $\Gamma$ spanned by
$\{ e^{i}\}$, 
the momenta $p_i$ lie in the  
reciprocal (dual) lattice $\tilde{\Gamma}$ 
spanned by
$\{ e^{\star}_i\}$. 
Observe also that, in a momentum eigenstate, 
the mode expansion (\ref{expansion}) implies
the boundary condition
\begin{align}
\varphi^i (t, x+2\pi)
&= \varphi^i (t, x) - 2\pi p_j (\eta^{-1})^{ij}
\nonumber \\
&= \varphi^i (t, x) + 2\pi e^{\star J}_j m_J (\eta^{-1})^{ij}
\nonumber \\
&= \varphi^i (t, x) + 2\pi e^{i}_I (K^{-1})^{IJ} m_J. 
\label{BC}
\end{align}
For generic integral values of $m_J$,
$(K^{-1})^{IJ} m_J$ are not integers and
hence 
the boson fields obey twisted boundary conditions.
The states corresponding to the momentum $p_j$ are 
represented by (by state-operator correspondence)  
the vertex operators  
\begin{align}
\ddagger \exp i p_i \varphi^i(t,x)\ddagger
=
\ddagger
\exp i m_I \phi_I (t,x)
\ddagger 
\end{align}
where $\ddagger \cdots \ddagger$ represents 
normal-ordering. 

Let us consider a subset of $\tilde{\Gamma}$,
that is obtained by choosing 
$m_I = K_{IJ}{\Lambda}^J$
with $\Lambda^J\in \mathbb{Z}^N$.  
For this choice, the momentum is given by 
\begin{align}
 p_i = \eta_{ij} e^{j}_J {\Lambda}^J
\end{align}
and the boson fields $\varphi^i$ obey untwisted boundary conditions. 
In the sector of the theory with this choice of momentum,
all excitations are local (excitations consisting of exciting electron-like particles). 
The corresponding vertex operators are 
\begin{align}
\ddagger \exp i \Theta (\boldsymbol{\Lambda}) \ddagger
=
\ddagger
\exp i \Lambda^I K_{IJ} \phi_J (t,x)
\ddagger. 
\end{align}

To summarize,
quantization of the K-matrix theory with the compactification conditions
(\ref{K-matrix}-\ref{compactification}) gives rise to
the spectrum of local (electrons) as well as 
non-local (quasiparticle) excitations, which 
are represented by untwisted and twisted boundary conditions, 
respectively. 
Once we specify the boundary condition 
by  some integer vector $\boldsymbol{m}$, 
we obtain the spectrum quantized within one sector 
(labeled by the equivalent class $[\boldsymbol{m}]$ with the relation $\boldsymbol{m}\equiv \boldsymbol{m}+K\boldsymbol{\Lambda}$) 
of the total spectrum. 
There are $|\det K|$ sectors in this compactified K-matrix theory.

The Hamiltonian and total momentum are  
\begin{align}
H_0&=\frac{1}{4\pi}\int^{2\pi}_0 dx\partial_x\varphi^iv_{ij}\partial_x\varphi^j \nonumber \\
&=
\frac{1}{2}\left(\eta^{-1}p \right)^iv_{ij}\left(\eta^{-1}p \right)^j 
-\frac{1}{24}\text{tr}\left(\eta^{-1}v\eta^{-1}\right)
\nonumber \\
&\quad+\sum^{\infty}_{n=1} n^2\left(\eta^{-1}b_{-n}\right)^iv_{ij}\left(\eta^{-1}b_n\right)^j
\label{Hamiltonian}
\end{align}
and 
\begin{align}
P_0&=\frac{1}{4\pi}\int^{2\pi}_0 dx\partial_x\varphi^i\eta_{ij}\partial_x\varphi^j \nonumber \\
&= 
\frac{1}{2}p_i(\eta^{-1})^{ij}p_j
-\frac{1}{24}\text{tr}\left(\eta^{-1}\right) 
+\sum^{\infty}_{n=1} n^2 b_{-ni}(\eta^{-1})^{ij}b_{nj}, 
\label{total_momentum}
\end{align}
respectively.
The eigenstates of $H_0$ and $P_0$ can be expressed as 
a direct product of their oscillator part
(the Fock states generated by $b_{ni}$)
and non-oscillator part 
(related to $\varphi^i_0$ and $p_i$).
For the non-oscillator part, 
one can choose to use the momentum eigenvalues $\{p_i\}$, which have values  $\{\eta_{ij} e^{j}_J {\Lambda}^J+e^{\star I}_i m_I\}$ as  the boundary condition $\phi^I(t, x+2\pi)=\phi^I(t, x)+2\pi (K^{-1})^{IJ}m_J$ [or $\varphi^i(t, x+2\pi)= \varphi^i (t, x) + 2\pi e^{i}_I (K^{-1})^{IJ} m_J$ 
in the rotated basis] is specified, to label the eigenstates. We denote these eigenstates of sector $[\boldsymbol{m}]$ as 
 $|{\boldsymbol{\Lambda}_{\boldsymbol{m}}}\rangle\equiv |\boldsymbol{\Lambda}+K^{-1}\boldsymbol{m}\rangle$.

The partition function for the sector $[\boldsymbol{m}]$ evaluated on a torus with modular parameter $\tau=\tau_1+i\tau_2$ is given by 
\begin{align}
& Z_{\boldsymbol{m}}(\tau)=\text{Tr}_{\boldsymbol{m}}
 \left[ e^{-2\pi i\tau_1P_0 }e^{-2\pi\tau_2 H_0} \right].
\end{align}

\subsubsection{Twisted boundary conditions by U(1) symmetries}
The K-matrix theory (\ref{K-matrix}) has $\mathrm{U}(1)^N$ symmetries. The corresponding conserved charges are given by
\begin{align}
\mathcal{C}^I\equiv \frac{1}{2\pi}\int^{2\pi}_0 dx\partial_x\phi^I=-2\pi (K^{-1})^{IJ}e^{j}_Jp_j.
\end{align}
The global U(1) transformations associated to these charge degrees of freedom are generated by 
\begin{align}
&\quad 
\mathcal{G}(\boldsymbol{\alpha})\equiv e^{-2\pi i\alpha_I\mathcal{C}^I} 
\nonumber \\
 \mbox{as}&\quad 
 \mathcal{G}(\boldsymbol{\alpha})\varphi^i(t,x) 
 \mathcal{G}(\boldsymbol{\alpha})^{-1}
=\varphi^i(t,x)+2\pi(\eta^{-1})^{ij} e^{\star J}_j\alpha_J,
\end{align}
where $\boldsymbol{\alpha}$ is 
a vector consisting of twisting phases.  

Now, 
starting from the original boundary condition for sector $[\boldsymbol{m}]$
(\ref{BC}), 
we can generate a new twisted boundary condition
by acting with $\mathcal{G}$
as 
\begin{align}
 \label{twisted bc, original}
\phi^I(t, x+2\pi)&=
\mathcal{G}(\boldsymbol{a}) \phi^I(t, x)\mathcal{G}(\boldsymbol{a})^{-1} +2\pi( K^{-1})^{IJ}m_J\nonumber \\
&= \phi^I(t, x)+2\pi (K^{-1})^{IJ}(m_J+a_J), 
\end{align}
or, in the rotated basis, 
\begin{align}
\varphi^i(t, x+2\pi)&=
 \mathcal{G}(\boldsymbol{a})\varphi^i(t,x)\mathcal{G}(\boldsymbol{a})^{-1}+2\pi (\eta^{-1})^{ij} e^{\star J}_j m_J \nonumber \\
&= \varphi(t,x)+2\pi(\eta^{-1})^{ij} e^{\star J}_j\left(m_J+a_J\right).
\label{twisted_BC}
\end{align}
With this twisted boundary condition,
the allowed values of the momenta $p$ are now shifted 
and given by
\begin{align}
& p_i=
e^{\star I}_i\left( m_I+K_{IJ}\Lambda^J+a_I\right) 
\equiv 
e^{\star I}_i K_{IJ}\Lambda^J_{\boldsymbol{m}+\boldsymbol{a}},
\end{align}
where
\begin{align}
 \label{def Lambda with a and m}
\Lambda^J_{\boldsymbol{m}+\boldsymbol{a}} 
: =
\Lambda^J 
+
(K^{-1})^{JI} \left(m_I+a_I \right), \  \Lambda^J\in \mathbb{Z}^N.
\end{align}
As in the untwisted case [in the absence of U(1) twisting phases], 
the eigenstates of the Hamiltonian and the total momentum
can be expressed as 
a direct product of their oscillator part
and non-oscillator part.
One can choose to use the momentum eigenvalues,
which are specified by a set of integers $\Lambda^J \in \mathbb{Z}^N$  
to label non-oscillator part of the eigenstates.
We denote these basis states as 
 $
 |{\boldsymbol{\Lambda}}_{\boldsymbol{m}+\boldsymbol{a}}\rangle 
$, 
which are given by the untwisted eigenstates with $\boldsymbol{m}$ shifted by $\boldsymbol{a}$.

\subsubsection{Twisted partition function}

The twisted partition function for sector $[\boldsymbol{m}]$  evaluated  on a torus with modular parameter $\tau=\tau_1+i\tau_2$ is given by 
\begin{align}
Z_{\boldsymbol{m}[\boldsymbol{a},\boldsymbol{b}]}(\tau)&=\text{Tr}_{\boldsymbol{m}+\boldsymbol{a}}
 \left[ \mathcal{G}(\boldsymbol{b})e^{-2\pi i\tau_1P_0 }e^{-2\pi\tau_2 H_0} \right],
\end{align}
where the trace is taken over the Hilbert space in the presence of
the twisted boundary condition generated by $\mathcal{G}(\boldsymbol{a})$. 
The operator insertion $\mathcal{G}(\boldsymbol{b})$
generates, in the path-integral picture, twisted boundary condition
in time direction. 
The partition function can be expressed as a product
of the oscillator part and the zero-mode part as 
\begin{align}
 \label{Z_m}
& Z_{\boldsymbol{m}[\boldsymbol{a},\boldsymbol{b}]}(\tau) = 
 \xi(\tau)\sum_{\boldsymbol{\Lambda}\in\mathbb{Z}^N}\zeta_{[\boldsymbol{m}+\boldsymbol{a},\boldsymbol{b}]}^{\boldsymbol{\Lambda}}(\tau)
 \langle \boldsymbol{\Lambda}_{\boldsymbol{m}+\boldsymbol{a}} |\boldsymbol{\Lambda}_{\boldsymbol{m}+\boldsymbol{a}}\rangle,
 \nonumber \\
 &\mbox{where}
 \nonumber \\
 &\zeta_{[\boldsymbol{m}+\boldsymbol{a},\boldsymbol{b}]}^{\boldsymbol{\Lambda}}(\tau)
 \equiv 
 \exp\left(2\pi i\boldsymbol{b}^T
  \boldsymbol{\Lambda}^{\ }_{\boldsymbol{m}+\boldsymbol{a}}\right. 
-\pi i\tau_1\boldsymbol{\Lambda}_{\boldsymbol{m}+\boldsymbol{a}}^TK\boldsymbol{\Lambda}^{\ }_{\boldsymbol{m}+\boldsymbol{a}}
\nonumber \\
&\quad \left.-\pi\tau_2\boldsymbol{\Lambda}_{\boldsymbol{m}+\boldsymbol{a}}^TV\boldsymbol{\Lambda}^{\ }_{\boldsymbol{m}+\boldsymbol{a}}\right).  
\end{align}
The oscillator part of the partition function
$\xi(\tau)$
is independent of the twisting angles $\boldsymbol{a}$ and $\boldsymbol{b}$ 
and 
will not play any important role in the following discussion.
The overlap 
$\langle \boldsymbol{\Lambda}_{\boldsymbol{m}+\boldsymbol{a}}|\boldsymbol{\Lambda}_{\boldsymbol{m}+\boldsymbol{a}}\rangle$
in Eq.\ (\ref{Z_m}) 
is simply 
$\langle \boldsymbol{\Lambda}_{\boldsymbol{m}+\boldsymbol{a}}|\boldsymbol{\Lambda}_{\boldsymbol{m}+\boldsymbol{a}}\rangle=1$, 
but we 
displayed 
$\langle \boldsymbol{\Lambda}_{\boldsymbol{m}+\boldsymbol{a}}|\boldsymbol{\Lambda}_{\boldsymbol{m}+\boldsymbol{a}}\rangle$
in Eq.\ (\ref{Z_m}) 
for the later comparison.  

\subsubsection{Large gauge transformations}

The large gauge transformations of U(1) symmetries are finite gauge transformations that preserve the spectrum of the theory. 
They are finite shifts of twisting 
phases $\boldsymbol{a}$ and $\boldsymbol{b}$ 
that preserve the U(1) operators $\mathcal{G}$ 
[or more precisely, 
preserve the (twisted) boundary conditions] 
and can be deduced from the compactification condition
of the K-matrix theory (\ref{compactification}). 
For $\mathrm{U(1)}^N$ symmetry, the large gauge transformations are given by
\begin{align}
\boldsymbol{a}\rightarrow \boldsymbol{a}+K\boldsymbol{\delta}, \ \ 
\boldsymbol{b}\rightarrow \boldsymbol{b}+K\boldsymbol{\delta}',\ \ \forall  \boldsymbol{\delta}, \boldsymbol{\delta}'\in\mathbb{Z}^N.
\end{align}
%
%
%

To discuss the behavior of the twisted partition
function under the large gauge transformation, 
let us consider $\widetilde{\mathrm{U(1)}}^2=\mathrm{U(1)}_c\times\mathrm{U(1)}_s$ symmetry: $\boldsymbol{a}=a_c\boldsymbol{Q}+a_s\boldsymbol{S}$ and  $\boldsymbol{b}=b_c\boldsymbol{Q}+b_s\boldsymbol{S}$, where $\boldsymbol{Q}$ and $\boldsymbol{S}$ are charge and spin vectors, respectively. The minimal shifts are given by
\begin{align}
\delta_c=1/\beta_c, \quad
\delta_s=1/\beta_s, \quad
\end{align}
where 
$\beta_c\equiv 
\min_{\boldsymbol{l}}|\boldsymbol{l}^TK^{-1}\boldsymbol{Q}|$ 
and $\beta_s\equiv \min_{\boldsymbol{l}}|\boldsymbol{l}^TK^{-1}\boldsymbol{S}|$ represent 
the elementary charge and spin (the smallest fractional charge and spin of 
quasiparticle excitations) of the system, respectively.
In other words, 
{\it classically},
the system is expected to be invariant 
under the following
large gauge transformation:  
\begin{align}
a_{c/s} \rightarrow a_{c/s} + \delta_{c/s}, \quad
b_{c/s} \rightarrow b_{c/s} + \delta_{c/s}.
\end{align}

The invariance under the large gauge transformation may, however, be 
violated at the quantum level.
From Eq.\ (\ref{Z_m}),
we see, under large gauge transformations for the charge $\mathrm{U}(1)_c$ symmetry,
\begin{align}
 & Z_{\boldsymbol{m}\left[a_c+\delta_c,b_c,a_s,b_s\right]}=Z_{\boldsymbol{m}[a_c,b_c,a_s,b_s]}, 
\nonumber \\
&Z_{\boldsymbol{m}\left[a_c,b_c+\delta_c,a_s,b_s\right]} 
=e^{2\pi i\delta_c \boldsymbol{Q}^T K^{-1} \left(a_c\boldsymbol{Q}+a_s\boldsymbol{S}\right)} 
\cdot 
Z_{\boldsymbol{m}[a_c,b_c,a_s,b_s]}.  
\end{align}
Similarly,
under large gauge transformations for the spin $\mathrm{U}(1)_s$ symmetry,
\begin{align}
&Z_{
 \boldsymbol{m}\left[a_c,b_c,a_s+\delta_s,b_s \right]}=Z_{\boldsymbol{m}[a_c,b_c,a_s,b_s]}, 
\nonumber \\
&Z_{\boldsymbol{m}\left[a_c,b_c, a_s,b_s+\delta_s\right]}
=e^{
 2\pi i\delta_s \boldsymbol{S}^T K^{-1}\left(a_c\boldsymbol{Q}+a_s\boldsymbol{S}\right)} 
\cdot
Z_{\boldsymbol{m}[a_c,b_c,a_s,b_s]}. 
\end{align}
Observe that 
the way the partition function changes 
under the large gauge transformations 
does not depend on the sector $\boldsymbol{m}$ we specify.

In the cases where there is only the charge U(1) symmetry, 
the above calculation tells us that $Z_{[a_c,b_c]}(\tau)$  is not invariant under the large-gauge transformations 
 if
\begin{align}
\boldsymbol{Q}^TK^{-1}\boldsymbol{Q} \neq 0. 
\end{align}
This large gauge anomaly is nothing but 
the quantum Hall effect.

\subsection{Symmetry projected partition functions: generalities}
\label{Symmetry projected partition functions: generalities}

Now let us move on to the situations of our main interest.
We consider the $K$-matrix theory
that preserves one of the $\mathrm{U}(1)$ symmetries,
$\mathrm{U}(1)_{c}$ or $\mathrm{U}(1)_{s}$,
but not both.
We denote this U(1) symmetry by $\mathscr{G}=\mathrm{U}(1)_{c,s}$.  
In addition, we assume
the $K$-matrix theory is 
invariant under yet another  
global unitary symmetry; we call the corresponding symmetry group
$\mathscr{G}'$.
In our examples below, $\mathscr{G}'$ consists of a single 
discrete unitary symmetry transformation such as
CP or P transformation. 
The total symmetry group is $\mathscr{G}\rtimes \mathscr{G}'$. 

Under the action of a symmetry generator $\mathcal{M}\in \mathscr{G}'$, 
the bosonic fields transform as:
\begin{align}
\mathcal{M}\boldsymbol{\phi}(t,x) \mathcal{M}^{-1}
=U_{{M}} \boldsymbol{\phi}(t, r_{{M}}x)+\pi K^{-1}\boldsymbol{\chi}_{{M}}, 
\label{G_trans_phi}
\end{align}
where $U_{{M}}$ is an integral $N\times N$ matrix, $r_{{M}}$ 
is a real number, and $\boldsymbol{\chi}_{{M}}$ is some $N$-component real vector. 
For on-site symmetry, $r_{{M}}=1$. 
For non-on-site symmetry 
(below we consider parity, P,  
or some on-site symmetry combined with parity,
such as CP),  
we have $r_{{M}}=-1$.
Assuming the $K$-matrix theory  
is invariant under group $\mathscr{G}'$,
$U_M$ and $r_M$ must satisfy
\begin{align}
 U^T_{{M}}K U^{\ }_{{M}}
=
r_{{M}}K, 
\quad
U^T_{{M}}V U^{\ }_{{M}}=r_{{M}}^2V=V,  
\label{M_constraint_1}
\end{align}
for any $\mathcal{M}\in \mathscr{G}'$. 
The invariance under $\mathscr{G}'$ also 
imposes constraints 
on the integer vector $\boldsymbol{Q}$ or $\boldsymbol{S}$ 
through the way the charge or spin current  
are transformed under $\mathscr{G}'$.

Following our discussion in the previous sections,
our strategy to diagnose the stability of the edge theory 
is to enforce the invariance under $\mathscr{G}'$ by projection,
and discuss the dependence of the projected partition function
on the twisting phases. 
In order for this strategy to work,
the twisted boundary conditions 
should be invariant under the symmetry $\mathscr{G}'$.  
Acting 
on the twisted boundary condition
with a symmetry generator
(\ref{twisted bc, original}), 
\begin{align}
\label{sym_comp_twisted_BCs}
&
\mathcal{M}K \boldsymbol{\phi}(t, x+2\pi)\mathcal{M}^{-1}
\nonumber \\
&\quad
=\mathcal{M}K\boldsymbol{\phi}(t, x)\mathcal{M}^{-1}+2\pi 
\left(\boldsymbol{m} + \boldsymbol{a}\right)\nonumber\\
\Rightarrow\, 
&
K\boldsymbol{\phi}(t, r_{{M}}(x+2\pi))
\nonumber\\
&\quad 
=K \boldsymbol{\phi}(t, r_{{M}}x)+2\pi r_{{M}} U^{T}_{M}
\left(\boldsymbol{m}+ \boldsymbol{a}\right).
\end{align}
In order for the twisted boundary condition
(\ref{twisted bc, original})
to be invariant under $\mathcal{M}$
for arbitrary value of $a_{c}$ ($a_{s}$),  
the charge vector $\boldsymbol{Q}$ (the spin vector $\boldsymbol{S}$) must satisfy
\begin{align}
 \label{condition on charge and spin vectors}
U_{{M}}^T\boldsymbol{m}=\boldsymbol{m} \quad \mbox{and} \quad U_{{M}}^T\boldsymbol{Q}=\boldsymbol{Q}\quad  (U_{{M}}^T\boldsymbol{S}=\boldsymbol{S}), 
\end{align}
respectively.  
In our discussion below,
we assume, for given $\mathscr{G}$ and $\mathscr{G}'$, 
this condition is satisfied.

Finally, from the group structure of $\mathscr{G}'$ 
and the statistics of vertex operators,  
which represent local excitations, 
there are further constrains on 
the possible form of $U_{{M}}$ and $\boldsymbol{\chi}_{{M}}$. 
This issue will be discussed in more details later with specific examples. 
In conclusion, a general $K$-matrix theory with 
symmetry group $\mathscr{G}\rtimes \mathscr{G}'$ 
is described by the data 
$\{K,\boldsymbol{Q}\, \mbox{or}\, \boldsymbol{S},\{U_{{M}},\boldsymbol{\chi}_{{M}}|\mathcal{M}\in \mathscr{G}'\}\}$ that satisfies the conditions discussed above.

The symmetry projected partition function for 
the sector $[\boldsymbol{m}]$ is defined by
\begin{align}
 \label{M_proj_Z_m}
& Z^{\mathrm{Proj}}_{\boldsymbol{m}[\boldsymbol{a},\boldsymbol{b}]} 
 \equiv
 \text{Tr}_{\boldsymbol{m}+\boldsymbol{a}} \left[\mathcal{P}_{\mathscr{G}'}
 \mathcal{G}(\boldsymbol{b}) e^{-2\pi i\tau_1P_0 }e^{-2\pi\tau_2 H_0} \right], 
 \nonumber \\
&\quad
\mbox{where}
\quad 
\mathcal{P}_{\mathscr{G}'}=|\mathscr{G}'|^{-1}\sum_{\mathcal{M}\in \mathscr{G}'}\mathcal{M} 
\end{align}
is the projection operator for the symmetry group $\mathscr{G}'$, 
satisfying 
$\mathcal{P}_{\mathscr{G}'}^2
=\mathcal{P}^{\ }_{\mathscr{G}'}$. 
The trace in 
Eq.\ (\ref{M_proj_Z_m}) 
is taken with respect to the Hilbert space 
in the presence of boundary conditions twisted by 
$\mathcal{G}(\boldsymbol{a})$,  
and the insertion of the operator $\mathcal{G}(\boldsymbol{b})$ 
inside the trace represents, in the path integral picture,  
the U(1) twisting phase in the temporal direction.
As mentioned earlier, the twisting should be invariant 
under $\mathscr{G}'$, and hence 
typically only the charge twisting angles $[a_c, b_c]$ or 
the spin twisting angles $[a_s, b_s]$ is allowed. 
In this section, we discuss some general properties of  
$Z^{\mathrm{Proj}}_{\boldsymbol{m}[\boldsymbol{a},\boldsymbol{b}]}$ 
keeping both charge and spin twisting angles.
Once $\mathscr{G}'$ is given,  
and the invariance of the twisting boundary condition by 
$\mathscr{G}'$ is taken into account, 
it is easy to ``switch off'' either one of charge or spin angle.

The twisted partition function,
that appears as a part of the projected partition function 
$Z^{\mathrm{Proj}}_{\boldsymbol{m}[\boldsymbol{a},\boldsymbol{b}]}$, 
can be evaluated as 
\begin{align}
 &Z^{M}_{\boldsymbol{m}[\boldsymbol{a},\boldsymbol{b}]}(\tau)
=\mathrm{Tr}_{\boldsymbol{m}+\boldsymbol{a}}
\left[\mathcal{M}\mathcal{G}(\boldsymbol{b})e^{-2\pi i\tau_1P_0 }e^{-2\pi\tau_2 H_0} \right]
\nonumber \\
&\quad\quad= \xi^{{M}}(\tau)
\sum_{\boldsymbol{\Lambda}\in\mathbb{Z}^N}
\zeta^{\boldsymbol{\Lambda}}_{[\boldsymbol{m}+\boldsymbol{a},\boldsymbol{b}]}
\langle \boldsymbol{\Lambda}_{\boldsymbol{m}+\boldsymbol{a}} |\mathcal{M}|\boldsymbol{\Lambda}_{\boldsymbol{m}+\boldsymbol{a}}\rangle, 
\label{M_twist_Z_m}
\end{align}
where the oscillator part of the partition function
$\xi^{{M}}(\tau)$ does not depend on the twisting phases 
$\boldsymbol{a}$ and 
$\boldsymbol{b}$.
The most crucial part of the calculations, 
as inferred from the previous examples of 
the Dirac fermions and the single-component boson,  
is the matrix element 
$\langle \boldsymbol{\Lambda}_{\boldsymbol{m}+\boldsymbol{m}+\boldsymbol{a}} |\mathcal{M}|\boldsymbol{\Lambda}_{\boldsymbol{m}+\boldsymbol{a}}
\rangle$ in Eq.\ (\ref{M_twist_Z_m}).
As one can read off from
Eq.\ (\ref{G_trans_phi}), 
the transformation 
$\mathcal{M}$ maps the momentum eigenvalues
$
\boldsymbol{\Lambda}_{\boldsymbol{m}+\boldsymbol{a}}
\to 
r_M U_M \boldsymbol{\Lambda}_{\boldsymbol{m}+\boldsymbol{a}}
$,
and hence 
$\mathcal{M}|\boldsymbol{\Lambda}_{\boldsymbol{m}+\boldsymbol{a}}\rangle$
should be equal to 
$|r_M U_M \boldsymbol{\Lambda}_{\boldsymbol{m}+\boldsymbol{a}}\rangle$
up to a phase factor.

To calculate this phase factor, 
in particular in the presence of the Klein factors, 
it is convenient to use 
the state-operator correspondence;
according to the state-operator correspondence,
for each sector of the Hilbert space 
constructed out of the 
zero-mode $|\boldsymbol{\Lambda}_{\boldsymbol{m}+\boldsymbol{a}}\rangle$,
we have a corresponding operator
\begin{align}
 \ddagger \, \exp{ i \boldsymbol{\Lambda}^T_{\boldsymbol{m}+\boldsymbol{a}} K \boldsymbol{\phi} }\,  \ddagger.
\end{align}

As a warm up, let us consider the untwisted 
($\boldsymbol{a}=0$)
counterpart when $\boldsymbol{m}=0$
\begin{align}
 \ddagger \, \exp{ i \boldsymbol{\Lambda}^T K \boldsymbol{\phi} } \, \ddagger. 
\end{align}
Now, using symmetry conditions (\ref{M_constraint_1}) we have
\begin{align}
&\mathcal{M}
\ddagger e^{i\boldsymbol{\Lambda}^TK\boldsymbol{\phi}(t,0)}\ddagger 
\mathcal{M}^{-1}
\nonumber \\
&= 
e^{i\Delta\boldsymbol{\phi}^{\boldsymbol{\Lambda}}_{{M}}}\ddagger 
e^{i\boldsymbol{\Lambda}^TK\left(\mathcal{M}\boldsymbol{\phi}(t,0) \ \mathcal{M}^{-1}\right)}
\ddagger 
\nonumber \\
&=e^{i\Delta\boldsymbol{\phi}^{\boldsymbol{\Lambda}}_{{M}}}
\ddagger 
e^{i\boldsymbol{\Lambda}^TK\left(U_{{M}}\boldsymbol{\phi}(t,0)+\pi K^{-1}\boldsymbol{\chi}_{{M}}\right)}
\ddagger 
\nonumber \\
&=e^{i\Delta\boldsymbol{\phi}^{\boldsymbol{\Lambda}}_{{M}}}
e^{i\pi \boldsymbol{\Lambda}^{T}\boldsymbol{\chi}_{{M}}}
\ddagger 
e^{i(r_{{M}}U_{{M}}\boldsymbol{\Lambda})^TK\boldsymbol{\phi}(t,0)}
\ddagger 
\label{vertex_under_M},
\end{align}
where $e^{i\Delta\boldsymbol{\phi}^{\boldsymbol{\Lambda}}_{{M}}}$ 
is the statistical phase factor of the vertex operator 
$\ddagger e^{i\boldsymbol{\Lambda}^TK\boldsymbol{\phi}} \ddagger$ under symmetry transformation $\mathcal{M}$, as explained in 
Appendix 
\ref{Statistical phase factor of the chiral boson field under symmetry transformation}. 
For bosonic systems, such phase factor $e^{i\Delta\boldsymbol{\phi}^{\boldsymbol{\Lambda}}_{{M}}}$ 
equals to $1$ because of the commutativity among bosons. 
For fermionic systems, however, 
we must take into account the anti-commutativity among fermions,
which may lead to an additional phase factor for the transformation 
of the vertex operator. 

In the presence of the twisting angles,
the action of $\mathcal{M}$ on non-oscillator state
$|\boldsymbol{\Lambda}_{\boldsymbol{m}+\boldsymbol{a}}\rangle$
may give rise to an additional phase factor which
in principle depends on $\boldsymbol{m}+\boldsymbol{a}$. 
Let us now take a close look at this. 
States $|\boldsymbol{\Lambda}_{\boldsymbol{m}+\boldsymbol{a}}\rangle$
labeled by shifted momentum
$\boldsymbol{\Lambda}_{\boldsymbol{m}+\boldsymbol{a}}$
can be viewed as generated 
from a ground state
$| \mathrm{GS}_{\boldsymbol{m}+\boldsymbol{a}} \rangle$
by acting on some raising operator. 
By assumption, 
$| \mathrm{GS}_{\boldsymbol{m}+\boldsymbol{a}} \rangle$ 
is invariant under $\mathcal{M}$, 
and hence 
\begin{align}
 \mathcal{M}| \mathrm{GS}_{\boldsymbol{m}+\boldsymbol{a}} \rangle=
 P^{{M}}_{[\boldsymbol{m}+\boldsymbol{a}]}| \mathrm{GS}_{\boldsymbol{m}+\boldsymbol{a}} \rangle, 
\end{align}
where the eigenvalue of $\mathcal{M}$ for the ground state, 
denoted by $P^{{M}}_{[\boldsymbol{m}+\boldsymbol{a}]}$,
depends on the spatial twisting phases.
The phase 
$P^{{M}}_{[\boldsymbol{m}+\boldsymbol{a}]}$  
together with
$e^{i\left[\Delta\boldsymbol{\phi}^{\boldsymbol{\Lambda}}_{{M}}+\pi \boldsymbol{\Lambda}^{T}\boldsymbol{\chi}_{{M}}\right]}$
in
Eq.\ (\ref{vertex_under_M}),
leads to 
\begin{align}
 \mathcal{M}|\boldsymbol{\Lambda}_{\boldsymbol{m}+\boldsymbol{a}} \rangle
=
P^{{M}}_{[\boldsymbol{m}+\boldsymbol{a}]}
e^{i\left[\Delta\boldsymbol{\phi}^{\boldsymbol{\Lambda}}_{{M}}+\pi \boldsymbol{\Lambda}^{T}\boldsymbol{\chi}_{{M}}\right]}
|r_{{M}}U_{{M}}\boldsymbol{\Lambda}_{\boldsymbol{m}+\boldsymbol{a}}\rangle.
\end{align}
In other words, 
the dependence on the twisting angles comes only from   
$P^{{M}}_{[\boldsymbol{m}+\boldsymbol{a}]}$, 
but not from  
$e^{i\left[\Delta\boldsymbol{\phi}^{\boldsymbol{\Lambda}}_{{M}}+\pi \boldsymbol{\Lambda}^{T}\boldsymbol{\chi}_{{M}}\right]}$. 
In fact, as we noted previously in 
the case of the Dirac fermion and the single-component boson,
once we insist on invariance under $\mathscr{G}'$, 
the eigenvalues of the symmetry transformation would not change
as we change $a_{s,c}$. 
If so, the phase 
$e^{i\left[\Delta\boldsymbol{\phi}^{\boldsymbol{\Lambda}}_{{M}}+\pi \boldsymbol{\Lambda}^{T}\boldsymbol{\chi}_{{M}}\right]}$,
which we get from the vertex operator of the untwisted theory, 
is the only phase factor that we need to keep track of.  


Since 
$\mathcal{M}$ maps the momentum eigenvalues
$
\boldsymbol{\Lambda}_{\boldsymbol{m}+\boldsymbol{a}}
\to 
r_M U_M \boldsymbol{\Lambda}_{\boldsymbol{m}+\boldsymbol{a}}
$,
in the summation in Eq.\ (\ref{M_twist_Z_m}), only those $\boldsymbol{\Lambda}$s
that satisfy 
$\boldsymbol{\Lambda}_{\boldsymbol{m}+\boldsymbol{a}} = r_M U_M \boldsymbol{\Lambda}_{\boldsymbol{m}+\boldsymbol{a}}$
contribute.
With the conditions (\ref{M_constraint_1})
and (\ref{condition on charge and spin vectors}),
this means
that 
the first term in
$\boldsymbol{\Lambda}_{\boldsymbol{m}+\boldsymbol{a}}$
in Eq.\ (\ref{def Lambda with a and m}) 
satisfies $\boldsymbol{\Lambda} = r_M U_M \boldsymbol{\Lambda}$. 
Then the twisted partition function (\ref{M_twist_Z_m}) is given by  
\begin{align}
\label{M_twist_Z_m_2}
&Z^{M}_{\boldsymbol{m}[\boldsymbol{a},\boldsymbol{b}]}(\tau) \nonumber\\
&= 
\xi^{{M}}(\tau)
P^{{M}}_{[\boldsymbol{m}+\boldsymbol{a}]}
\sum_{
 \substack{
  \boldsymbol{\Lambda} \in \mathbb{Z}^{N},  \\
  \boldsymbol{\Lambda}=r_{{M}}U_{M}\boldsymbol{\Lambda}
 }
}
e^{i\left[ \Delta\boldsymbol{\phi}^{{\boldsymbol{\Lambda}}}_{{M}}+\pi {\boldsymbol{\Lambda}}^{T}\boldsymbol{\chi}_{{M}}\right]}
\zeta_{[\boldsymbol{m}+\boldsymbol{a},\boldsymbol{b}]}^{{\boldsymbol{\Lambda}}}(\tau). 
\end{align}

From Eq.\ (\ref{M_twist_Z_m_2}) 
we observe that the symmetry projected partition function 
for the sector $[\boldsymbol{m}]$ depends only on parameters 
$\boldsymbol{m}+\boldsymbol{a}$ and $\boldsymbol{b}$. 
This means that the way the projected partition function changes 
under large gauge transformation does not depend on 
$\boldsymbol{m}$ (i.e., it is independent of sector). 
For compactness, we will drop the label $\boldsymbol{m}$ 
(or just set $\boldsymbol{m}=\boldsymbol{0}$) on partition functions in the following discussion.

\subsection{$\mathscr{G}\rtimes \mathscr{G}'=\mathrm{U}(1)_c\rtimes Z^{\mathrm{CP}}_2$}

Now we consider the non-on-site  CP symmetry.
Let
\begin{align}
(\mathcal{CP})\boldsymbol{\phi}(t,x) (\mathcal{CP})^{-1}
=U_{\mathrm{CP}} \boldsymbol{\phi}(t,-x)+\pi K^{-1}\boldsymbol{\chi}_{\mathrm{CP}}, 
\label{CP_trans_phi}
\end{align}
where $U_{\mathrm{CP}}$ is an integer $N\times N$ matrix (the same as the dimension of $K$) and $\boldsymbol{\chi}_{\mathrm{CP}}$ is some $N$-component real vector. 
In order for the system to be CP invariant, we require 
\begin{align}
& U^T_{\mathrm{CP}}K U^{\ }_{\mathrm{CP}}=-K,
 \quad
U^T_{\mathrm{CP}}V U_{\mathrm{CP}}=V,  
\quad
U^2_{\mathrm{CP}}=I_N,
\nonumber \\
&U_{\mathrm{CP}}^T\boldsymbol{\boldsymbol{Q}}=\boldsymbol{\boldsymbol{Q}},
\quad 
(I_N-U_{\mathrm{CP}}^T)\boldsymbol{\chi}_{\mathrm{CP}}
=
2\epsilon 
 \boldsymbol{\boldsymbol{Q}} \mod 2, \label{CP_constraint}
\end{align}
where $I_N$ is the $N\times N$ identity matrix and the value 
$\epsilon=0, 1/2$ represents the sign of 
the CP operator squared for fermionic systems, 
with the relation 
\begin{align}
 (\mathcal{CP})^2= e^{i2 \pi \epsilon {N}_f},
\end{align}
where $N_f$ is the total fermion number operator. 

In fact, 
these constraints on $(K, \boldsymbol{Q}, U_{\mathrm{CP}}, \boldsymbol{\chi}_{\mathrm{CP}})$ 
are identical to the corresponding data in $K$-matrix theories  
with time-reversal invariance. 
\cite{LevinStern2012} 
The most general gauge inequivalent solution (which exists for a non-chiral K-matrix theory; $N$ must be even) is of the form 
\begin{align}
K&=\left(
    \begin{array}{cccc}
      0 & A & B & B\\
      A^T & 0 & C & -C \\
      B^T & C^T & \Gamma & W \\
      B^T & -C^T & W^T & -\Gamma
    \end{array}
  \right), \
\boldsymbol{Q}=\left(
    \begin{array}{c}
      0\\
      \boldsymbol{q}'\\
      \boldsymbol{q}\\
      \boldsymbol{q}
    \end{array}
  \right), \nonumber \\ 
U_{\mathrm{CP}}
&=\left(
    \begin{array}{cccc}
      -I_M & 0 & 0 & 0\\
      0 & I_M & 0 & 0 \\
      0 & 0 & 0 & I_{N/2-M} \\
      0 & 0 & I_{N/2-M} & 0
    \end{array}
  \right),   
  \nonumber \\
\boldsymbol{\chi}_{\mathrm{CP}}&=\left(
    \begin{array}{c}
      \boldsymbol{x}\\
      0\\
      (1-2\epsilon)\boldsymbol{x}'\\
     (1-2\epsilon)\boldsymbol{x}'+2 \epsilon \boldsymbol{q}
    \end{array} \label{CP_solution}
  \right).
\end{align}
Here, the matrix $A$ is $M \times M$, 
while the matrices $B$, $C$ are $M \times (N-M)$. 
The matrices $\Gamma$, $W$ are both $({N}/{2}-M) \times ({N}/{2}-M)$. 
Similarly, $\boldsymbol{q'}$ is of dimension $M$ and $\boldsymbol{\boldsymbol{Q}}$ 
is of dimension $({N}/{2}-M)$. 
Finally, $\boldsymbol{x}$ 
is a $M$-dimensional vector 
consisting of 1's and 0's, 
while $\boldsymbol{x}'$ is a 
$({N}/{2}-M)$-dimensional vector 
consisting of 1's and 0's. 
There are only a few constraints on $(A, B, C, \Gamma, W, \boldsymbol{q}, \boldsymbol{q}', \boldsymbol{x},\boldsymbol{x}')$. First, $W$ must be antisymmetric: $W =-W^T$. 
This requirement follows from CP symmetry (\ref{CP_constraint}). Second, $\boldsymbol{q}'$ must be even-valued.
This constraint comes from $Q_I=K_{II} \mod 2$,
which means the insulator is composed out of electrons. For the same reason, the parity of $Q_I$ must match with that of $K_{II}$, but can be either even or odd. Finally, the greatest common factor of $\{Q_I\}$ must be 1.

Once these data are given, 
we now calculate the CP symmetry projected partition function with charge U(1) symmetry ($\boldsymbol{a}=a_c\boldsymbol{Q}$, $\boldsymbol{b}=b_c\boldsymbol{Q}$):
\begin{align}
 &Z^{\mathrm{Proj}}_{[a_c,b_c]}(\tau)
 =\text{Tr}_{a_c}\left[\mathcal{P}_{\mathrm{CP}}
 \mathcal{G}(b_c)e^{-2\pi i\tau_1P_0 }e^{-2\pi\tau_2 H_0} \right], 
 \nonumber \\
 &\quad 
 \mbox{with}
 \quad 
 \mathcal{P}_{\mathrm{CP}}=
   \frac{1+
    \mathcal{CP}}{2}. 
\end{align}

\subsubsection{Bosonic systems}

In the case where the system is composed of bosons, 
the most general data $(K, \boldsymbol{Q}, U_{\mathrm{CP}}, \boldsymbol{\chi}_{\mathrm{CP}})$ is given by (\ref{CP_solution}), 
with an additional condition that 
the charge vector $\boldsymbol{Q}$ is even valued (and thus the diagonal elements of $\Gamma$ are also even). 
In this CP invariant theory, the function $\zeta_{[a_c,b_c]}^{{\boldsymbol{\Lambda}}}(\tau)$ 
with the constraint ${\boldsymbol{\Lambda}}=-U_{\mathrm{CP}}^T{\boldsymbol{\Lambda}}$  is given by
\begin{align}
\zeta_{[a_c,b_c]}^{{\boldsymbol{\Lambda}}}(\tau) &= 
\exp\left(-\pi\tau_2{\boldsymbol{\Lambda}}_{a_c}^TV{\boldsymbol{\Lambda}}^{\ }_{a_c}\right),
\end{align}
where ${\boldsymbol{\Lambda}}_{a_c}\equiv {\boldsymbol{\Lambda}} + a_cK^{-1}\boldsymbol{Q}$ and the fact $\boldsymbol{Q}^T\boldsymbol{\Lambda}_{a_c}=\boldsymbol{\Lambda}_{a_c}^TK\boldsymbol{\Lambda}_{a_c}=0$ (by CP symmetry) is used.
Therefore, the partition function
\begin{align}
 Z^{\mathrm{CP}}_{[a_c,b_c]}=\text{Tr}_{a_c}
 \left[(\mathcal{CP})
 \mathcal{G}(b_c)e^{-2\pi i\tau_1P_0 }e^{-2\pi\tau_2 H_0} \right]
\end{align}
is calculated as [Eq. (\ref{M_twist_Z_m_2})]
\begin{align}
 Z^{\mathrm{CP}}_{[a_c,b_c]}(\tau) 
 &= P^{\mathrm{CP}}_{[a_c]}
 \xi^{\mathrm{CP}}(\tau)
 \sum_{
  \substack{
   \boldsymbol{\Lambda} \in \mathbb{Z}^N \\
 \boldsymbol{\Lambda}=-U^T_{\mathrm{CP}}\boldsymbol{\Lambda} 
}
}
 e^{-\pi\tau_2 {\boldsymbol{\Lambda}}_{a_c}^T {\boldsymbol{\Lambda}}^{\ }_{a_c}+i\pi{\boldsymbol{\Lambda}}^T\chi_{\mathrm{CP}}}, 
 \label{Z^{CP}_U_boson}
\end{align}
where 
$P^{\mathrm{CP}}_{[a_c]}$ is the CP eigenvalue of the ground state.
Observe that 
the charge U(1) transformation operator
$\mathcal{G}(b_c)=e^{-2\pi ib_c{N}_f}$ 
and 
the spatial translation operator (in space coordinate $x$) 
$e^{-2\pi i\tau_1P_0}$ 
in the partition function are both projected out, 
leading to the independence of $a_c$ and $\tau_1$ 
in $Z^{\mathrm{CP}}$. 
This can also be argued by the fact that the total charge 
$J^0_c$ and momentum $P_0$ are odd under CP, 
while the Hamiltonian $H_0$ is even. 
For the same reason, 
the function $\xi^{\mathrm{CP}}$ just depends on $\tau_2$.
[See similar discussion 
near Eqs.\ (\ref{fermionic CP projected part fn})
and
(\ref{bosonic cp projected part fn}).]

The bosonic CP symmetry projected partition function is given by
\begin{align}
 Z^{\mathrm{Proj}}_{[a_c,b_c]}(\tau) 
 =\frac{1}{2}
 \left[
 Z_{[a_c,b_c]}(\tau)+Z^{\mathrm{CP}}_{[a_c]}(\tau_2)  \label{CP_proj_Z_boson}
\right],
\end{align}
with the form of $Z_{[a_c,b_c]}(\tau)$ 
given by (\ref{Z_m}). 
Under a large gauge transformation 
$a_c \rightarrow a_c + \delta_c$ and $b_c \rightarrow b_c + \delta_c$,
where 
$\delta_c\equiv 
(\min_{\boldsymbol{l}}|\boldsymbol{l}^TK^{-1}\boldsymbol{Q}|)^{-1}$ , we have
\begin{align}
 Z_{\left[a_c+\delta_c,b_c\right]}(\tau)
&= 
Z_{[a_c,b_c]}(\tau), 
\\
Z^{\mathrm{CP}}_{\left[a_c+\delta_c\right]}(\tau_2)
&= 
\frac{ P^{\mathrm{CP}}_{[a_c+ \delta_c]} }
{P^{\mathrm{CP}}_{[a_c]}}
\cdot 
e^{-i\pi\boldsymbol{\Lambda}^T_c\boldsymbol{\chi}_{\mathrm{CP}}}
\cdot
Z^{\mathrm{CP}}_{[a_c]}(\tau_2),
\nonumber 
\end{align}
where $\boldsymbol{\Lambda}_c \equiv \delta_c K^{-1}\boldsymbol{Q}$ is an integer vector, and
\begin{align}
 Z_{\left[
 a_c,b_c+\delta_c
\right]}(\tau)
=
e^{2\pi i\delta_c a_c\boldsymbol{Q}^TK^{-1}\boldsymbol{Q}}
\cdot
Z_{[a_c,b_c]}(\tau). 
\end{align}
Since $\boldsymbol{Q}^TK^{-1}\boldsymbol{Q}=0$ by CP symmetry, 
the total projected partition function is invariant 
under $b_c \to b_c+\delta_c$.
The crucial part is the behavior of the partition function
under $a_c \rightarrow a_c + \delta_c$.
If we demand the CP eigenvalue be invariant under $a_c \rightarrow a_c + \delta_c$,
{\it i.e.}, $P^{\mathrm{CP}}_{[a_c+ \delta_c]}=P^{\mathrm{CP}}_{[a_c]}$, 
then the partition function is (not) large gauge invariant if the value of $\boldsymbol{\Lambda}^T_c\boldsymbol{\chi}_{\mathrm{CP}}$ is even (odd).
Therefore, the quantity
\begin{align}
\boldsymbol{\Lambda}^T_c\boldsymbol{\chi}_{\mathrm{CP}}=\delta_c\boldsymbol{\chi}_{\mathrm{CP}}^TK^{-1}\boldsymbol{Q}
\end{align}
gives the criterion: 
"$\boldsymbol{\Lambda}^T_c\boldsymbol{\chi}_{\mathrm{CP}}=$ odd number" corresponds to theory with anomaly (topological phase), 
while "$\boldsymbol{\Lambda}^T_c\boldsymbol{\chi}_{\mathrm{CP}}=$ even number" corresponds to  theory without anomaly (trivial phase).

\subsubsection{Fermionic systems}

For fermionic systems, 
the most general data $(K, \boldsymbol{Q}, U_{\mathrm{CP}}, \boldsymbol{\chi}_{\mathrm{CP}})$ is given by  
\begin{align}
K&=\left(
    \begin{array}{cc}
       \Gamma & W \\
      W^T & -\Gamma
    \end{array}
  \right),
  \quad
\boldsymbol{Q}=\left(
    \begin{array}{c}
      \boldsymbol{q}\\
      \boldsymbol{q}
    \end{array}
  \right), 
  \nonumber \\ 
U_{\mathrm{CP}}&=\left(
    \begin{array}{cc}
       0 & I_{N/2} \\
       I_{N/2} & 0
    \end{array}
  \right), 
  \quad 
\boldsymbol{\chi}_{\mathrm{CP}}=\left(
    \begin{array}{c}
      2(1/2-\epsilon)\boldsymbol{x}'\\
     2(1/2-\epsilon)\boldsymbol{x}'+2 \epsilon \boldsymbol{q}
    \end{array} \label{CP_solution_fermion}
  \right).
\end{align}
The calculation of the CP symmetry projected partition function 
in this theory can be done in the same way as the bosonic case, 
except the statistical phase factors
that arise in:
\begin{align}
(\mathcal{CP})
\ddagger
e^{i\Theta(\boldsymbol{\Lambda})}
\ddagger 
(\mathcal{CP})^{-1}
&=e^{i\Delta\phi^{\boldsymbol{\Lambda}}_{\mathrm{CP}}}
e^{i\pi \boldsymbol{\Lambda}^{T}\boldsymbol{\chi}_{\mathrm{CP}}}
\ddagger
e^{i\Theta(-U_{\mathrm{CP}}\boldsymbol{\Lambda})}
\ddagger, 
\label{vertex_under_CP_fermion}
\end{align}
where 
\begin{align}
 i\Delta\phi^{\boldsymbol{\Lambda}}_{\mathrm{CP}}&
 =i\pi \left(\sum^{N/2}_{I=1}\Lambda_IQ_I \right)
 \left(\sum^{N}_{J={N}/{2}+1}\Lambda_JQ_J \right)
 \mod 2\pi i \label{statistical_CP_phase}
\end{align}
is the statistical phase factor 
due to Fermi statistics 
derived in Appendix 
\ref{Statistical phase factor of the chiral boson field under symmetry transformation}. 
For CP invariant vectors 
$\boldsymbol{\Lambda}$
satisfying $\boldsymbol{\Lambda}=-U_{\mathrm{CP}}\boldsymbol{\Lambda}$, we can express $\boldsymbol{\Lambda}$ as $(\boldsymbol{\lambda}, -\boldsymbol{\lambda})^T$,
where $\boldsymbol{\lambda}$ is an $N/2$ dimensional integer vector. Then the statistical phase can be expressed as 
\begin{align}
i\Delta\phi^{\boldsymbol{\Lambda}}_{\mathrm{CP}}
&=-i\pi \sum^{{N}/{2}}_{I=1}\lambda_Iq_I 
=-i\pi \boldsymbol{\lambda}^T\boldsymbol{q} \mod 2\pi i.
\end{align}
On the other hand, 
\begin{align}
i\pi \boldsymbol{\Lambda}^{T}\boldsymbol{\chi}_{\mathrm{CP}}=-i\epsilon\pi\boldsymbol{\lambda}^T\boldsymbol{q} \mod 2\pi i&
\end{align}
Writing 
$\boldsymbol{\Lambda}_{a_c}^T \boldsymbol{\Lambda}^{\ }_{a_c}
=2\boldsymbol{\lambda}^T_{a_c}\boldsymbol{\lambda}^{\ }_{a_c}$, where $\boldsymbol{\lambda}_{a_c}\equiv \boldsymbol{\lambda} + \frac{a_c}{\delta_c}\boldsymbol{\lambda}_c$ and $\boldsymbol{\lambda}_c$ is defined as 
\begin{align}
 \delta_cK^{-1}\boldsymbol{Q}
 =\boldsymbol{\Lambda}_c\equiv \left(
    \begin{array}{c}
      \boldsymbol{\lambda}_c\\
     -\boldsymbol{\lambda}_c
    \end{array} 
  \right), 
\end{align}
(remember that $\boldsymbol{\Lambda}_c=-U_{\mathrm{CP}}\boldsymbol{\Lambda}_c$, so $\boldsymbol{\lambda}_c$ is well-defined), then we have 
\begin{align}
 &\quad
 Z^{\mathrm{CP}}_{[a_c,b_c]}(\tau) \nonumber \\
 &= 
 P^{\mathrm{CP}}_{[a_c]}
 \xi^{\mathrm{CP}}
 (\tau)
 \sum_{
  \substack{
   \boldsymbol{\Lambda} \in \mathbb{Z}^N \\
 \boldsymbol{\Lambda}=-U^T_{\mathrm{CP}}\boldsymbol{\Lambda} 
}
}
 e^{-\pi\tau_2 \boldsymbol{\Lambda}_{a_c}^T \boldsymbol{\Lambda}^{\ }_{a_c}
 +i\pi\boldsymbol{\Lambda}^T\chi_{\mathrm{CP}}+i\Delta\phi^{\boldsymbol{\Lambda}}_{CP}} \nonumber \\
&= 
P^{\mathrm{CP}}_{[a_c]}
\xi^{\mathrm{CP}}(\tau)
\sum_{\boldsymbol{\lambda}\in\mathbb{Z}^{{N}/{2}}}
e^{-2\pi\tau_2 \boldsymbol{\lambda}_{a_c}^T \boldsymbol{\lambda}^{\ }_{a_c}-2\pi i (\epsilon+1/2)\boldsymbol{\lambda}^T\boldsymbol{q}}. 
\label{Z^{CP}_U_fermion}
\end{align}
As in the case of the bosonic systems discussed previously, 
here $Z^{\mathrm{CP}}_{[a_c,b_c]}(\tau)$
depends only on $a_c$ and $\tau_2$.
The fermionic CP symmetry projected partition functions are given by
\begin{align}
 Z^{\mathrm{Proj}}_{[a_c,b_c]}(\tau) =\frac{1}{2}\left[
  Z_{[a_c,b_c]}(\tau)+Z^{\mathrm{CP}}_{[a_c]}(\tau_2) 
  \right] 
  \label{CP_proj_Z_fermion}
\end{align}
with $Z_{[a_c,b_c]}(\tau)$ given by Eq.\ (\ref{Z_m}).
Under the large gauge transformation
$a_c \rightarrow a_c + \delta_c$ and $b_c \rightarrow b_c + \delta_c$
\begin{align}
Z_{
 \left[a_c+\delta_c,b_c\right]}(\tau)
&=
Z_{\left[a_c,b_c+\delta_c\right]}(\tau)=Z_{[a_c,b_c]}(\tau), 
\nonumber\\
Z^{\mathrm{CP}}_{\left[
 a_c+\delta_c
\right]}(\tau_2)
&=
\frac{P^{\mathrm{CP}}_{[a_c+ \delta_c]}}
{P^{\mathrm{CP}}_{[a_c]}} 
\cdot 
e^{i2 \pi (\epsilon-1/2) \boldsymbol{\lambda}^T_c\boldsymbol{q}}
\cdot 
Z^{\mathrm{CP}}_{[a_c]}(\tau_2), 
\end{align}
where $\boldsymbol{Q}^TK^{-1}\boldsymbol{Q}=0$ (by CP symmetry) is used. 
Therefore, the fermionic theory is always anomaly-free if $\epsilon=1/2$ 
[$(\mathcal{CP})^2=(-1)^{{N}_f}$]. 
For $\epsilon=0$ [$(\mathcal{CP})^2=1$], 
the quantity $\boldsymbol{\lambda}^T_c\boldsymbol{q}$ gives the stability criterion: 
"$\boldsymbol{\lambda}^T_c\boldsymbol{q}=$ odd number" 
corresponds to an anomalous theory (topological phase), 
while 
"$\boldsymbol{\lambda}^T_c\boldsymbol{q}=$ even number" corresponds to theory without anomaly (trivial phase):
\begin{align}
 \boldsymbol{\lambda}^T_c\boldsymbol{q} &= \mbox{odd}\quad   \Longrightarrow  \quad \mbox{stable edge ("topological")}, 
 \nonumber \\
 \boldsymbol{\lambda}^T_c\boldsymbol{q} &= \mbox{even}\quad  \Longrightarrow  \quad \mbox{unstable edge ("trivial")}.  
\end{align}

%

 \subsection{Examples}
\subsubsection{The double Laughlin edge state}

As an example, let us now consider the case of the doubled fermionic Laughlin state
described by
\begin{align}
K&=\left(
    \begin{array}{cc}
      \frac{1}{\nu} & 0\\
      0 & -\frac{1}{\nu}
    \end{array}
  \right), 
\quad
U_{\mathrm{CP}}=\left(
    \begin{array}{cc}
      0 & 1\\
      1 & 0
    \end{array}
  \right),
  \nonumber \\
\boldsymbol{Q}&=\left(
    \begin{array}{c}
      1\\
      1
    \end{array}
  \right), 
\quad 
\boldsymbol{\chi}_{\mathrm{CP}}=\left(
    \begin{array}{c}
      0\\
      2\epsilon 
    \end{array}
  \right),
\end{align}
where $\nu^{-1}$ is an odd integer and 
$\epsilon$ can be either $0$ or $1/2$. 
The elementary charge in the system is 
$\beta_c= \min_{\boldsymbol{l}} |
\boldsymbol{l}^TK^{-1}\boldsymbol{Q}|=\nu$, 
and the quantity 
$\boldsymbol{\Lambda}_c$ is given by 
$\boldsymbol{\Lambda}_c= K^{-1}\boldsymbol{Q}/m_c=(1, -1)^T=(\boldsymbol{\lambda}_c, -\boldsymbol{\lambda}_c)^T$. 
From the previous discussion, the criterion for topological phases is: 
if $\epsilon=1/2$, the system is in the trivial phase; 
if $\epsilon=0$, since $\boldsymbol{\lambda}^T_c\boldsymbol{q}=1$, the system is in the topological phase.

In this theory, we have $\xi^{\mathrm{CP}}(\tau)=\eta(2i\nu\tau_2)^{-1}$ and thus 
$Z^{\mathrm{CP}}$  is given by 
[Eq.\ (\ref{Z^{CP}_U_fermion})]
\begin{align}
Z^{\mathrm{CP}}_{[a_c]}(\tau) 
&= 
\frac{P^{\mathrm{CP}}_{[a_c]}}
{\eta(2i\nu\tau_2)}
\sum_{\boldsymbol{\lambda} \in \mathbb{Z}}
e^{-2\pi\tau_2 \left(\boldsymbol{\lambda}+\nu a_c \right)^2
-i2\pi(\epsilon+1/2)\boldsymbol{\lambda}} 
\nonumber \\
&= 
e^{2 \pi i\nu a_c\left(\epsilon-1/2\right)}
\frac{P^{\mathrm{CP}}_{[a_c]}}{\eta(2i\nu\tau_2)}
\vartheta\left[
    \begin{matrix}
      \nu a_c\\
      -\left(\epsilon-1/2\right)
    \end{matrix}
  \right] (0, 2i\tau_2). 
\end{align}
The total CP symmetry projected partition function is given by 
Eq.\ (\ref{CP_proj_Z_fermion}).
For $\nu=1$, which corresponds to the "integer" CP symmetric system (without ground-state degeneracy), 
the results here agree exactly with   
the CP projected partition function obtained for the free fermion theory.
Under a large gauge transformation $a_c \rightarrow a_c+1/\nu$, we have
\begin{align}
 Z^{\mathrm{CP}}_{\left[a_c+\frac{1}{\nu}\right]}(\tau_2)&=
 e^{2\pi i(\epsilon-1/2)} 
 \cdot 
 \frac{P^{\mathrm{CP}}_{[a_c+\frac{1}{\nu}]}}
 {P^{\mathrm{CP}}_{[a_c]}}
 \cdot Z^{\mathrm{CP}}_{[a_c]}(\tau_2).
\end{align}


Alternatively,
the stability of the edge state of this theory can also be analyzed 
by enumerating potential (interaction) terms 
that can potentially gap the edge state without breaking CP and charge U(1)
symmetries.
They are given by 
\begin{align}
U(x)\cos\left[
 \Theta(\boldsymbol{\Lambda})-\alpha(x) \right]
 =U(x)\cos\left[
  \frac{n}{\nu}(\phi_1+\phi_2)-\alpha(x) \right],
\end{align}
where $U(x)$ and $\alpha(x)$ represent 
the strength and phase of the potential, respectively,
which are allowed to be spatially inhomogeneous, 
and 
$\boldsymbol{\Lambda}^T=(n, -n), \ n \in \mathbb{Z}$
is a charge conserving vector. 
Under the CP transformation, 
\begin{align}
&
(\mathcal{CP})
\left[ U(x)\cos\left(\Theta(\boldsymbol{\Lambda})-\alpha(x) \right)\right]
(\mathcal{CP})^{-1} 
\nonumber \\
&= U(x)\cos\left[
 \Theta(\boldsymbol{\Lambda})-(2\epsilon+1) n\pi -\alpha(x) \right],
\end{align}
where Eqs.\ (\ref{vertex_under_CP_fermion}) and (\ref{statistical_CP_phase}) are used.
 For $\epsilon=1/2$, the scattering term is CP invariant for any integer $n$. 
 Such perturbation can gap out the edge without breaking CP symmetry 
 of the ground state $\frac{1}{\nu}\langle \phi_1 + \phi_2\rangle$. 
On the other hand, for $\epsilon=0$ the scattering term is CP invariant just for even $n$. In this case, however, the gapping perturbation also spontaneously breaks the CP symmetry of the ground state: $\frac{1}{\nu}\langle \phi_1 + \phi_2\rangle \rightarrow \frac{1}{\nu}\langle \phi_1 + \phi_2\rangle-\pi$. 
The argument here agrees with our generalized Laughlin
argument based on the CP projected partition function.

\subsubsection{The fermionic $4\times 4$ K-matrix theory}

As yet another example, 
let us consider the fermionic K-matrix theory described by 
the following $4\times 4$ K-matrix:
\begin{align}
K&=\left(
    \begin{array}{cc}
     \Gamma & 0\\
      0 & -\Gamma
    \end{array}
  \right),
  \quad 
U_{\mathrm{CP}}=\left(
    \begin{array}{cc}
      0 & I_2\\
      I_2 & 0
    \end{array}
  \right), 
  \nonumber \\
\boldsymbol{Q}&=\left(1, 1, 1, 1\right)^T, 
\quad
\boldsymbol{\chi}_{\mathrm{CP}}
=\left(0, 0, 2\epsilon, 2\epsilon\right)^T,  
\label{fermionic_4*4_K_matrix}
\end{align}
where $\Gamma$ is a $2\times 2$ matrix. It is convenient to parameterize the matrix as \cite{LevinStern2012}
\begin{align}
K=\left(
    \begin{array}{cccc}
     b+us & b & 0 & 0\\
     b & b+vs & 0 & 0\\
     0 & 0& -b-us & -b\\
     0 & 0& -b & -b-vs
    \end{array}
  \right), 
\end{align}
where $b, u ,v, s$ are integers 
and $u$ and $v$ have no common factor. In terms of these parameters, the elementary charge is
\begin{align}
\beta_c= \min_{\boldsymbol{l}}|\boldsymbol{l}^TK^{-1}\boldsymbol{Q}|=\frac{1}{(u+v)b+uvs},
\end{align}
and the quantities $\boldsymbol{\Lambda}_c$ and $\boldsymbol{\lambda}_c$ are given by
\begin{align}
\boldsymbol{\Lambda}_c=\frac{1}{m_c}K^{-1}\boldsymbol{Q}=\left(
    \begin{array}{c}
      v\\
      u\\
      -v\\
      -u
    \end{array}
  \right)= \left(
    \begin{array}{c}
     \boldsymbol{\lambda}_c\\
      -\boldsymbol{\lambda}_c
    \end{array}
   \right). 
\end{align}
From these, the criterion for the presence/absence of SPT phases is: 
if $\epsilon=1/2$, the system is always in the trivial phase; 
if $\epsilon=0$,  since $\boldsymbol{\lambda}_c^T \boldsymbol{q}=u+v$, the parity of $u+v$ determines whether the phase is trivial ($u+v$ is even) or topological  ($u+v$ is odd).

The CP twisted partition function $Z^{\mathrm{CP}}$ is given by [Eq. (\ref{Z^{CP}_U_fermion})]
\begin{align}
&Z^{\mathrm{CP}}_{[a_c]}(\tau) 
= \xi^{\mathrm{CP}}(\tau)
P^{\mathrm{CP}}_{[a_c]}
\nonumber \\
&\qquad
\times 
\sum_{\lambda_1, \lambda_2 \in \mathbb{Z}}e^{-2\pi\tau_2\left[ \left(\lambda_1+m_ca_cv \right)^2+\left(\lambda_2+m_ca_cu \right)^2 \right]} 
\nonumber \\
&\qquad \times  
e^{-2\pi i(\epsilon+1/2)(\lambda_1 +\lambda_2)}.
\end{align}
Under a large gauge transformation $a_c \rightarrow a_c+1/\beta_c$, 
the CP-twisted partition function transforms as
\begin{align}
Z^{\mathrm{CP}}_{[a_c+1/\beta_c]}(\tau_2)
&=e^{i2 \pi(\epsilon+1/2)(u+v)}
\cdot 
\frac{P^{\mathrm{CP}}_{[a_c+1/\beta_c]}}
{P^{\mathrm{CP}}_{
[a_c] }
}\cdot Z^{\mathrm{CP}}_{[a_c]}(\tau_2).
\end{align}

We can also look for gapping potentials and see if we can gap the edge without breaking CP and charge U(1) symmetries.  
\cite{LevinStern2009}
To gap out the 4 edge modes of theory (\ref{fermionic_4*4_K_matrix}), we need to find two linearly independent and charge conserving vectors $\boldsymbol{\Lambda}_1$ and $\boldsymbol{\Lambda}_2$ that satisfy Haldane's null vector criterion:
\begin{align}
\boldsymbol{\Lambda}^T_1K\boldsymbol{\Lambda}_1=\boldsymbol{\Lambda}^T_2K\boldsymbol{\Lambda}_2=\boldsymbol{\Lambda}^T_1K\boldsymbol{\Lambda}_2=0. \label{Haldane_criterion}
\end{align}
Such $\boldsymbol{\Lambda}_1$ and $\boldsymbol{\Lambda}_2$ can be the following cases:

\paragraph{$\boldsymbol{\Lambda}_1=U_{\mathrm{CP}}^T\boldsymbol{\Lambda}_1$ and $\boldsymbol{\Lambda}_2=-U_{\mathrm{CP}}^T\boldsymbol{\Lambda}_2$:}
In this case, the charge conserving conditions $\boldsymbol{\Lambda}^T_1\boldsymbol{Q}=\boldsymbol{\Lambda}^T_2\boldsymbol{Q}=0$ and  Eq. (\ref{Haldane_criterion}) give that
\begin{align}
&\boldsymbol{\Lambda}_1 =n_1  (1, -1, 1, -1)^T\equiv n_1\boldsymbol{\Lambda}_-, \nonumber \\
&\boldsymbol{\Lambda}_2 =n_2  (v, u, -v, -u)^T\equiv n_2\boldsymbol{\Lambda}_+,
\end{align}
where $n_1, n_2 \in \mathbb{Z}$. 
Under CP the scattering term 
$\sum^2_{i=1} U_i(x)\cos\left[\Theta(\boldsymbol{\Lambda}_i)-\alpha_i(x) \right]$ 
transforms as
\begin{align}
&\quad
(\mathcal{CP})
\left[ \sum^2\nolimits_{i=1} U_i(x)
\cos\left[
\Theta(\boldsymbol{\Lambda}_i)-\alpha_i(x) \right] \right]
(\mathcal{CP})^{-1} 
 \nonumber \\
&= 
U_1(x)\cos\left[-\Theta(\boldsymbol{\Lambda}_1)-\alpha_1(x) \right] 
\nonumber\\
&\quad
+U_2(x)\cos\left[
 \Theta(\boldsymbol{\Lambda}_2)
 -n_2(2\epsilon+1)(u+v)\pi-\alpha_2(x) \right]. 
\end{align}
By choosing $\alpha_1(x)=k\pi, \ k=0, 1$,
then, for $\epsilon=1/2$, the scattering term is CP invariant for any integers $n_1$ and $n_2$. 
Such perturbation can gap out the edge without breaking CP symmetry 
of the ground state 
$\{\left\langle \Theta(\boldsymbol{\Lambda}_-)\right\rangle, \left\langle\Theta(\boldsymbol{\Lambda}_+)\right\rangle\}$. 
For $\epsilon=0$, the scattering term is CP invariant 
if $n_2(u+v)$ is even. Under CP transformation the ground state transforms as 
\begin{align}
&\{\left\langle \Theta(\boldsymbol{\Lambda}_-)\right\rangle, \left\langle\Theta(\boldsymbol{\Lambda}_+)\right\rangle\} 
\nonumber \\
&
\rightarrow \{-\left\langle \Theta(\boldsymbol{\Lambda}_-)\right\rangle, \left\langle\Theta(\boldsymbol{\Lambda}_+)\right\rangle-(u+v)\pi\}.
\end{align}
Therefore, the perturbation will gap out the edge with ($u+v$ is odd)/without ($u+v$ is even) breaking the CP symmetry of the ground state spontaneously.

\paragraph{$\boldsymbol{\Lambda}_2=-U_{\mathrm{CP}}^T\boldsymbol{\Lambda}_1$:} 
In this case the CP invariant scattering term is
\begin{align}
&U(x)\big[
 \cos\left(\Theta(\boldsymbol{\Lambda}_1)-\alpha(x) \right) 
 \nonumber\\
&\quad 
+\cos(\Theta(\boldsymbol{\Lambda}_2)+\pi\boldsymbol{\Lambda}_1^T\chi
+\Delta\phi_{\mathrm{CP}}^{\boldsymbol{\Lambda}_1}-\alpha(x) ) 
\big],
\end{align}
where we used the fact that 
$\boldsymbol{\Lambda}_i^T\boldsymbol{Q}=0$ 
(the charge neutrality condition) and 
$\Delta\phi_{\mathrm{CP}}^{\boldsymbol{\Lambda}_1}=\Delta\phi_{\mathrm{CP}}^{\boldsymbol{\Lambda}_2}$. 
Defining $\boldsymbol{\Lambda}'_{\pm}=\boldsymbol{\Lambda}_1\pm\boldsymbol{\Lambda}_2$, 
we can then find that $\boldsymbol{\Lambda}'_+$ is an integer multiple of $(v,u,-v,-u)$ and $\boldsymbol{\Lambda}'_-$ is an integer multiple of $(1,-1,1,-1)$. From the analysis in (i), we know that $\langle\Theta(v,u,-v,-u)\rangle$ spontaneously breaks CP for odd $u+v$ ($\epsilon=0$), so it is impossible that $\langle\Theta(\boldsymbol{\Lambda}_1)\rangle$ and $\langle\Theta(\boldsymbol{\Lambda}_2)\rangle$ (thus $\langle\Theta(\boldsymbol{\Lambda}_1)+\Theta(\boldsymbol{\Lambda}_2)\rangle$) can condensate without 
spontaneously breaking CP symmetry. The result for $\epsilon=1/2$ 
is the same in (i): the perturbation can gap out the edge without breaking CP symmetry of the ground state.
On the other hand, for even $u+v$ we can take 
\begin{align}
\boldsymbol{\Lambda}'^T_- =(1, -1, 1, -1), 
\quad 
\boldsymbol{\Lambda}'^T_+=(v, u, -v, -u),
\end{align}
so that
\begin{align}
&\boldsymbol{\Lambda}^T_1=\frac{1}{2}(1+v, -1+u, 1-v, -1-u), \nonumber \\
&\boldsymbol{\Lambda}^T_2 =\frac{1}{2}(-1+v, 1+u, -1-v, 1-u).
\end{align}
Under CP the ground state transforms as
\begin{align}
&
\{\left\langle \Theta(\boldsymbol{\Lambda}_1)\right\rangle, \left\langle\Theta(\boldsymbol{\Lambda}_2)\right\rangle\} 
\nonumber\\
&\rightarrow 
\{\left\langle \Theta(\boldsymbol{\Lambda}_2)\right\rangle
-({u+v})(\epsilon+1/2)\pi, 
\nonumber \\
&\qquad
\left\langle\Theta(\boldsymbol{\Lambda}_1)\right\rangle
-({u+v})(\epsilon+1/2)\pi\},
\end{align}
which does not break CP spontaneously. 
Also, 
since there are no non-primitive linear combinations
\footnote{
 The linear combination $a_1\boldsymbol{\Lambda}_1+a_2\boldsymbol{\Lambda}_2$ is non-primitive 
 if there are some integer vector $\boldsymbol{\Lambda}$ and integer $k>1$ such that 
 $a_1\boldsymbol{\Lambda}_1+a_2\boldsymbol{\Lambda}_2=k\boldsymbol{\Lambda}$.} 
 $a_1\boldsymbol{\Lambda}_1+a_2\boldsymbol{\Lambda}_2$ for any integers $a_1$ and $a_2$, 
 the perturbation does not break CP for the whole family of condensations 
 $\langle\Theta(a_1\boldsymbol{\Lambda}_1+a_2\boldsymbol{\Lambda}_2)\rangle$ 
 (for both $\epsilon=0, 1/2$).
Therefore, the argument of the stability 
of the edge state by the microscopic analysis is consistent with the one by the large gauge 
invariance of the CP symmetry-projected partition function.

\section{Discussion}
\label{Discussion}

We have developed a general theoretical framework
that allows us to determine under which conditions
a given edge CFT is gappable/ingappable. 
While we have worked out particular examples
with CP or P symmetry, our theoretical framework 
is applicable to other examples
with local and non-local symmetries. 
For example, 
our methodology can be applicable 
to 
reflection symmetric fermionic SPT phases 
that are classified and tabulated in  
Refs.\ \onlinecite{Chiu13,Morimoto13}.

Our consideration in this paper is limited
to an anomaly associated to global U(1) symmetries
(once CP is enforced)
and thus limited to systems with conserved U(1) symmetry
(such as conservation of the particle number or
$z$-component of SU(2) spin).
We use a gauge flux of these U(1) symmetries 
as an adiabatic parameter 
in developing
Laughlin's  gauge argument.
For systems that do not have such continuous symmetry, 
one may need to consider an anomaly associated to
gravitational degrees of freedom 
such as modular invariance.    
\cite{Ryu2012}
As one can see from the fact that
the real part of the modular parameter $\tau_1$ is projected 
out by orientifolding [recall discussion near  
Eq.\ (\ref{fermionic CP projected part fn})],
the modular group of the torus $\mathrm{PSL}(2,\mathbb{Z})$
cannot be used to study conformal field theories 
on the Klein bottle. 
Nevertheless, an analogue of the $S$-modular transformation  
(which exchanges the space and time direction on the torus)
still plays a role in 
orientifold conformal field theories.
To be more precise, the ``loop channel'' calculations 
presented in this paper can be cast into an 
equivalent calculation in the ``tree channel'' by using 
crosscap states.
We plan to visit these tree channel pictures 
in a forthcoming publication. 

Related to this question, 
in this paper, we discussed partition 
functions of edge theories on the Klein bottle, 
but not the other spacetime manifolds such as 
annulus or M\"obius strip. 
It may be interesting to ask if there is any 
role played by these other geometries. 
In oriented cases, the modular invariance 
on the torus 
is believed to be enough to define the conformal field
theory on any (oriented) world sheet.
For the unoriented cases, one may wonder if considering
the consistency of the theory on the Klein bottle
would be enough to define the conformal field theory
on all (unoriented) worldsheets.

We will defer detailed studies of this issue for the future,
but it may be worth pointing out the following: 
in string theory, 
conformal field theories on the Klein bottle 
appear in Type I superstring theory.
There are, in addition to the Klein bottle, 
other worldsheet geometries such 
as a strip and the M\"obius strip. Some properties are physically constrained by 
tadpole cancellation, and worldsheet geometries with boundaries come hand-in-hand
with D-branes. It would be interesting to explore what role these and other consistency conditions might play in a condensed matter setting. 

\acknowledgements 

We thank 
Po-Yao Chang, 
Takahiro Morimoto,
and Christopher Mudry
for discussion and
collaborations in closely related works. 
SR thanks participants
in ``Topological Phases of Matter'' Workshop (June 10-14, 2013)
at  Simons Center for Geometry and Physics, 
in ``IAS Program on Topological Matter, Superconductvity and Majorana''
(January 6-30, 2014)
at Jocky Club Institute for Advanced Study at Hong Kong University of Science 
and Technology, 
in 
PCTS workshop, 
``Symmetry in Topological Phases'' 
(March 17-18, 2014) 
at Princeton Center for Theoretical Science, 
where the results of the paper were presented.
SR thanks Thomas Quella for useful discussion.  
GYC is supported by NSF grant under DMR-1064319.

\appendix

\section{The CP eigenvalue of the ground state}
\label{Discussion on the CP eigenvalue of the ground state}

In the bulk of the paper, 
we have enforced CP invariance at all steps of 
an adiabatic evolution
(for all values of the flux $a$). 
In fact, the system (defined by Lagrangian with boundary conditions)
is classically CP invariant, and hence one would assume this is so
even at quantum level.  
What we discovered, under this assumption, is the violation of the electromagnetic U(1) symmetry.
As we mentioned in Sec.\ \ref{Generalized Laughlin's argument}, 
an alternative point of view is possible; 
if we were to enforce the electromagnetic U(1) symmetry, 
CP would then be violated
(when the CP symmetry in question is ``topological'' -- the one
which leads topological insulators). 
Therefore, while the system preserves, at the classical level,
both the electromagnetic U(1) and CP symmetries, 
there is a tension between these symmetries once we quantize the system. 
Once we demand the electromagnetic U(1) symmetry be 
strictly conserved, instead of enforcing CP,  
$P_{[a]}$ may not be independent of $a$. 
In the following, we determine 
$P_{[a]}$ under the assumption of the U(1) conservation.  

\paragraph{The IQHE}

As a warm up, let us start from 
an edge state of the (integer) quantum Hall system; 
it suffers from an anomaly, and hence cannot exist on its own.
(We follow closely Ref.\ \onlinecite{Polchinski98}). 
The edge state of the IQHE is a chiral fermion $\psi_R$. 
We consider an edge of circumference $2\pi$, and 
impose the twisted boundary condition:
 $\psi_R(x+2\pi) = e^{2\pi i \nu} \psi_R(x)$.  
From the state-operator correspondence,
there is an operator associated to the ground state
for a given $\nu$,
which we call $\mathcal{A}_{\nu}$. 
The operator can be determined from the following general principle: 
(i) any unitary on-site symmetry in field theories can be used to 
generate a twisting boundary condition;
(ii) in CFT, Hilbert space with twisted boundary condition form
an independent sector (Virasoro module);
(iii) due to the state-operator correspondence, there is an operator
that corresponds to a ground state of the twisted Hilbert space. 
The identification of the ground state operator can be 
done conveniently in terms of bosonization:
\begin{align}
 \psi_R \simeq e^{ i \varphi_R}. 
\end{align}
One could then infer the operator corresponding to 
the ground state: 
\begin{align}
 \mathcal{A}_{\nu}
 \equiv
 e^{ i (-\nu +1/2)\varphi_R}, 
\end{align}
where $\varphi_R$ is a chiral boson field. 
From this expression for the ground state operator 
one infers that the charge of the ground state is
\begin{align}
 \label{charge of GS}
 F_R = 1/2-\nu. 
\end{align}
Thus, we conclude that the 
ground state fermion number at the edge of the
quantum Hall system changes as a function of twisting angle.
Because of the spectral flow, 
as one changes $a\to a+1$, the fermion number jumps by one
(discontinuously).
Had the charge been conserved ({\it i.e.}, had there been no anomaly),
the ground state fermion number should 
be independent of the twisting angle. 
The ground state charge (\ref{charge of GS}) 
is the origin of 
the factor 
$e^{-2\pi i (b-1/2) (a-1/2)}$ 
in 
the partition function (\ref{chiral part fn Dirac}). 
\footnote{
While we have determined the ground state and its charge as above,  
we could take an alternative point of view.
Let us assume that we actually do not know, a priori, that the U(1) symmetry
is anomalous. We would like to test if this symmetry is anomalous
or not. For this purpose, we pretend the charge U(1) is conserved. 
We do so since the charge U(1) is classically conserved, and if so,
one would guess naively that the ground state fermion number does not
change as we adiabatically change $a$ and $b$. 
Under this assumption, what one would discover is that 
the partition function is not invariant under $a\to a+1$. 
Therefore, even though we started from the assumption that 
the U(1) is conserved, we run into the ``inconsistency''
in that the partition function is not invariant under $a\to a+1$,
in stead of $b\to b+1$ 
-- we then conclude we cannot conserve the U(1) at the quantum level. 
}

\paragraph{The QSHE with conserved $S_z$}

Let us now consider 
the edge theory of a bulk quantum spin Hall system
with conserving $S_z$. 
The edge state now consists of
both left- and right-movers,
$\psi_L$ and $\psi_R$. 
These fermion fields can be bosonized as 
\begin{align}
 \psi_L
 \sim
 e^{ i \varphi_L},
 \quad
 \psi_R 
 \sim 
 e^{- i \varphi_R}. 
\end{align}
(Here, we do not include Klein factors while 
they are important in discussing CP symmetric topological
insulators.)
Following the same argument as in the case of the QHE, 
the operator corresponding the 
ground state of the left-moving sector is
$
 e^{ i (-\nu_L +1/2) \varphi_L }
$
where $\nu_L$ is the twisting angle for the left movers. 
Similarly, 
the ground state for the right moving sector can be represented
as 
$
 e^{ i (-\nu_R +1/2) \varphi_R }
$
By combining the left- and right-moving parts of the ground 
state properly, we have a ground state for the combined 
non-chiral system. 

Below, we put more emphasis on U(1) charge conservation
than $S_z$ conservation; we first give a priority to 
the U(1) charge conservation and see this necessary leads
to violation of $S_z$ conservation --
this is nothing but chiral anomaly. 
Since charge U(1) is conserved, it makes sense to 
twist boundary conditions by charge U(1) symmetry,
$\nu_L=\nu_R$. 
We are thus lead to the ground state vertex operator
\begin{align}
 e^{ i (-\nu +1/2) \varphi_L }
 e^{ +i (-\nu +1/2) \varphi_R }
 =
 e^{ i (-\nu+1/2) \phi }. 
\end{align}
Here, the non-chiral field $\phi= \varphi_L + \varphi_R$
is charge neutral. 
One could combine the left- and right-moving sectors differently
to get 
$ 
 e^{ i (-\nu+1/2) \theta } 
$
with $\theta=\varphi_L - \varphi_R$. 
This choice, however, is not consistent with 
charge U(1) conservation since 
$\theta$ is not charge neutral and the 
ground state fermion number 
$F_V=F_L+F_R$
changes as a function of $\nu$,
$\nu= \nu+1/2$. 

While the ground state 
$e^{ i (-\nu+1/2) \phi }$
is consistent with charge U(1) conservation, 
the price we paid is that
the ground state is charged under spin $S_z$ 
conservation. 
This means that as one adiabatically inserts
charge flux, the $S_z$ quantum number of the 
edge state changes -- the spin is ``pumped''
from the edge in question to other edges, 
or vice versa. 

\paragraph{CP symmetric bosonic topological insulators}

Let us now break the continuous the U(1) spin $S_z$ conservation
and instead impose CP symmetry;
we consider the case of CP symmetric topological insulators. 
The relevant symmetries are charge U(1) and CP. 
In particular, we focus on the bosonic version of the 
topological insulator that we discuss in
Sec.\ \ref{Bosonic CP symmetry protected topological phase}. 
The CP acts on the bosonic field as
in Eq.\ (\ref{def: CP acting of boson}). 
Following above discussion, we consider the ground state 
that preserves the electromagnetic U(1) symmetry
as a function of twisting angle $\nu$.  
The ground state is then given by
\begin{align}
 e^{i \nu \theta R/\alpha'}  
\end{align}
The CP eigenvalue of the ground state is
\begin{align}
 &
 (\mathcal{CP}) e^{i\nu \theta R/\alpha'} (\mathcal{CP})^{-1}
 =
 P_{[\nu]}
 e^{i\nu \theta R/\alpha'}
 \nonumber \\
 &
 \quad 
 \mbox{where} 
 \quad
  P_{[\nu]} = e^{ i2 \pi \nu \epsilon}. 
\end{align}
Thus, for the topologically trivial case $\epsilon=0$,
the CP eigenvalue is independent of $\epsilon$, 
where as when $\epsilon=1/2$ (topological), 
the ground state CP eigenvalue evolves as a function of $\nu$. 
This signals the conflict of the symmetry;
once we choose to preserve the U(1), CP is necessarily broken.

\section{Statistical phase factor of the chiral boson field under symmetry transformation}
\label{Statistical phase factor of the chiral boson field under symmetry transformation}

For any local quasiparticle excitation 
$\ddagger \exp{i\boldsymbol{\Lambda}^TK\boldsymbol{\phi}}\ddagger$, 
where 
$\boldsymbol{\Lambda}^TK\boldsymbol{\phi}=
\sum_I \Lambda_I(K\boldsymbol{\phi})_I\equiv \sum_I \theta_I$, the symmetry transformation $\mathcal{G}$ acts as
\begin{align}
&\quad 
\mathcal{G}
 \ddagger 
 e^{i\boldsymbol{\Lambda}^TK\boldsymbol{\phi}}
 \ddagger 
 \mathcal{G}^{-1} 
=\mathcal{G}
\ddagger e^{\sum_J i\theta_I}
\ddagger \mathcal{G}^{-1} 
\nonumber \\
&=
\mathcal{G}\ddagger 
{\prod_I}'e^{i\theta_I}\cdot 
e^{-\frac{1}{2}\sum_{I<J} [i\theta_I, i\theta_J]} 
\ddagger \mathcal{G}^{-1} 
\nonumber \\
&=
\ddagger
{\prod_I}'e^{\mathcal{G}i\theta_I\mathcal{G}^{-1}}
\ddagger 
\cdot e^{-\frac{1}{2}\sum_{I<J} \mathcal{G}[i\theta_I, i\theta_J]\mathcal{G}^{-1}}  
\nonumber \\
&\equiv  
\ddagger 
{\prod_I}'e^{\mathcal{G}i\theta_I\mathcal{G}^{-1}}
\ddagger 
\cdot e^{-\frac{1}{2}\sum_{I<J} [\mathcal{G}i\theta_I\mathcal{G}^{-1},\mathcal{G} i\theta_J\mathcal{G}^{-1}]} 
\cdot 
e^{ i \Delta \phi^{\boldsymbol{\Lambda}}_{\mathrm{G}} }
\nonumber \\
&=
\ddagger 
e^{\sum_j\mathcal{G} i\theta_I\mathcal{G}^{-1}
+i\Delta\phi^{\boldsymbol{\Lambda}}_\mathrm{G} }
\ddagger ,
\end{align}
where we have used the Baker-Campbell-Hausdorff formula 
(with the commutator $[i\theta_I, i\theta_J]$ being a $c$-number), 
the ordered-product "${\prod}'_I$" is defined 
as an ordered product 
in the ascending order of indices, 
and
\begin{align}
i\Delta\phi^{\boldsymbol{\Lambda}}_\mathrm{G}
\equiv \frac{1}{2}\sum_{I<J} \left([\mathcal{G}i\theta_I\mathcal{G}^{-1},\mathcal{G} i\theta_J\mathcal{G}^{-1}]- \mathcal{G}[i\theta_I, i\theta_J]\mathcal{G}^{-1}\right). \label{extra_phase_1}
\end{align}
Note that we keep the form $\mathcal{G}[i\theta_I, i\theta_J]\mathcal{G}^{-1}$ even if $[i\theta_I, i\theta_J]$ is a $c$-number, since in general $\mathcal{G}$ can be an antiunitary operator 
({\it e.g.} T symmetry).
On the other hand, 
\begin{align}
\mathcal{G}e^{i\boldsymbol{\Lambda}^TK\boldsymbol{\phi}} \mathcal{G}^{-1}&=e^{\mathcal{G}i\boldsymbol{\Lambda}^TK\boldsymbol{\phi}\mathcal{G}^{-1}}=e^{\mathcal{G}\left(\sum_I i\theta_I\right)\mathcal{G}^{-1}},
\end{align}
so we have
\begin{align}
\mathcal{G}\left(i\boldsymbol{\Lambda}^TK\boldsymbol{\phi}\right)\mathcal{G}^{-1}
&=
\sum_I\mathcal{G} i\theta_I\mathcal{G}^{-1}
+i\Delta\phi^{\boldsymbol{\Lambda}}_\mathrm{G}  \mod 2\pi i. 
\label{extra_phase_2}
\end{align}
This means the way that the operator $\mathcal{G}$ acts on the chiral boson field $\phi$ is not always linear, because some nontrivial phase factor 
$\Delta\phi^{\boldsymbol{\Lambda}}_\mathrm{G}$ ($\neq 2n\pi$) might arise. 
In bosonic system, the phase factor is always the multiple of $2\pi i$, corresponding to Bose statistics, and thus we can ignore it (in this case $\mathcal{G}$ is linear in $\phi$). 
In fermionic systems,
however, we must be careful with the phase factor, 
which might be nontrivial, 
because of the Fermi statistics.

In the following we take CP and T symmetries as examples. 

\paragraph{CP symmetry}

From the canonical commutation relation 
(\ref{CCR K matrix 1}),
when $x\neq x'$ 
(but $x \rightarrow x'$ is taken when we consider the 
operator product expansion of vertex operators) 
and $I\neq J$, we have 
\begin{align}
&
[(K \boldsymbol{\phi})_I(t, x), 
(K\boldsymbol{\phi})_J(t, x')]
\nonumber \\
&\quad
=-i\pi\text{sgn}(I-J)Q_IQ_J+2\pi i N_{IJ}, \label{commutator_KPhi}
\end{align}
where $N_{IJ}$ is the component of an integer matrix. Now for CP symmetry defined in Sec.\  
\ref{Symmetry projected partition functions: generalities},
the extra phase is given by
\begin{align}
i\Delta\phi^{\boldsymbol{\Lambda}}_\mathrm{CP}
&= 
-\frac{1}{2}\sum^{N}_{I<J}\Lambda_I\Lambda_J
\Big\{\left[(U_{\mathrm{CP}}K\boldsymbol{\phi})_I,
(U_{\mathrm{CP}}K\boldsymbol{\phi})_J \right] 
\nonumber \\
&\quad 
-  \left[(K\boldsymbol{\phi})_I, (K\boldsymbol{\phi})_J \right]
\Big\}.
\end{align}
Since $U_{\mathrm{CP}}$ has the form $\bigl(\begin{smallmatrix}
0&I_{N/2}\\ I_{N/2}&0
\end{smallmatrix} \bigr)$, where $N$ is an even integer, 
we have, for $1\leq I<J \leq N$,
\begin{align}
 &\left[(U_{\mathrm{CP}}K\boldsymbol{\phi})_I,
 (U_{\mathrm{CP}}K\boldsymbol{\phi})_J \right] 
 \nonumber \\
=&\left\{ 
  \begin{array}{l l}
   +i\pi(U_{\mathrm{CP}}\boldsymbol{Q})_I
   (U_{\mathrm{CP}}\boldsymbol{Q})_J 
   &  \text{if $1\leq I <J \leq {N}/{2}$}\\
    &  \text{or ${N}/{2}+1\leq I <J \leq N$}\\
    -i\pi(U_{\mathrm{CP}}\boldsymbol{Q})_I
    (U_{\mathrm{CP}}\boldsymbol{Q})_J & \text{if $1\leq I \leq {N}/{2}$}\\
    &   \text{and ${N}/{2}+1\leq J \leq N$} 
  \end{array} \right. \nonumber \\
&\mod 2\pi i.
\end{align}
Then
\begin{align}
&\quad 
i\Delta\phi^{\boldsymbol{\Lambda}}_\mathrm{CP} 
\nonumber\\
&= -\frac{i\pi}{2}\sum_{1\leq I <J \leq {N}/{2}}
\Lambda_I\Lambda_J
\left[(U_{\mathrm{CP}}\boldsymbol{Q})_I
(U_{\mathrm{CP}}\boldsymbol{Q})_J -Q_IQ_J \right] 
\nonumber \\
&\quad-\frac{i\pi}{2}
\sum_{{N}/{2}+1\leq I <J \leq N}
\Lambda_I\Lambda_J
\left[(U_{\mathrm{CP}}\boldsymbol{Q})_I
(U_{\mathrm{CP}}\boldsymbol{Q})_J -Q_IQ_J \right] 
\nonumber \\
&\quad
+\frac{i\pi}{2}
\sum_{\substack{1\leq I \leq {N}/{2} \\ {N}/{2}+1\leq J \leq N}}i
\Lambda_I\Lambda_J
\left[(U_{\mathrm{CP}}\boldsymbol{Q})_I
(U_{\mathrm{CP}}\boldsymbol{Q})_J +Q_IQ_J \right] 
\nonumber \\
&=\frac{i\pi}{2}
\sum_{\substack{1\leq I \leq {N}/{2} \\ {N}/{2}+1\leq J \leq N}}\Lambda_I\Lambda_J
\left[(U_{\mathrm{CP}}\boldsymbol{Q})_I
(U_{\mathrm{CP}}\boldsymbol{Q})_J +Q_IQ_J \right] 
\nonumber \\
&=
i\pi \left(\sum^{{N}/{2}}_{I=1}\Lambda_IQ_I \right)\left(\sum^{N}_{J={N}/{2}+1}\Lambda_JQ_J \right) \mod 2\pi i, \label{extra_phase_CP}
\end{align}
where the second equality holds since the sum of the first two terms in the first equality vanishes. 
For CP invariant vector $\boldsymbol{\Lambda}$, 
with $\boldsymbol{\Lambda}=-U_{\mathrm{CP}}\boldsymbol{\Lambda}$, we can express $\boldsymbol{\Lambda}$ as $(\boldsymbol{\lambda}, -\boldsymbol{\lambda})^T$,
where $\boldsymbol{\lambda}$ is any $N/2$ dimensional integer vector. 
Then the statistical phase can be expressed as 
\begin{align}
i\Delta\phi^{\boldsymbol{\Lambda}}_{\mathrm{CP}}=&
-i\pi \left(\sum^{{N}/{2}}_{I=1}\lambda_Iq_I \right)^2 
=-i\pi \sum^{{N}/{2}}_{I=1}\lambda_Iq_I 
\nonumber \\
=&-i\pi \boldsymbol{\lambda}^T\boldsymbol{q} \mod 2\pi i. \label{statistical_phase_CP}
\end{align}

\paragraph{T symmetry}
The set of data $\{K, \boldsymbol{Q}, U_{\mathrm{T}}, \boldsymbol{\chi}_{\mathrm{T}}\}$ for 
the T symmetric K-matrix theory 
is the same as the case of CP. 
The only difference is that T is an antiunitary operator, 
which results in [from Eq.\ (\ref{commutator_KPhi})]
\begin{align}
 \mathcal{T}[(K\boldsymbol{\phi})_I, (K\boldsymbol{\phi})_J]\mathcal{T}^{-1}
 =-[(K\boldsymbol{\phi})_I, (K\boldsymbol{\phi})_J], 
\end{align}
and hence reverses the sign in front of $Q_IQ_J$ in 
Eq.\ (\ref{extra_phase_CP})
\begin{align}
&\quad
i\Delta\phi^{\boldsymbol{\Lambda}}_\mathrm{T}
\nonumber \\
&= -\frac{i\pi}{2}\sum_{1\leq I <J \leq N/2}
\Lambda_I\Lambda_J
\left[(U_{\mathrm{T}}\boldsymbol{Q})_I
(U_{\mathrm{T}}\boldsymbol{Q})_J +Q_IQ_J \right] 
\nonumber \\
&\quad
-\frac{i\pi}{2}\sum_{N/2+1\leq I <J \leq N}
\Lambda_I\Lambda_J
\left[(U_{\mathrm{T}}\boldsymbol{Q})_I
(U_{\mathrm{T}}\boldsymbol{Q})_J +Q_IQ_J \right] 
\nonumber \\
&\quad
+\frac{i\pi}{2}\sum_{\substack{1\leq I \leq N/2 \\ N/2+1\leq J \leq N}}\Lambda_I\Lambda_J
\left[(U_{\mathrm{T}}\boldsymbol{Q})_I
(U_{\mathrm{T}}\boldsymbol{Q})_J -Q_IQ_J \right] 
\nonumber \\
&=
-i\pi 
\left(
\sum_{1\leq I <J \leq N/2}+\sum_{N/2+1\leq I <J \leq N}
\right)
\Lambda_I\Lambda_JQ_IQ_J
\nonumber \\
& \mod 2\pi i, \label{extra_phase_T}
\end{align}
where the second equality holds since the third term in the first equality vanishes. 
For a T-invariant vector $\boldsymbol{\Lambda}$
satisfying $\boldsymbol{\Lambda}=-U_{T}\boldsymbol{\Lambda}$, 
we can express $\boldsymbol{\Lambda}$ as $(\boldsymbol{\lambda}, -\boldsymbol{\lambda})^T$, 
where $\boldsymbol{\lambda}$ is any $N/2$ dimensional integer vector. 
Then the statistical phase is 
\begin{align}
i\Delta\phi^{\boldsymbol{\Lambda}}_{\mathrm{T}}&=
-2\pi i \sum_{1\leq I <J \leq {N/2}}\lambda_I\lambda_Jq_Iq_J 
=0 \mod 2\pi i.
\end{align}
Therefore, in discussion of 
the K-matrix theory with T symmetry, 
statistical phases are irrelevant and can safely 
be ignored, 
as pointed out in Ref.\ \onlinecite{LevinStern2012}.

\bibliography{reference}

\begin{thebibliography}{84}%
\makeatletter
\providecommand \@ifxundefined [1]{%
 \@ifx{#1\undefined}
}%
\providecommand \@ifnum [1]{%
 \ifnum #1\expandafter \@firstoftwo
 \else \expandafter \@secondoftwo
 \fi
}%
\providecommand \@ifx [1]{%
 \ifx #1\expandafter \@firstoftwo
 \else \expandafter \@secondoftwo
 \fi
}%
\providecommand \natexlab [1]{#1}%
\providecommand \enquote  [1]{``#1''}%
\providecommand \bibnamefont  [1]{#1}%
\providecommand \bibfnamefont [1]{#1}%
\providecommand \citenamefont [1]{#1}%
\providecommand \href@noop [0]{\@secondoftwo}%
\providecommand \href [0]{\begingroup \@sanitize@url \@href}%
\providecommand \@href[1]{\@@startlink{#1}\@@href}%
\providecommand \@@href[1]{\endgroup#1\@@endlink}%
\providecommand \@sanitize@url [0]{\catcode `\\12\catcode `\$12\catcode
  `\&12\catcode `\#12\catcode `\^12\catcode `\_12\catcode `\%12\relax}%
\providecommand \@@startlink[1]{}%
\providecommand \@@endlink[0]{}%
\providecommand \url  [0]{\begingroup\@sanitize@url \@url }%
\providecommand \@url [1]{\endgroup\@href {#1}{\urlprefix }}%
\providecommand \urlprefix  [0]{URL }%
\providecommand \Eprint [0]{\href }%
\providecommand \doibase [0]{http://dx.doi.org/}%
\providecommand \selectlanguage [0]{\@gobble}%
\providecommand \bibinfo  [0]{\@secondoftwo}%
\providecommand \bibfield  [0]{\@secondoftwo}%
\providecommand \translation [1]{[#1]}%
\providecommand \BibitemOpen [0]{}%
\providecommand \bibitemStop [0]{}%
\providecommand \bibitemNoStop [0]{.\EOS\space}%
\providecommand \EOS [0]{\spacefactor3000\relax}%
\providecommand \BibitemShut  [1]{\csname bibitem#1\endcsname}%
\let\auto@bib@innerbib\@empty
\bibitem [{\citenamefont {Laughlin}(1981)}]{Laughlin1981}%
  \BibitemOpen
  \bibfield  {author} {\bibinfo {author} {\bibfnamefont {R.~B.}\ \bibnamefont
  {Laughlin}},\ }\href@noop {} {\bibfield  {journal} {\bibinfo  {journal}
  {Phys. Rev. B}\ }\textbf {\bibinfo {volume} {23}},\ \bibinfo {pages} {5632}
  (\bibinfo {year} {1981})}\BibitemShut {NoStop}%
\bibitem [{\citenamefont {Prange}\ and\ \citenamefont
  {Girvin}(1987)}]{review_QHE}%
  \BibitemOpen
  \bibfield  {author} {\bibinfo {author} {\bibfnamefont {R.~E.}\ \bibnamefont
  {Prange}}\ and\ \bibinfo {author} {\bibfnamefont {S.~M.}\ \bibnamefont
  {Girvin}},\ }\href@noop {} {\emph {\bibinfo {title} {The Quantum Hall
  Effect}}}\ (\bibinfo  {publisher} {Springer, New York},\ \bibinfo {year}
  {1987})\BibitemShut {NoStop}%
\bibitem [{\citenamefont {Hasan}\ and\ \citenamefont
  {Kane}(2010)}]{HasanKane2010}%
  \BibitemOpen
  \bibfield  {author} {\bibinfo {author} {\bibfnamefont {M.~Z.}\ \bibnamefont
  {Hasan}}\ and\ \bibinfo {author} {\bibfnamefont {C.~L.}\ \bibnamefont
  {Kane}},\ }\href@noop {} {\bibfield  {journal} {\bibinfo  {journal} {Rev.
  Mod. Phys.}\ }\textbf {\bibinfo {volume} {82}},\ \bibinfo {pages} {3045}
  (\bibinfo {year} {2010})}\BibitemShut {NoStop}%
\bibitem [{\citenamefont {Qi}\ and\ \citenamefont {Zhang}(2011)}]{QiZhang2011}%
  \BibitemOpen
  \bibfield  {author} {\bibinfo {author} {\bibfnamefont {X.-L.}\ \bibnamefont
  {Qi}}\ and\ \bibinfo {author} {\bibfnamefont {S.-C.}\ \bibnamefont {Zhang}},\
  }\href@noop {} {\bibfield  {journal} {\bibinfo  {journal} {Rev. Mod. Phys.}\
  }\textbf {\bibinfo {volume} {83}},\ \bibinfo {pages} {1057} (\bibinfo {year}
  {2011})}\BibitemShut {NoStop}%
\bibitem [{\citenamefont {Hasan}\ and\ \citenamefont
  {Moore}(2011)}]{HasanMoore2011}%
  \BibitemOpen
  \bibfield  {author} {\bibinfo {author} {\bibfnamefont {M.~Z.}\ \bibnamefont
  {Hasan}}\ and\ \bibinfo {author} {\bibfnamefont {J.~E.}\ \bibnamefont
  {Moore}},\ }\href@noop {} {\bibfield  {journal} {\bibinfo  {journal} {Annu.
  Rev. Condens. Matter Phys.}\ }\textbf {\bibinfo {volume} {2}},\ \bibinfo
  {pages} {55} (\bibinfo {year} {2011})}\BibitemShut {NoStop}%
\bibitem [{\citenamefont {Schnyder}\ \emph {et~al.}(2008)\citenamefont
  {Schnyder}, \citenamefont {Ryu}, \citenamefont {Furusaki},\ and\
  \citenamefont {Ludwig}}]{Schnyder2008}%
  \BibitemOpen
  \bibfield  {author} {\bibinfo {author} {\bibfnamefont {A.~P.}\ \bibnamefont
  {Schnyder}}, \bibinfo {author} {\bibfnamefont {S.}~\bibnamefont {Ryu}},
  \bibinfo {author} {\bibfnamefont {A.}~\bibnamefont {Furusaki}}, \ and\
  \bibinfo {author} {\bibfnamefont {A.~W.~W.}\ \bibnamefont {Ludwig}},\
  }\href@noop {} {\bibfield  {journal} {\bibinfo  {journal} {Phys. Rev. B}\
  }\textbf {\bibinfo {volume} {78}},\ \bibinfo {pages} {195125} (\bibinfo
  {year} {2008})}\BibitemShut {NoStop}%
\bibitem [{\citenamefont {Ryu}\ \emph {et~al.}(2010)\citenamefont {Ryu},
  \citenamefont {Schnyder}, \citenamefont {Furusaki},\ and\ \citenamefont
  {Ludwig}}]{SRFLnewJphys}%
  \BibitemOpen
  \bibfield  {author} {\bibinfo {author} {\bibfnamefont {S.}~\bibnamefont
  {Ryu}}, \bibinfo {author} {\bibfnamefont {A.}~\bibnamefont {Schnyder}},
  \bibinfo {author} {\bibfnamefont {A.}~\bibnamefont {Furusaki}}, \ and\
  \bibinfo {author} {\bibfnamefont {A.~W.~W.}\ \bibnamefont {Ludwig}},\
  }\href@noop {} {\bibfield  {journal} {\bibinfo  {journal} {New J. Phys.}\
  }\textbf {\bibinfo {volume} {12}},\ \bibinfo {pages} {065010} (\bibinfo
  {year} {2010})}\BibitemShut {NoStop}%
\bibitem [{\citenamefont {Kitaev}(2009)}]{Kitaev2009}%
  \BibitemOpen
  \bibfield  {author} {\bibinfo {author} {\bibfnamefont {A.~Y.}\ \bibnamefont
  {Kitaev}},\ }\href@noop {} {\bibfield  {journal} {\bibinfo  {journal} {AIP
  Conf.\ Proc.}\ }\textbf {\bibinfo {volume} {1134}},\ \bibinfo {pages} {22}
  (\bibinfo {year} {2009})}\BibitemShut {NoStop}%
\bibitem [{\citenamefont {Pollmann}\ \emph {et~al.}(2010)\citenamefont
  {Pollmann}, \citenamefont {Berg}, \citenamefont {Turner},\ and\ \citenamefont
  {Oshikawa}}]{Pollmann2010}%
  \BibitemOpen
  \bibfield  {author} {\bibinfo {author} {\bibfnamefont {F.}~\bibnamefont
  {Pollmann}}, \bibinfo {author} {\bibfnamefont {E.}~\bibnamefont {Berg}},
  \bibinfo {author} {\bibfnamefont {A.~M.}\ \bibnamefont {Turner}}, \ and\
  \bibinfo {author} {\bibfnamefont {M.}~\bibnamefont {Oshikawa}},\ }\href@noop
  {} {\bibfield  {journal} {\bibinfo  {journal} {Phys. Rev. B}\ }\textbf
  {\bibinfo {volume} {81}},\ \bibinfo {pages} {064439} (\bibinfo {year}
  {2010})}\BibitemShut {NoStop}%
\bibitem [{\citenamefont {Chen}\ \emph {et~al.}(2011)\citenamefont {Chen},
  \citenamefont {Gu},\ and\ \citenamefont {Wen}}]{Chen2011}%
  \BibitemOpen
  \bibfield  {author} {\bibinfo {author} {\bibfnamefont {X.}~\bibnamefont
  {Chen}}, \bibinfo {author} {\bibfnamefont {Z.-C.}\ \bibnamefont {Gu}}, \ and\
  \bibinfo {author} {\bibfnamefont {X.-G.}\ \bibnamefont {Wen}},\ }\href@noop
  {} {\bibfield  {journal} {\bibinfo  {journal} {Phys. Rev. B}\ }\textbf
  {\bibinfo {volume} {83}},\ \bibinfo {pages} {035107} (\bibinfo {year}
  {2011})}\BibitemShut {NoStop}%
\bibitem [{\citenamefont {Schuch}\ \emph {et~al.}(2011)\citenamefont {Schuch},
  \citenamefont {Perez-Garcia},\ and\ \citenamefont {Cirac}}]{Schuch2011}%
  \BibitemOpen
  \bibfield  {author} {\bibinfo {author} {\bibfnamefont {N.}~\bibnamefont
  {Schuch}}, \bibinfo {author} {\bibfnamefont {D.}~\bibnamefont
  {Perez-Garcia}}, \ and\ \bibinfo {author} {\bibfnamefont {I.}~\bibnamefont
  {Cirac}},\ }\href@noop {} {\bibfield  {journal} {\bibinfo  {journal} {Phys.
  Rev. B}\ }\textbf {\bibinfo {volume} {84}},\ \bibinfo {pages} {165139}
  (\bibinfo {year} {2011})}\BibitemShut {NoStop}%
\bibitem [{\citenamefont {Fidkowski}\ and\ \citenamefont
  {Kitaev}(2010)}]{Fidkowski2010}%
  \BibitemOpen
  \bibfield  {author} {\bibinfo {author} {\bibfnamefont {L.}~\bibnamefont
  {Fidkowski}}\ and\ \bibinfo {author} {\bibfnamefont {A.}~\bibnamefont
  {Kitaev}},\ }\href@noop {} {\bibfield  {journal} {\bibinfo  {journal} {Phys.
  Rev. B}\ }\textbf {\bibinfo {volume} {81}},\ \bibinfo {pages} {134509}
  (\bibinfo {year} {2010})}\BibitemShut {NoStop}%
\bibitem [{\citenamefont {Fidkowski}\ and\ \citenamefont
  {Kitaev}(2011)}]{Fidkowski2011}%
  \BibitemOpen
  \bibfield  {author} {\bibinfo {author} {\bibfnamefont {L.}~\bibnamefont
  {Fidkowski}}\ and\ \bibinfo {author} {\bibfnamefont {A.}~\bibnamefont
  {Kitaev}},\ }\href@noop {} {\bibfield  {journal} {\bibinfo  {journal} {Phys.
  Rev. B}\ }\textbf {\bibinfo {volume} {83}},\ \bibinfo {pages} {075103}
  (\bibinfo {year} {2011})}\BibitemShut {NoStop}%
\bibitem [{\citenamefont {Turner}\ \emph {et~al.}(2011)\citenamefont {Turner},
  \citenamefont {Pollmann},\ and\ \citenamefont {Berg}}]{Turner2011}%
  \BibitemOpen
  \bibfield  {author} {\bibinfo {author} {\bibfnamefont {A.~M.}\ \bibnamefont
  {Turner}}, \bibinfo {author} {\bibfnamefont {F.}~\bibnamefont {Pollmann}}, \
  and\ \bibinfo {author} {\bibfnamefont {E.}~\bibnamefont {Berg}},\ }\href@noop
  {} {\bibfield  {journal} {\bibinfo  {journal} {Phys. Rev. B}\ }\textbf
  {\bibinfo {volume} {83}},\ \bibinfo {pages} {075102} (\bibinfo {year}
  {2011})}\BibitemShut {NoStop}%
\bibitem [{\citenamefont {Tang}\ and\ \citenamefont {Wen}(2012)}]{Tang2012}%
  \BibitemOpen
  \bibfield  {author} {\bibinfo {author} {\bibfnamefont {E.}~\bibnamefont
  {Tang}}\ and\ \bibinfo {author} {\bibfnamefont {X.-G.}\ \bibnamefont {Wen}},\
  }\href@noop {} {\bibfield  {journal} {\bibinfo  {journal} {Phys. Rev. Lett.}\
  }\textbf {\bibinfo {volume} {109}},\ \bibinfo {pages} {096403} (\bibinfo
  {year} {2012})}\BibitemShut {NoStop}%
\bibitem [{\citenamefont {Chen}\ \emph {et~al.}(2013)\citenamefont {Chen},
  \citenamefont {Gu}, \citenamefont {Lin},\ and\ \citenamefont
  {Wen}}]{Chen2013}%
  \BibitemOpen
  \bibfield  {author} {\bibinfo {author} {\bibfnamefont {X.}~\bibnamefont
  {Chen}}, \bibinfo {author} {\bibfnamefont {Z.-C.}\ \bibnamefont {Gu}},
  \bibinfo {author} {\bibfnamefont {Z.-X.}\ \bibnamefont {Lin}}, \ and\
  \bibinfo {author} {\bibfnamefont {X.-G.}\ \bibnamefont {Wen}},\ }\href@noop
  {} {\bibfield  {journal} {\bibinfo  {journal} {Phys. Rev. B}\ }\textbf
  {\bibinfo {volume} {87}},\ \bibinfo {pages} {155114} (\bibinfo {year}
  {2013})}\BibitemShut {NoStop}%
\bibitem [{\citenamefont {Chen}\ \emph {et~al.}(2012)\citenamefont {Chen},
  \citenamefont {Gu}, \citenamefont {Liu},\ and\ \citenamefont
  {Wen}}]{Chen2012}%
  \BibitemOpen
  \bibfield  {author} {\bibinfo {author} {\bibfnamefont {X.}~\bibnamefont
  {Chen}}, \bibinfo {author} {\bibfnamefont {Z.-C.}\ \bibnamefont {Gu}},
  \bibinfo {author} {\bibfnamefont {Z.-X.}\ \bibnamefont {Liu}}, \ and\
  \bibinfo {author} {\bibfnamefont {X.-G.}\ \bibnamefont {Wen}},\ }\href@noop
  {} {\bibfield  {journal} {\bibinfo  {journal} {Science}\ }\textbf {\bibinfo
  {volume} {338}},\ \bibinfo {pages} {1604} (\bibinfo {year}
  {2012})}\BibitemShut {NoStop}%
\bibitem [{\citenamefont {Gu}\ and\ \citenamefont {Wen}(2012)}]{Gu2012}%
  \BibitemOpen
  \bibfield  {author} {\bibinfo {author} {\bibfnamefont {Z.-C.}\ \bibnamefont
  {Gu}}\ and\ \bibinfo {author} {\bibfnamefont {X.-G.}\ \bibnamefont {Wen}},\
  }\href@noop {} {\  (\bibinfo {year} {2012})},\ \Eprint
  {http://arxiv.org/abs/1200.2648} {1200.2648} \BibitemShut {NoStop}%
\bibitem [{\citenamefont {Qi}(2013)}]{Qi2013}%
  \BibitemOpen
  \bibfield  {author} {\bibinfo {author} {\bibfnamefont {X.-L.}\ \bibnamefont
  {Qi}},\ }\href@noop {} {\bibfield  {journal} {\bibinfo  {journal} {New J.
  Phys.}\ }\textbf {\bibinfo {volume} {15}},\ \bibinfo {pages} {065002}
  (\bibinfo {year} {2013})}\BibitemShut {NoStop}%
\bibitem [{\citenamefont {Ryu}\ and\ \citenamefont {Zhang}(2012)}]{Ryu2012}%
  \BibitemOpen
  \bibfield  {author} {\bibinfo {author} {\bibfnamefont {S.}~\bibnamefont
  {Ryu}}\ and\ \bibinfo {author} {\bibfnamefont {S.-C.}\ \bibnamefont
  {Zhang}},\ }\href@noop {} {\bibfield  {journal} {\bibinfo  {journal} {Phys.
  Rev. B}\ }\textbf {\bibinfo {volume} {85}},\ \bibinfo {pages} {245132}
  (\bibinfo {year} {2012})}\BibitemShut {NoStop}%
\bibitem [{\citenamefont {Yao}\ and\ \citenamefont {Ryu}(2013)}]{Yao2012}%
  \BibitemOpen
  \bibfield  {author} {\bibinfo {author} {\bibfnamefont {H.}~\bibnamefont
  {Yao}}\ and\ \bibinfo {author} {\bibfnamefont {S.}~\bibnamefont {Ryu}},\
  }\href@noop {} {\bibfield  {journal} {\bibinfo  {journal} {Phys. Rev. B}\
  }\textbf {\bibinfo {volume} {88}},\ \bibinfo {pages} {064507} (\bibinfo
  {year} {2013})}\BibitemShut {NoStop}%
\bibitem [{\citenamefont {{Gu}}\ and\ \citenamefont
  {{Levin}}(2013)}]{GuLevin2013}%
  \BibitemOpen
  \bibfield  {author} {\bibinfo {author} {\bibfnamefont {Z.-C.}\ \bibnamefont
  {{Gu}}}\ and\ \bibinfo {author} {\bibfnamefont {M.}~\bibnamefont {{Levin}}},\
  }\href@noop {} {\bibfield  {journal} {\bibinfo  {journal} {ArXiv e-prints}\ }
  (\bibinfo {year} {2013})},\ \Eprint {http://arxiv.org/abs/1304.4569}
  {arXiv:1304.4569 [cond-mat.str-el]} \BibitemShut {NoStop}%
\bibitem [{\citenamefont {Lu}\ and\ \citenamefont {Vishwanath}(2012)}]{Lu2012}%
  \BibitemOpen
  \bibfield  {author} {\bibinfo {author} {\bibfnamefont {Y.-M.}\ \bibnamefont
  {Lu}}\ and\ \bibinfo {author} {\bibfnamefont {A.}~\bibnamefont
  {Vishwanath}},\ }\href@noop {} {\bibfield  {journal} {\bibinfo  {journal}
  {Phys. Rev. B}\ }\textbf {\bibinfo {volume} {86}},\ \bibinfo {pages} {125119}
  (\bibinfo {year} {2012})}\BibitemShut {NoStop}%
\bibitem [{\citenamefont {{Ryu}}\ \emph {et~al.}(2012)\citenamefont {{Ryu}},
  \citenamefont {{Moore}},\ and\ \citenamefont
  {{Ludwig}}}]{RyuMooreLudwig2012}%
  \BibitemOpen
  \bibfield  {author} {\bibinfo {author} {\bibfnamefont {S.}~\bibnamefont
  {{Ryu}}}, \bibinfo {author} {\bibfnamefont {J.~E.}\ \bibnamefont {{Moore}}},
  \ and\ \bibinfo {author} {\bibfnamefont {A.~W.~W.}\ \bibnamefont
  {{Ludwig}}},\ }\href {\doibase 10.1103/PhysRevB.85.045104} {\bibfield
  {journal} {\bibinfo  {journal} {\prb}\ }\textbf {\bibinfo {volume} {85}},\
  \bibinfo {eid} {045104} (\bibinfo {year} {2012})},\ \Eprint
  {http://arxiv.org/abs/1010.0936} {arXiv:1010.0936 [cond-mat.str-el]}
  \BibitemShut {NoStop}%
\bibitem [{\citenamefont {Ringel}\ and\ \citenamefont
  {Stern}(2012)}]{Ringel2012}%
  \BibitemOpen
  \bibfield  {author} {\bibinfo {author} {\bibfnamefont {Z.}~\bibnamefont
  {Ringel}}\ and\ \bibinfo {author} {\bibfnamefont {A.}~\bibnamefont {Stern}},\
  }\href@noop {} {\  (\bibinfo {year} {2012})},\ \Eprint
  {http://arxiv.org/abs/1212.3796} {1212.3796} \BibitemShut {NoStop}%
\bibitem [{\citenamefont {{Koch-Janusz}}\ and\ \citenamefont
  {{Ringel}}(2013)}]{Koch-Janusz2013}%
  \BibitemOpen
  \bibfield  {author} {\bibinfo {author} {\bibfnamefont {M.}~\bibnamefont
  {{Koch-Janusz}}}\ and\ \bibinfo {author} {\bibfnamefont {Z.}~\bibnamefont
  {{Ringel}}},\ }\href@noop {} {\bibfield  {journal} {\bibinfo  {journal}
  {ArXiv e-prints}\ } (\bibinfo {year} {2013})},\ \Eprint
  {http://arxiv.org/abs/1311.6507} {arXiv:1311.6507 [cond-mat.str-el]}
  \BibitemShut {NoStop}%
\bibitem [{\citenamefont {Cappelli}\ and\ \citenamefont
  {Randellini}(2013)}]{Cappelli13}%
  \BibitemOpen
  \bibfield  {author} {\bibinfo {author} {\bibfnamefont {A.}~\bibnamefont
  {Cappelli}}\ and\ \bibinfo {author} {\bibfnamefont {E.}~\bibnamefont
  {Randellini}},\ }\href@noop {} {\bibfield  {journal} {\bibinfo  {journal}
  {JHEP}\ }\textbf {\bibinfo {volume} {12}},\ \bibinfo {pages} {101} (\bibinfo
  {year} {2013})}\BibitemShut {NoStop}%
\bibitem [{\citenamefont {{Wang}}\ and\ \citenamefont
  {{Wen}}(2013)}]{Wang2013}%
  \BibitemOpen
  \bibfield  {author} {\bibinfo {author} {\bibfnamefont {J.}~\bibnamefont
  {{Wang}}}\ and\ \bibinfo {author} {\bibfnamefont {X.-G.}\ \bibnamefont
  {{Wen}}},\ }\href@noop {} {\bibfield  {journal} {\bibinfo  {journal} {ArXiv
  e-prints}\ } (\bibinfo {year} {2013})},\ \Eprint
  {http://arxiv.org/abs/1307.7480} {arXiv:1307.7480 [hep-lat]} \BibitemShut
  {NoStop}%
\bibitem [{Note1()}]{Note1}%
  \BibitemOpen
  \bibinfo {note} {As a clarifying remark, SPT phases are not
  topologically-ordered phases of matter in that they do not have defining
  properties of topological order such as non-trivial ground state degeneracy,
  fractional statistics, etc. Nevertheless, such phases are topologically
  non-trivial in the sense that they are not adiabatically connected to a
  trivial phase. On the other hand, there is a class of topologically-ordered
  phases that have some interesting properties in the presence of symmetries --
  they are called symmetry-enriched topological phases. The methodology
  proposed in this paper (a generalization of Laughlin's argument) works both
  for SPT phases and symmetry-enriched topological phases.}\BibitemShut {Stop}%
\bibitem [{\citenamefont {Sule}\ \emph {et~al.}(2013)\citenamefont {Sule},
  \citenamefont {Chen},\ and\ \citenamefont {Ryu}}]{Sule13}%
  \BibitemOpen
  \bibfield  {author} {\bibinfo {author} {\bibfnamefont {O.~M.}\ \bibnamefont
  {Sule}}, \bibinfo {author} {\bibfnamefont {X.}~\bibnamefont {Chen}}, \ and\
  \bibinfo {author} {\bibfnamefont {S.}~\bibnamefont {Ryu}},\ }\href@noop {}
  {\bibfield  {journal} {\bibinfo  {journal} {Phys.\ Rev.\ B}\ }\textbf
  {\bibinfo {volume} {88}},\ \bibinfo {pages} {075125} (\bibinfo {year}
  {2013})}\BibitemShut {NoStop}%
\bibitem [{\citenamefont {Levin}(2013)}]{Levin2013}%
  \BibitemOpen
  \bibfield  {author} {\bibinfo {author} {\bibfnamefont {M.}~\bibnamefont
  {Levin}},\ }\href@noop {} {\bibfield  {journal} {\bibinfo  {journal} {Phys.
  Rev. X}\ }\textbf {\bibinfo {volume} {3}},\ \bibinfo {pages} {021009}
  (\bibinfo {year} {2013})},\ \Eprint {http://arxiv.org/abs/1301.7355}
  {1301.7355} \BibitemShut {NoStop}%
\bibitem [{\citenamefont {Levin}\ and\ \citenamefont {Gu}(2012)}]{LevinGu12}%
  \BibitemOpen
  \bibfield  {author} {\bibinfo {author} {\bibfnamefont {M.}~\bibnamefont
  {Levin}}\ and\ \bibinfo {author} {\bibfnamefont {Z.-C.}\ \bibnamefont {Gu}},\
  }\href@noop {} {\bibfield  {journal} {\bibinfo  {journal} {Phys. Rev. B.}\
  }\textbf {\bibinfo {volume} {86}},\ \bibinfo {pages} {115109} (\bibinfo
  {year} {2012})}\BibitemShut {NoStop}%
\bibitem [{\citenamefont {Turner}\ \emph {et~al.}(2010)\citenamefont {Turner},
  \citenamefont {Zhang},\ and\ \citenamefont {Vishwanath}}]{Turner11}%
  \BibitemOpen
  \bibfield  {author} {\bibinfo {author} {\bibfnamefont {A.~M.}\ \bibnamefont
  {Turner}}, \bibinfo {author} {\bibfnamefont {Y.}~\bibnamefont {Zhang}}, \
  and\ \bibinfo {author} {\bibfnamefont {A.}~\bibnamefont {Vishwanath}},\
  }\href@noop {} {\bibfield  {journal} {\bibinfo  {journal} {Phys. Rev. B}\
  }\textbf {\bibinfo {volume} {82}},\ \bibinfo {pages} {241102} (\bibinfo
  {year} {2010})}\BibitemShut {NoStop}%
\bibitem [{\citenamefont {Turner}\ \emph {et~al.}(2012)\citenamefont {Turner},
  \citenamefont {Zhang}, \citenamefont {Mong},\ and\ \citenamefont
  {Vishwanath}}]{Turner12}%
  \BibitemOpen
  \bibfield  {author} {\bibinfo {author} {\bibfnamefont {A.~M.}\ \bibnamefont
  {Turner}}, \bibinfo {author} {\bibfnamefont {Y.}~\bibnamefont {Zhang}},
  \bibinfo {author} {\bibfnamefont {R.~S.~K.}\ \bibnamefont {Mong}}, \ and\
  \bibinfo {author} {\bibfnamefont {A.}~\bibnamefont {Vishwanath}},\
  }\href@noop {} {\bibfield  {journal} {\bibinfo  {journal} {Phys. Rev. B}\
  }\textbf {\bibinfo {volume} {85}},\ \bibinfo {pages} {165120} (\bibinfo
  {year} {2012})}\BibitemShut {NoStop}%
\bibitem [{\citenamefont {Hughes}\ \emph {et~al.}(2011)\citenamefont {Hughes},
  \citenamefont {Prodan},\ and\ \citenamefont {Bernevig}}]{Hughes11}%
  \BibitemOpen
  \bibfield  {author} {\bibinfo {author} {\bibfnamefont {T.~L.}\ \bibnamefont
  {Hughes}}, \bibinfo {author} {\bibfnamefont {E.}~\bibnamefont {Prodan}}, \
  and\ \bibinfo {author} {\bibfnamefont {B.~A.}\ \bibnamefont {Bernevig}},\
  }\href@noop {} {\bibfield  {journal} {\bibinfo  {journal} {Phys. Rev. B}\
  }\textbf {\bibinfo {volume} {83}},\ \bibinfo {pages} {245132} (\bibinfo
  {year} {2011})}\BibitemShut {NoStop}%
\bibitem [{\citenamefont {Fu}(2011)}]{Fu11}%
  \BibitemOpen
  \bibfield  {author} {\bibinfo {author} {\bibfnamefont {L.}~\bibnamefont
  {Fu}},\ }\href@noop {} {\bibfield  {journal} {\bibinfo  {journal} {Phys. Rev.
  Lett.}\ }\textbf {\bibinfo {volume} {106}},\ \bibinfo {pages} {106802}
  (\bibinfo {year} {2011})}\BibitemShut {NoStop}%
\bibitem [{\citenamefont {Slager}\ \emph {et~al.}(2013)\citenamefont {Slager},
  \citenamefont {Mesaros}, \citenamefont {Juricic},\ and\ \citenamefont
  {Zaanen}}]{Slager13}%
  \BibitemOpen
  \bibfield  {author} {\bibinfo {author} {\bibfnamefont {R.-J.}\ \bibnamefont
  {Slager}}, \bibinfo {author} {\bibfnamefont {A.}~\bibnamefont {Mesaros}},
  \bibinfo {author} {\bibfnamefont {V.}~\bibnamefont {Juricic}}, \ and\
  \bibinfo {author} {\bibfnamefont {J.}~\bibnamefont {Zaanen}},\ }\href@noop {}
  {\bibfield  {journal} {\bibinfo  {journal} {Nat. Phys}\ }\textbf {\bibinfo
  {volume} {9}},\ \bibinfo {pages} {98} (\bibinfo {year} {2013})}\BibitemShut
  {NoStop}%
\bibitem [{\citenamefont {Fang}\ \emph
  {et~al.}(2012{\natexlab{a}})\citenamefont {Fang}, \citenamefont {Gilbert},\
  and\ \citenamefont {Bernevig}}]{Fang12a}%
  \BibitemOpen
  \bibfield  {author} {\bibinfo {author} {\bibfnamefont {C.}~\bibnamefont
  {Fang}}, \bibinfo {author} {\bibfnamefont {M.}~\bibnamefont {Gilbert}}, \
  and\ \bibinfo {author} {\bibfnamefont {A.}~\bibnamefont {Bernevig}},\
  }\href@noop {} {\  (\bibinfo {year} {2012}{\natexlab{a}})}\BibitemShut
  {NoStop}%
\bibitem [{\citenamefont {Fang}\ \emph
  {et~al.}(2012{\natexlab{b}})\citenamefont {Fang}, \citenamefont {Gilbert},\
  and\ \citenamefont {Bernevig}}]{Fang12b}%
  \BibitemOpen
  \bibfield  {author} {\bibinfo {author} {\bibfnamefont {C.}~\bibnamefont
  {Fang}}, \bibinfo {author} {\bibfnamefont {M.}~\bibnamefont {Gilbert}}, \
  and\ \bibinfo {author} {\bibfnamefont {A.}~\bibnamefont {Bernevig}},\
  }\href@noop {} {\  (\bibinfo {year} {2012}{\natexlab{b}})}\BibitemShut
  {NoStop}%
\bibitem [{\citenamefont {Hsieh}\ \emph {et~al.}(2012)\citenamefont {Hsieh},
  \citenamefont {Lin}, \citenamefont {Liu}, \citenamefont {Duan}, \citenamefont
  {Bansil},\ and\ \citenamefont {Fu}}]{Hsieh12}%
  \BibitemOpen
  \bibfield  {author} {\bibinfo {author} {\bibfnamefont {T.~H.}\ \bibnamefont
  {Hsieh}}, \bibinfo {author} {\bibfnamefont {H.}~\bibnamefont {Lin}}, \bibinfo
  {author} {\bibfnamefont {J.}~\bibnamefont {Liu}}, \bibinfo {author}
  {\bibfnamefont {W.}~\bibnamefont {Duan}}, \bibinfo {author} {\bibfnamefont
  {A.}~\bibnamefont {Bansil}}, \ and\ \bibinfo {author} {\bibfnamefont
  {L.}~\bibnamefont {Fu}},\ }\href@noop {} {\bibfield  {journal} {\bibinfo
  {journal} {Nat. Commun.}\ }\textbf {\bibinfo {volume} {3}},\ \bibinfo {pages}
  {982} (\bibinfo {year} {2012})}\BibitemShut {NoStop}%
\bibitem [{\citenamefont {Teo}\ \emph {et~al.}(2008)\citenamefont {Teo},
  \citenamefont {Fu},\ and\ \citenamefont {Kane}}]{Teo08}%
  \BibitemOpen
  \bibfield  {author} {\bibinfo {author} {\bibfnamefont {J.~C.~Y.}\
  \bibnamefont {Teo}}, \bibinfo {author} {\bibfnamefont {L.}~\bibnamefont
  {Fu}}, \ and\ \bibinfo {author} {\bibfnamefont {C.~L.}\ \bibnamefont
  {Kane}},\ }\href@noop {} {\bibfield  {journal} {\bibinfo  {journal} {Phys.
  Rev. B}\ }\textbf {\bibinfo {volume} {78}},\ \bibinfo {pages} {045426}
  (\bibinfo {year} {2008})}\BibitemShut {NoStop}%
\bibitem [{\citenamefont {Tanaka}\ \emph {et~al.}(2012)\citenamefont {Tanaka},
  \citenamefont {Ren}, \citenamefont {Sato}, \citenamefont {Nakayama},
  \citenamefont {Souma}, \citenamefont {Takahashi}, \citenamefont {Segawa},\
  and\ \citenamefont {Ando}}]{Tanaka12}%
  \BibitemOpen
  \bibfield  {author} {\bibinfo {author} {\bibfnamefont {Y.}~\bibnamefont
  {Tanaka}}, \bibinfo {author} {\bibfnamefont {Z.}~\bibnamefont {Ren}},
  \bibinfo {author} {\bibfnamefont {T.}~\bibnamefont {Sato}}, \bibinfo {author}
  {\bibfnamefont {K.}~\bibnamefont {Nakayama}}, \bibinfo {author}
  {\bibfnamefont {S.}~\bibnamefont {Souma}}, \bibinfo {author} {\bibfnamefont
  {T.}~\bibnamefont {Takahashi}}, \bibinfo {author} {\bibfnamefont
  {K.}~\bibnamefont {Segawa}}, \ and\ \bibinfo {author} {\bibfnamefont
  {Y.}~\bibnamefont {Ando}},\ }\href@noop {} {\bibfield  {journal} {\bibinfo
  {journal} {Nature Physics}\ }\textbf {\bibinfo {volume} {8}},\ \bibinfo
  {pages} {800} (\bibinfo {year} {2012})}\BibitemShut {NoStop}%
\bibitem [{\citenamefont {Dziawa}\ \emph {et~al.}(2012)\citenamefont {Dziawa},
  \citenamefont {Kowalski}, \citenamefont {Dybko}, \citenamefont {Buczko},
  \citenamefont {Szczerbakow}, \citenamefont {Szot}, \citenamefont
  {Lusakowska}, \citenamefont {Balasubramanian}, \citenamefont {Wojek},
  \citenamefont {Berntsen}, \citenamefont {Tjernberg},\ and\ \citenamefont
  {Story}}]{Dziawa12}%
  \BibitemOpen
  \bibfield  {author} {\bibinfo {author} {\bibfnamefont {P.}~\bibnamefont
  {Dziawa}}, \bibinfo {author} {\bibfnamefont {B.~J.}\ \bibnamefont
  {Kowalski}}, \bibinfo {author} {\bibfnamefont {K.}~\bibnamefont {Dybko}},
  \bibinfo {author} {\bibfnamefont {R.}~\bibnamefont {Buczko}}, \bibinfo
  {author} {\bibfnamefont {A.}~\bibnamefont {Szczerbakow}}, \bibinfo {author}
  {\bibfnamefont {M.}~\bibnamefont {Szot}}, \bibinfo {author} {\bibfnamefont
  {E.}~\bibnamefont {Lusakowska}}, \bibinfo {author} {\bibfnamefont
  {T.}~\bibnamefont {Balasubramanian}}, \bibinfo {author} {\bibfnamefont
  {B.~M.}\ \bibnamefont {Wojek}}, \bibinfo {author} {\bibfnamefont {M.~H.}\
  \bibnamefont {Berntsen}}, \bibinfo {author} {\bibfnamefont {O.}~\bibnamefont
  {Tjernberg}}, \ and\ \bibinfo {author} {\bibfnamefont {T.}~\bibnamefont
  {Story}},\ }\href@noop {} {\bibfield  {journal} {\bibinfo  {journal} {Nature
  Materials}\ }\textbf {\bibinfo {volume} {11}},\ \bibinfo {pages} {1023}
  (\bibinfo {year} {2012})}\BibitemShut {NoStop}%
\bibitem [{\citenamefont {Chiu}\ \emph {et~al.}(2013)\citenamefont {Chiu},
  \citenamefont {Yao},\ and\ \citenamefont {Ryu}}]{Chiu13}%
  \BibitemOpen
  \bibfield  {author} {\bibinfo {author} {\bibfnamefont {C.-K.}\ \bibnamefont
  {Chiu}}, \bibinfo {author} {\bibfnamefont {H.}~\bibnamefont {Yao}}, \ and\
  \bibinfo {author} {\bibfnamefont {S.}~\bibnamefont {Ryu}},\ }\href@noop {}
  {\bibfield  {journal} {\bibinfo  {journal} {Phys. Rev. B}\ }\textbf {\bibinfo
  {volume} {88}},\ \bibinfo {pages} {075142} (\bibinfo {year}
  {2013})}\BibitemShut {NoStop}%
\bibitem [{\citenamefont {Morimoto}\ and\ \citenamefont
  {Furusaki}(2013)}]{Morimoto13}%
  \BibitemOpen
  \bibfield  {author} {\bibinfo {author} {\bibfnamefont {T.}~\bibnamefont
  {Morimoto}}\ and\ \bibinfo {author} {\bibfnamefont {A.}~\bibnamefont
  {Furusaki}},\ }\href@noop {} {\bibfield  {journal} {\bibinfo  {journal}
  {Phys. Rev. B}\ }\textbf {\bibinfo {volume} {88}},\ \bibinfo {pages} {125129}
  (\bibinfo {year} {2013})},\ \Eprint {http://arxiv.org/abs/1306.2505}
  {1306.2505} \BibitemShut {NoStop}%
\bibitem [{\citenamefont {Zhang}\ \emph {et~al.}(2013)\citenamefont {Zhang},
  \citenamefont {Kane},\ and\ \citenamefont {Mele}}]{Zhang13}%
  \BibitemOpen
  \bibfield  {author} {\bibinfo {author} {\bibfnamefont {F.}~\bibnamefont
  {Zhang}}, \bibinfo {author} {\bibfnamefont {C.~L.}\ \bibnamefont {Kane}}, \
  and\ \bibinfo {author} {\bibfnamefont {E.~J.}\ \bibnamefont {Mele}},\
  }\href@noop {} {\bibfield  {journal} {\bibinfo  {journal} {Phys. Rev. Lett.}\
  }\textbf {\bibinfo {volume} {111}},\ \bibinfo {pages} {056403} (\bibinfo
  {year} {2013})},\ \Eprint {http://arxiv.org/abs/1303.4144} {1303.4144}
  \BibitemShut {NoStop}%
\bibitem [{\citenamefont {Callan}\ \emph {et~al.}(1987)\citenamefont {Callan},
  \citenamefont {Lovelace}, \citenamefont {Nappi},\ and\ \citenamefont
  {Yost}}]{Callan1987}%
  \BibitemOpen
  \bibfield  {author} {\bibinfo {author} {\bibfnamefont {C.~G.}\ \bibnamefont
  {Callan}}, \bibinfo {author} {\bibfnamefont {C.}~\bibnamefont {Lovelace}},
  \bibinfo {author} {\bibfnamefont {C.~R.}\ \bibnamefont {Nappi}}, \ and\
  \bibinfo {author} {\bibfnamefont {S.~A.}\ \bibnamefont {Yost}},\ }\href@noop
  {} {\bibfield  {journal} {\bibinfo  {journal} {Nucl.\ Phys.\ B}\ }\textbf
  {\bibinfo {volume} {293}},\ \bibinfo {pages} {83} (\bibinfo {year}
  {1987})}\BibitemShut {NoStop}%
\bibitem [{\citenamefont {Polchinski}\ and\ \citenamefont
  {Cai}(1988)}]{PolchinskiCai1988}%
  \BibitemOpen
  \bibfield  {author} {\bibinfo {author} {\bibfnamefont {J.}~\bibnamefont
  {Polchinski}}\ and\ \bibinfo {author} {\bibfnamefont {Y.}~\bibnamefont
  {Cai}},\ }\href@noop {} {\bibfield  {journal} {\bibinfo  {journal} {Nucl.\
  Phys.\ B}\ }\textbf {\bibinfo {volume} {296}},\ \bibinfo {pages} {91}
  (\bibinfo {year} {1988})}\BibitemShut {NoStop}%
\bibitem [{\citenamefont {Angelantonj}\ and\ \citenamefont
  {Sagnotti}(2002)}]{Angelantonj02}%
  \BibitemOpen
  \bibfield  {author} {\bibinfo {author} {\bibfnamefont {C.}~\bibnamefont
  {Angelantonj}}\ and\ \bibinfo {author} {\bibfnamefont {A.}~\bibnamefont
  {Sagnotti}},\ }\href@noop {} {\bibfield  {journal} {\bibinfo  {journal}
  {Phys. Rept.}\ }\textbf {\bibinfo {volume} {371}},\ \bibinfo {pages} {1}
  (\bibinfo {year} {2002})}\BibitemShut {NoStop}%
\bibitem [{\citenamefont {Sagnotti}(1988)}]{Sagnotti1988}%
  \BibitemOpen
  \bibfield  {author} {\bibinfo {author} {\bibfnamefont {A.}~\bibnamefont
  {Sagnotti}},\ }\href@noop {} {\bibfield  {journal} {\bibinfo  {journal}
  {Cargese '87, ``Non-perturbative Quantum Field Theory,'' ed.\ G. Mack et al.
  (Pergamon Press)}\ ,\ \bibinfo {pages} {521}} (\bibinfo {year}
  {1988})}\BibitemShut {NoStop}%
\bibitem [{\citenamefont {Dai}\ \emph {et~al.}(1989)\citenamefont {Dai},
  \citenamefont {Leigh},\ and\ \citenamefont {Polchinski}}]{Dai1989}%
  \BibitemOpen
  \bibfield  {author} {\bibinfo {author} {\bibfnamefont {J.}~\bibnamefont
  {Dai}}, \bibinfo {author} {\bibfnamefont {R.~G.}\ \bibnamefont {Leigh}}, \
  and\ \bibinfo {author} {\bibfnamefont {J.}~\bibnamefont {Polchinski}},\
  }\href@noop {} {\bibfield  {journal} {\bibinfo  {journal} {Mod. Phys. Lett.
  A}\ }\textbf {\bibinfo {volume} {4}},\ \bibinfo {pages} {2073} (\bibinfo
  {year} {1989})}\BibitemShut {NoStop}%
\bibitem [{Note2()}]{Note2}%
  \BibitemOpen
  \bibinfo {note} {For a connection between orientifolds and time-reversal
  symmetric topological insulators and superconductors from the spacetime point
  of view (as opposed to the worldsheet point of view presented in this paper),
  see Refs.\ \protect \rev@citealpnum {RyuTakayanagi2010a,
  RyuTakayanagi2010b}.}\BibitemShut {Stop}%
\bibitem [{\citenamefont {Hsieh}\ \emph {et~al.}()\citenamefont {Hsieh},
  \citenamefont {Morimoto},\ and\ \citenamefont {Ryu}}]{Hsieh13}%
  \BibitemOpen
  \bibfield  {author} {\bibinfo {author} {\bibfnamefont {C.-T.}\ \bibnamefont
  {Hsieh}}, \bibinfo {author} {\bibfnamefont {T.}~\bibnamefont {Morimoto}}, \
  and\ \bibinfo {author} {\bibfnamefont {S.}~\bibnamefont {Ryu}},\ }\href@noop
  {} {\bibinfo  {journal} {unpublished}\ }\BibitemShut {NoStop}%
\bibitem [{Note3()}]{Note3}%
  \BibitemOpen
\bibfield  {journal} {  }\bibinfo {note} {The fermionic models with CP and
  charge U(1) symmetries can also be interpreted/realized as a BdG system with
  parity and spin U(1) symmeries ({\protect \it e.g.}, $z$-component of spin,
  $S_z$, is conserved).}\BibitemShut {Stop}%
\bibitem [{\citenamefont {Chang}\ \emph {et~al.}()\citenamefont {Chang},
  \citenamefont {Mudry},\ and\ \citenamefont {Ryu}}]{Chang13}%
  \BibitemOpen
  \bibfield  {author} {\bibinfo {author} {\bibfnamefont {P.-Y.}\ \bibnamefont
  {Chang}}, \bibinfo {author} {\bibfnamefont {C.}~\bibnamefont {Mudry}}, \ and\
  \bibinfo {author} {\bibfnamefont {S.}~\bibnamefont {Ryu}},\ }\href@noop {}
  {\bibinfo  {journal} {unpublished}\ }\BibitemShut {NoStop}%
\bibitem [{\citenamefont {Chen}\ and\ \citenamefont {Vishwanath}()}]{Chen2014}%
  \BibitemOpen
\bibfield  {journal} {  }\bibfield  {author} {\bibinfo {author} {\bibfnamefont
  {X.}~\bibnamefont {Chen}}\ and\ \bibinfo {author} {\bibfnamefont
  {A.}~\bibnamefont {Vishwanath}},\ }\href@noop {} {\bibinfo  {journal}
  {\texttt{arXiv:1401.3736}}\ }\BibitemShut {NoStop}%
\bibitem [{\citenamefont {{Kapustin}}(2014)}]{Kapustin2014}%
  \BibitemOpen
\bibfield  {journal} {  }\bibfield  {author} {\bibinfo {author} {\bibfnamefont
  {A.}~\bibnamefont {{Kapustin}}},\ }\href@noop {} {\bibfield  {journal}
  {\bibinfo  {journal} {ArXiv e-prints}\ } (\bibinfo {year} {2014})},\ \Eprint
  {http://arxiv.org/abs/1403.1467} {arXiv:1403.1467 [cond-mat.str-el]}
  \BibitemShut {NoStop}%
\bibitem [{\citenamefont {Levin}\ and\ \citenamefont
  {Stern}(2009)}]{LevinStern2009}%
  \BibitemOpen
  \bibfield  {author} {\bibinfo {author} {\bibfnamefont {M.}~\bibnamefont
  {Levin}}\ and\ \bibinfo {author} {\bibfnamefont {A.}~\bibnamefont {Stern}},\
  }\href@noop {} {\bibfield  {journal} {\bibinfo  {journal} {Phys. Rev. Lett.}\
  }\textbf {\bibinfo {volume} {103}},\ \bibinfo {pages} {196803} (\bibinfo
  {year} {2009})}\BibitemShut {NoStop}%
\bibitem [{\citenamefont {Levin}\ and\ \citenamefont
  {Stern}(2012)}]{LevinStern2012}%
  \BibitemOpen
  \bibfield  {author} {\bibinfo {author} {\bibfnamefont {M.}~\bibnamefont
  {Levin}}\ and\ \bibinfo {author} {\bibfnamefont {A.}~\bibnamefont {Stern}},\
  }\href@noop {} {\bibfield  {journal} {\bibinfo  {journal} {Phys.\ Rev.\ B}\
  }\textbf {\bibinfo {volume} {86}},\ \bibinfo {pages} {115131} (\bibinfo
  {year} {2012})},\ \Eprint {http://arxiv.org/abs/1205.1244} {1205.1244}
  \BibitemShut {NoStop}%
\bibitem [{\citenamefont {Neupert}\ \emph {et~al.}(2011)\citenamefont
  {Neupert}, \citenamefont {Santos}, \citenamefont {Ryu}, \citenamefont
  {Chamon},\ and\ \citenamefont {Mudry}}]{Neupert2011}%
  \BibitemOpen
  \bibfield  {author} {\bibinfo {author} {\bibfnamefont {T.}~\bibnamefont
  {Neupert}}, \bibinfo {author} {\bibfnamefont {L.}~\bibnamefont {Santos}},
  \bibinfo {author} {\bibfnamefont {S.}~\bibnamefont {Ryu}}, \bibinfo {author}
  {\bibfnamefont {C.}~\bibnamefont {Chamon}}, \ and\ \bibinfo {author}
  {\bibfnamefont {C.}~\bibnamefont {Mudry}},\ }\href@noop {} {\bibfield
  {journal} {\bibinfo  {journal} {Phys.\ Rev.\ B}\ }\textbf {\bibinfo {volume}
  {84}},\ \bibinfo {pages} {165107} (\bibinfo {year} {2011})},\ \Eprint
  {http://arxiv.org/abs/1106.3989} {1106.3989} \BibitemShut {NoStop}%
\bibitem [{\citenamefont {Cappelli}\ \emph {et~al.}(2002)\citenamefont
  {Cappelli}, \citenamefont {Huerta},\ and\ \citenamefont
  {Zemba}}]{Cappelli01}%
  \BibitemOpen
  \bibfield  {author} {\bibinfo {author} {\bibfnamefont {A.}~\bibnamefont
  {Cappelli}}, \bibinfo {author} {\bibfnamefont {M.}~\bibnamefont {Huerta}}, \
  and\ \bibinfo {author} {\bibfnamefont {G.~R.}\ \bibnamefont {Zemba}},\
  }\href@noop {} {\bibfield  {journal} {\bibinfo  {journal} {Nucl.\ Phys.\ B}\
  }\textbf {\bibinfo {volume} {636}},\ \bibinfo {pages} {568} (\bibinfo {year}
  {2002})}\BibitemShut {NoStop}%
\bibitem [{\citenamefont {Cappelli}\ and\ \citenamefont
  {Zemba}(1997)}]{Cappelli96}%
  \BibitemOpen
  \bibfield  {author} {\bibinfo {author} {\bibfnamefont {A.}~\bibnamefont
  {Cappelli}}\ and\ \bibinfo {author} {\bibfnamefont {G.~R.}\ \bibnamefont
  {Zemba}},\ }\href@noop {} {\bibfield  {journal} {\bibinfo  {journal} {Nucl.
  Phys. B}\ }\textbf {\bibinfo {volume} {490}},\ \bibinfo {pages} {595}
  (\bibinfo {year} {1997})}\BibitemShut {NoStop}%
\bibitem [{\citenamefont {Cappelli}\ \emph {et~al.}(2010)\citenamefont
  {Cappelli}, \citenamefont {Viola},\ and\ \citenamefont {Zemba}}]{Cappelli09}%
  \BibitemOpen
  \bibfield  {author} {\bibinfo {author} {\bibfnamefont {A.}~\bibnamefont
  {Cappelli}}, \bibinfo {author} {\bibfnamefont {G.}~\bibnamefont {Viola}}, \
  and\ \bibinfo {author} {\bibfnamefont {G.~R.}\ \bibnamefont {Zemba}},\
  }\href@noop {} {\bibfield  {journal} {\bibinfo  {journal} {Annals of
  Physics}\ }\textbf {\bibinfo {volume} {325}},\ \bibinfo {pages} {465}
  (\bibinfo {year} {2010})}\BibitemShut {NoStop}%
\bibitem [{\citenamefont {Cappelli}\ and\ \citenamefont
  {Viola}(2011)}]{Cappelli09b}%
  \BibitemOpen
  \bibfield  {author} {\bibinfo {author} {\bibfnamefont {A.}~\bibnamefont
  {Cappelli}}\ and\ \bibinfo {author} {\bibfnamefont {G.}~\bibnamefont
  {Viola}},\ }\href@noop {} {\bibfield  {journal} {\bibinfo  {journal} {J.
  Phys. A}\ }\textbf {\bibinfo {volume} {44}},\ \bibinfo {pages} {075401}
  (\bibinfo {year} {2011})}\BibitemShut {NoStop}%
\bibitem [{\citenamefont {Hori}\ \emph {et~al.}(2003)\citenamefont {Hori},
  \citenamefont {Katz}, \citenamefont {Klemm}, \citenamefont {Pandharipande},
  \citenamefont {Thomas}, \citenamefont {Vafa}, \citenamefont {Vakil},\ and\
  \citenamefont {Zaslow}}]{Mirror}%
  \BibitemOpen
  \bibfield  {author} {\bibinfo {author} {\bibfnamefont {K.}~\bibnamefont
  {Hori}}, \bibinfo {author} {\bibfnamefont {S.}~\bibnamefont {Katz}}, \bibinfo
  {author} {\bibfnamefont {A.}~\bibnamefont {Klemm}}, \bibinfo {author}
  {\bibfnamefont {R.}~\bibnamefont {Pandharipande}}, \bibinfo {author}
  {\bibfnamefont {R.}~\bibnamefont {Thomas}}, \bibinfo {author} {\bibfnamefont
  {C.}~\bibnamefont {Vafa}}, \bibinfo {author} {\bibfnamefont {R.}~\bibnamefont
  {Vakil}}, \ and\ \bibinfo {author} {\bibfnamefont {E.}~\bibnamefont
  {Zaslow}},\ }\href@noop {} {\emph {\bibinfo {title} {{Mirror symmetry}}}}\
  (\bibinfo  {publisher} {Clay mathematics monographs, Cambridge, MA},\
  \bibinfo {year} {2003})\BibitemShut {NoStop}%
\bibitem [{Note4()}]{Note4}%
  \BibitemOpen
  \bibinfo {note} {In the above calculations, the invariance is violated only
  for the temporal boundary condition. One in fact has a choice: by redefining
  $Z_{[a,b]}\to Z_{[a,b]}e^{-2\pi {i}ab}$, the partition function is now
  anomalous for spatial boundary condition. Such multiplication of the phase is
  related to (re-) assignment the U(1) charge to the ground state. See later
  discussion for more details.}\BibitemShut {Stop}%
\bibitem [{Note5()}]{Note5}%
  \BibitemOpen
  \bibinfo {note} {In the presence of both U(1) and CP symmetries, by combining
  these two symmetries, one can generate a series of CP transformations, \cite
  {Weinberg} $$ \begin {array}{lll} \protect \mathcal {U}_{\alpha } \psi ^{\
  }_L(x) \protect \mathcal {U}^{-1}_{\alpha } &=& e^{i\alpha } \psi ^{\protect
  \dag }_{R}(-x), \\ \protect \mathcal {U}_{\alpha } \psi ^{\ }_R(x) \protect
  \mathcal {U}^{-1}_{\alpha } &=& \eta e^{i\alpha } \psi ^{\protect \dag
  }_{L}(-x), \end {array} $$ where $\protect \mathcal {U}_{\alpha } = e^{i
  \alpha F_V} \protect \mathcal {CP}$. While these transformations are each
  qualified to be called CP, the inclusion of such a U(1) phase factor does not
  play any essential role in our discussion.}\BibitemShut {Stop}%
\bibitem [{\citenamefont {Qi}\ \emph {et~al.}(2008)\citenamefont {Qi},
  \citenamefont {Hughes},\ and\ \citenamefont {Zhang}}]{QiHughesZhang2007}%
  \BibitemOpen
  \bibfield  {author} {\bibinfo {author} {\bibfnamefont {X.-L.}\ \bibnamefont
  {Qi}}, \bibinfo {author} {\bibfnamefont {T.~L.}\ \bibnamefont {Hughes}}, \
  and\ \bibinfo {author} {\bibfnamefont {S.-C.}\ \bibnamefont {Zhang}},\
  }\href@noop {} {\bibfield  {journal} {\bibinfo  {journal} {Nat. Phys.}\
  }\textbf {\bibinfo {volume} {4}},\ \bibinfo {pages} {273} (\bibinfo {year}
  {2008})},\ \Eprint {http://arxiv.org/abs/0710.0730} {0710.0730} \BibitemShut
  {NoStop}%
\bibitem [{\citenamefont {{Brunner}}\ and\ \citenamefont
  {{Hori}}(2004)}]{Brunner2003}%
  \BibitemOpen
  \bibfield  {author} {\bibinfo {author} {\bibfnamefont {I.}~\bibnamefont
  {{Brunner}}}\ and\ \bibinfo {author} {\bibfnamefont {K.}~\bibnamefont
  {{Hori}}},\ }\href {\doibase 10.1088/1126-6708/2004/11/005} {\bibfield
  {journal} {\bibinfo  {journal} {Journal of High Energy Physics}\ }\textbf
  {\bibinfo {volume} {11}},\ \bibinfo {eid} {005} (\bibinfo {year} {2004})},\
  \Eprint {http://arxiv.org/abs/hep-th/0303135} {hep-th/0303135} \BibitemShut
  {NoStop}%
\bibitem [{\citenamefont {Horava}(1996)}]{Horava1996}%
  \BibitemOpen
  \bibfield  {author} {\bibinfo {author} {\bibfnamefont {P.}~\bibnamefont
  {Horava}},\ }\href@noop {} {\bibfield  {journal} {\bibinfo  {journal} {J.
  Geom. Phys.}\ }\textbf {\bibinfo {volume} {21}},\ \bibinfo {pages} {1}
  (\bibinfo {year} {1996})}\BibitemShut {NoStop}%
\bibitem [{Note6()}]{Note6}%
  \BibitemOpen
  \bibinfo {note} {In addition to the Hamiltonian, there is a chemical
  potential which appears as an operator insertion $e^{-2\pi i (b-1/2)F_V}$ in
  the partition function. Viewing this operator as a part of the partition
  function, the system with the chemical potential is in general not invariant
  under CP since $(\protect \mathcal {CP}) {F}_V {(\protect \mathcal
  {CP})}^{-1} = -{F}_V$. The only exceptions are the cases when $b=0,1/2$. When
  $b\not =0, 1/2$, the system is not invariant under CP and so we cannot make a
  projection by CP symmetry. We therefore limit ourselves to
  $b=0,1/2$.}\BibitemShut {Stop}%
\bibitem [{\citenamefont {Oshikawa}(2000{\natexlab{a}})}]{Oshikawa2000a}%
  \BibitemOpen
  \bibfield  {author} {\bibinfo {author} {\bibfnamefont {M.}~\bibnamefont
  {Oshikawa}},\ }\href@noop {} {\bibfield  {journal} {\bibinfo  {journal}
  {Phys. Rev. Lett.}\ }\textbf {\bibinfo {volume} {84}},\ \bibinfo {pages}
  {1535} (\bibinfo {year} {2000}{\natexlab{a}})}\BibitemShut {NoStop}%
\bibitem [{\citenamefont {Oshikawa}(2000{\natexlab{b}})}]{Oshikawa2000b}%
  \BibitemOpen
  \bibfield  {author} {\bibinfo {author} {\bibfnamefont {M.}~\bibnamefont
  {Oshikawa}},\ }\href@noop {} {\bibfield  {journal} {\bibinfo  {journal}
  {Phys. Rev. Lett.}\ }\textbf {\bibinfo {volume} {84}},\ \bibinfo {pages}
  {3370} (\bibinfo {year} {2000}{\natexlab{b}})}\BibitemShut {NoStop}%
\bibitem [{\citenamefont {Oshikawa}(2003)}]{Oshikawa2003}%
  \BibitemOpen
  \bibfield  {author} {\bibinfo {author} {\bibfnamefont {M.}~\bibnamefont
  {Oshikawa}},\ }\href@noop {} {\bibfield  {journal} {\bibinfo  {journal}
  {Phys. Rev. Lett.}\ }\textbf {\bibinfo {volume} {90}},\ \bibinfo {pages}
  {236401} (\bibinfo {year} {2003})}\BibitemShut {NoStop}%
\bibitem [{Note7()}]{Note7}%
  \BibitemOpen
  \bibinfo {note} {The equivalence of the two pictures, one in terms of
  twisting boundary conditions, and the other in terms of background gauge
  fields, can be established by a gauge transformation that ``unwinds'' the
  boundary conditions, and vice versa. When the electromagnetic U(1) symmetry
  happens to be anomalous, care may be required in invoking such equivalence.
  (See, for example, Ref.\ \protect \rev@citealpnum {Landsteiner2013}.) In our
  approach, when an ambiguity such as the CP eigenvalue of the ground state
  arises, we follow what we expect in the absence of anomalies. We test the
  consistency of such an assumption arising from enforcement of CP symmetry
  with the electromagnetic U(1) symmetry by inspecting the behavior of the
  partition function under the adiabatic process of flux
  insertion.}\BibitemShut {Stop}%
\bibitem [{Note8()}]{Note8}%
  \BibitemOpen
  \bibinfo {note} {It is instructive to compare the CP projected partition
  function with the partition function with P projection. Parity transformation
  acts on fermion fields as $$ \begin {array}{ccc} \protect \mathcal {P}\psi
  _L(x) \protect \mathcal {P}^{-1}&=& \eta e^{i\alpha } \psi _R(-x), \\
  \protect \mathcal {P}\psi _R(x) \protect \mathcal {P}^{-1}&=& e^{i\alpha }
  \psi _L(-x). \end {array} $$ In our fermionic edge theory, by analyzing mass
  terms, one can check that there is no topological phase protected by parity
  symmetry (of any kind) and the electromagnetic U(1) symmetry. The absence of
  topological phases can be seen from the fact that the Klein bottle partition
  function with parity projection is anomaly-free. First recall that P symmetry
  is consistent with the twisting boundary condition only when $\nu _L=-\nu
  _R$. As we require only the U(1) charge conservation, this means only $\nu
  _{L}=\nu _{R}=0$ (periodic boundary condition) or $\nu _{L}=-\nu _R=1/2$
  (antiperiodic boundary condition) are allowed. With this in mind, the
  projection works, for a given $r>0$, as $$ \begin {array}{l} \displaystyle
  \protect \mathrm {Tr}\protect \tmspace +\thinmuskip {.1667em} \left [
  \protect \mathcal {P}\protect \tmspace +\thinmuskip {.1667em} e^{-2\pi i
  (b-1/2) F_V} q^{H_R} \protect \mathaccentV {bar}016{q}^{H_L} \right ] \\
  \propto \displaystyle \DOTSB \prod@ \slimits@ _r \left [ 1 +\eta e^{4\pi i
  (b-1/2)} e^{- 2 i \alpha } (q \protect \mathaccentV {bar}016{q})^r \right ].
  \end {array} $$ Thus, the phase $\alpha $ as well as $\eta $ simply shifts
  the chemical potential $b$. In the case of P, we can freely change the time
  boundary condition $b$, but not the spatial boundary condition. Observe that
  this situation is opposite to what we had for CP symmetry. In the case of CP
  symmetry, we can freely change the space boundary condition, but not the time
  boundary condition. As before, we change $b\to b+1$ and ask if the theory is
  invariant under this large gauge transformation or not. Depending on the
  spatial boundary conditions, (periodic/antiperiodic), the partition function
  may pick up an anomalous phase. However, observe that the chemical potential
  enters in the partition function as $e^{ 4\pi i (b-1/2)}$ not $e^{2\pi i
  (b-1/2)}$. Due to this doubling, there are no anomalous phases.}\BibitemShut
  {Stop}%
\bibitem [{Note9()}]{Note9}%
  \BibitemOpen
  \bibinfo {note} {Here and in the following, the Dirac delta function $\delta
  (x-x')$ and $\protect \mathrm {sgn}\protect \tmspace +\thinmuskip {.1667em}
  (x-x')$ in the commutator should be interpreted as its periodic counter part,
  such as $\DOTSB \sum@ \slimits@ _{m \in \protect \mathbb {Z}} \delta
  (x-x'-2m\pi )$, when the system is put on a circle of circumference $2\pi
  $.}\BibitemShut {Stop}%
\bibitem [{Note10()}]{Note10}%
  \BibitemOpen
  \bibinfo {note} {The linear combination $a_1\protect \boldsymbol {\Lambda
  }_1+a_2\protect \boldsymbol {\Lambda }_2$ is non-primitive if there are some
  integer vector $\protect \boldsymbol {\Lambda }$ and integer $k>1$ such that
  $a_1\protect \boldsymbol {\Lambda }_1+a_2\protect \boldsymbol {\Lambda
  }_2=k\protect \boldsymbol {\Lambda }$.}\BibitemShut {Stop}%
\bibitem [{\citenamefont {Polchinski}(1998)}]{Polchinski98}%
  \BibitemOpen
  \bibfield  {author} {\bibinfo {author} {\bibfnamefont {J.}~\bibnamefont
  {Polchinski}},\ }\href@noop {} {\emph {\bibinfo {title} {String Theory}}}\
  (\bibinfo  {publisher} {Cambridge University Press (Cambridge, UK)},\
  \bibinfo {year} {1998})\BibitemShut {NoStop}%
\bibitem [{Note11()}]{Note11}%
  \BibitemOpen
  \bibinfo {note} {While we have determined the ground state and its charge as
  above, we could take an alternative point of view. Let us assume that we
  actually do not know, a priori, that the U(1) symmetry is anomalous. We would
  like to test if this symmetry is anomalous or not. For this purpose, we
  pretend the charge U(1) is conserved. We do so since the charge U(1) is
  classically conserved, and if so, one would guess naively that the ground
  state fermion number does not change as we adiabatically change $a$ and $b$.
  Under this assumption, what one would discover is that the partition function
  is not invariant under $a\to a+1$. Therefore, even though we started from the
  assumption that the U(1) is conserved, we run into the ``inconsistency'' in
  that the partition function is not invariant under $a\to a+1$, in stead of
  $b\to b+1$ -- we then conclude we cannot conserve the U(1) at the quantum
  level.}\BibitemShut {Stop}%
\bibitem [{\citenamefont {{Ryu}}\ and\ \citenamefont
  {{Takayanagi}}(2010{\natexlab{a}})}]{RyuTakayanagi2010a}%
  \BibitemOpen
  \bibfield  {author} {\bibinfo {author} {\bibfnamefont {S.}~\bibnamefont
  {{Ryu}}}\ and\ \bibinfo {author} {\bibfnamefont {T.}~\bibnamefont
  {{Takayanagi}}},\ }\href {\doibase 10.1016/j.physletb.2010.08.019} {\bibfield
   {journal} {\bibinfo  {journal} {Physics Letters B}\ }\textbf {\bibinfo
  {volume} {693}},\ \bibinfo {pages} {175} (\bibinfo {year}
  {2010}{\natexlab{a}})},\ \Eprint {http://arxiv.org/abs/1001.0763}
  {arXiv:1001.0763 [hep-th]} \BibitemShut {NoStop}%
\bibitem [{\citenamefont {{Ryu}}\ and\ \citenamefont
  {{Takayanagi}}(2010{\natexlab{b}})}]{RyuTakayanagi2010b}%
  \BibitemOpen
  \bibfield  {author} {\bibinfo {author} {\bibfnamefont {S.}~\bibnamefont
  {{Ryu}}}\ and\ \bibinfo {author} {\bibfnamefont {T.}~\bibnamefont
  {{Takayanagi}}},\ }\href {\doibase 10.1103/PhysRevD.82.086014} {\bibfield
  {journal} {\bibinfo  {journal} {\prd}\ }\textbf {\bibinfo {volume} {82}},\
  \bibinfo {eid} {086014} (\bibinfo {year} {2010}{\natexlab{b}})},\ \Eprint
  {http://arxiv.org/abs/1007.4234} {arXiv:1007.4234 [hep-th]} \BibitemShut
  {NoStop}%
\bibitem [{\citenamefont {Weinberg}(1996)}]{Weinberg}%
  \BibitemOpen
  \bibfield  {author} {\bibinfo {author} {\bibfnamefont {S.}~\bibnamefont
  {Weinberg}},\ }\href@noop {} {\emph {\bibinfo {title} {{The quantum theory of
  fields}}}}\ (\bibinfo  {publisher} {Cambridge University Press},\ \bibinfo
  {year} {1996})\BibitemShut {NoStop}%
\bibitem [{\citenamefont {{Landsteiner}}\ \emph {et~al.}(2013)\citenamefont
  {{Landsteiner}}, \citenamefont {{Meg{\'{\i}}as}},\ and\ \citenamefont
  {{Pe{\~n}a-Benitez}}}]{Landsteiner2013}%
  \BibitemOpen
  \bibfield  {author} {\bibinfo {author} {\bibfnamefont {K.}~\bibnamefont
  {{Landsteiner}}}, \bibinfo {author} {\bibfnamefont {E.}~\bibnamefont
  {{Meg{\'{\i}}as}}}, \ and\ \bibinfo {author} {\bibfnamefont {F.}~\bibnamefont
  {{Pe{\~n}a-Benitez}}},\ }in\ \href {\doibase 10.1007/978-3-642-37305-3_17}
  {\emph {\bibinfo {booktitle} {Lecture Notes in Physics, Berlin Springer
  Verlag}}},\ \bibinfo {series} {Lecture Notes in Physics, Berlin Springer
  Verlag}, Vol.\ \bibinfo {volume} {871},\ \bibinfo {editor} {edited by\
  \bibinfo {editor} {\bibfnamefont {D.}~\bibnamefont {{Kharzeev}}}, \bibinfo
  {editor} {\bibfnamefont {K.}~\bibnamefont {{Landsteiner}}}, \bibinfo {editor}
  {\bibfnamefont {A.}~\bibnamefont {{Schmitt}}}, \ and\ \bibinfo {editor}
  {\bibfnamefont {H.-U.}\ \bibnamefont {{Yee}}}}\ (\bibinfo {year} {2013})\ p.\
  \bibinfo {pages} {433},\ \Eprint {http://arxiv.org/abs/1207.5808}
  {arXiv:1207.5808 [hep-th]} \BibitemShut {NoStop}%
\end{thebibliography}%

\end{document}